\documentclass[11pt]{article}
\usepackage[usenames,dvipsnames,svgnames,table]{xcolor} 

\usepackage{enumitem} 
	\usepackage{rotfloat} 
	\usepackage{graphicx}
	\usepackage{pgfplots}
	\usepackage{epsfig}
	\usepackage{amssymb, amsmath}
	\usepackage{verbatim}
	\usepackage[comma]{natbib}
	\usepackage{kotex}
	\usepackage{multirow}
	\usepackage{authblk}
	\usepackage[colorlinks]{hyperref} 
	\usepackage[colorinlistoftodos, textsize=scriptsize]{todonotes} 
	\usepackage{subcaption} 

		\newtheorem{theorem}{Theorem}[section]

		\newenvironment{proof}[1][Proof]{\begin{trivlist}
				\item[\hskip \labelsep {\bfseries #1}]}{\end{trivlist}}

		\DeclareMathOperator*{\argmax}{argmax}

		\newcommand{\bea}{\begin{eqnarray*}}
			\newcommand{\eea}{\end{eqnarray*}}
		\newcommand{\bean}{\begin{eqnarray}}
			\newcommand{\eean}{\end{eqnarray}}
		
		\newcommand{\bfX}{{\bf X}}

		\newcommand{\sg}{\Sigma}
		\newcommand{\what}{\widehat}

		\newcommand{\lra}{\longrightarrow}

		\newcommand{\calG}{\mathcal{G}}

		\newcommand{\calP}{\mathcal{P}}
		\newcommand{\calQ}{\mathcal{Q}}
		\newcommand{\calS}{\mathcal{S}}

		\newcommand{\bbP}{\mathbb{P}} 
		\newcommand{\bbR}{\mathbb{R}}
		\newcommand{\bbE}{\mathbb{E}}
		
		\allowdisplaybreaks
		
\begin{document}
			
			\title{Bayesian optimal change point detection in high-dimensions}
			\author[1]{Jaehoon Kim}
			\author[1]{Kyoungjae Lee}
			\author[2]{Lizhen Lin}
			\affil[1]{Department of Statistics, Sungkyunkwan University}
			\affil[2]{Department of Mathematics, The University of Maryland}
			
			\maketitle
			\begin{abstract}

We propose the first Bayesian methods for detecting change points in high-dimensional mean and covariance structures. These methods are constructed using pairwise Bayes factors, leveraging modularization to identify significant changes in individual components efficiently. We establish that the proposed methods consistently detect and estimate change points under much  milder conditions than existing approaches in the literature. Additionally, we demonstrate that their localization rates are nearly optimal in terms of rates.  The practical performance of the proposed methods is evaluated through extensive simulation studies, where they are compared to state-of-the-art techniques. The results show comparable or superior performance across most scenarios. Notably, the methods effectively detect change points whenever signals of sufficient magnitude are present, irrespective of the number of signals. Finally, we apply the proposed methods to genetic and financial datasets, illustrating their practical utility in real-world applications.	
			\end{abstract}
			
			
			\section{Introduction}\label{sec:intro}

			Detecting change points in the mean vector or covariance matrix of multivariate data can be crucial in various scenarios.
			Changes in the mean vector may indicate shifts in the central tendency of the data,
			while changes in the covariance matrix signify alterations in the relationships among variables. 
			For instance, in financial analysis, representing the returns of various assets as a multivariate time series allows the detection of changes in the mean or covariance during significant events or financial crises. Similarly, in manufacturing, monitoring multivariate data related to product characteristics helps ensure product quality. Identifying change points in the mean vector or covariance matrix can signal deviations in production processes, enabling timely interventions to maintain quality and minimize defects.

			Suppose that we observe a sequence of random vectors $X_1, \ldots , X_n$ independently from $p$-dimensional Gaussian distributions, say $X_i \overset{ind.}{\sim} N_p(\mu_i, \sg_i)$ for $i=1,\ldots,n$.
			We say that there exist $K$ change points, $\{i_1,\ldots, i_K\}$, if $1< i_1 < \cdots < i_K < n$ and  
			\begin{align*}
				( \mu_{i_k }, \sg_{i_k }) = &\cdots = (\mu_{i_{k+1}-1} ,\sg_{i_{k+1}-1}) , \quad k=0,\ldots, K , \\
				(\mu_{i_k}, \sg_{i_k}) &\neq (\mu_{i_{k+1}}, \sg_{i_{k+1}}) , \quad k=1,\ldots, K,
			\end{align*}
			where $i_0=1$ and $i_{K+1}=n+1$.
			In general, the number of change points  $K$ and their locations are unknown. We focus on scenarios where changes occur in either the mean vector or the covariance matrix. Our primary objectives are (i) testing for the existence of change points and (ii) estimating the number and locations of change points if they are present. Notably, we consider a high-dimensional setting where the number of variables  $p$  increases with the sample size  $n$ and may even  be much larger than $n$.

			In this paper, we develop two Bayesian methods for  detecting change points in in mean and covariance structures, respectively.
			These methods build on the maximum pairwise Bayes factor approach introduced by \cite{lee2021maximum}   and \cite{lee2024bayesian}, adapting it to address the challenges of high-dimensional change point detection.
			Specifically, we select a window size $n_w > 1$ and compute the maximum pairwise Bayes factors using only the $n_w$ points before and after each central point $l$.
			Choosing an appropriate window size $n_w$ is crucial, and the optimal choices  depends on the data characteristics. 
To address this, we propose multiscale methods that consider multiple window sizes simultaneously, thereby mitigating the impact of suboptimal window size selection and achieving stable performance. Furthermore, we extend our methods to handle scenarios where both mean and covariance changes occur simultaneously, providing a comprehensive solution for change point detection in complex data structures. It is worth noting that while the pairwise Bayes factor approach has been previously applied to two-sample hypothesis testing, as in \cite{lee2021maximum} and \cite{lee2024bayesian}, extending such an approach to high-dimensional change point detection poses significant challenges, particularly on the theoretical front. Our work introduces technical innovations to establish minimax-optimal localization rates, making this the first Bayesian method to achieve such results in high-dimensional change point detection. 			
			
%

While several frequentist procedures have been proposed for change point detection in high-dimensional settings (e.g., \cite{dette2022estimating}, \cite{wang2021optimal}, \cite{grundy2020high}, \cite{wang2018high}, \cite{avanesov2018change}, and \cite{matteson2014nonparametric}), to the best of our knowledge, no Bayesian methods have been developed to address high-dimensional  change point detections.  The Bayesian methods proposed in this paper offer significant advantages, requiring weaker conditions compared to existing frequentist approaches—details of which will be elaborated later. Moreover, Bayesian change point detection inherently differs from its frequentist counterparts, with unique characteristics, making the development of such methods both important and of independent interest.

			The main contributions of this paper are threefold.
			Firstly, we propose the first Bayesian tests for change point detection in either the mean or the covariance structure in high-dimensional settings, supported by theoretical foundations. 
			We establish the consistency of the proposed Bayesian tests under both null and alternative hypotheses with mild conditions (Theorems \ref{thm:mean_changepoint} and \ref{thm:cov_changepoint}).
			Secondly, building on the proposed Bayesian tests, we develop Bayesian  change-point detection methods that estimate the location and number of change points, which consistently estimate both the number and locations of change points (Theorems \ref{thm:changepoint_est_mean} and \ref{thm:changepoint_est}). 
			The proposed methods require weaker assumptions  than existing approaches, while demonstrating nearly minimax localization rates (Theorems \ref{thm:changepoint_est_mean_lb} and \ref{thm:changepoint_est_lowerbound}).
		Thirdly, the proposed Bayesian tests and change point detection methods are scalable to high-dimensional settings and feature straightforward implementation. Leveraging the modularization framework from \cite{lee2021maximum} and \cite{lee2024bayesian}, these methods significantly enhance computational efficiency by avoiding the inversion of singular matrices, making them practical and efficient for real-world applications.

			The remaining of the paper are structured as follows: 
			Section \ref{sec:change_mean} (Section \ref{sec:change_cov}) introduces a Bayesian test for detecting change points and estimation methods for estimating the locations of these points in the mean (covariance) structure. 
			Sections \ref{sec:simulation} and \ref{sec:real} assess the practical performance of the proposed methods through simulation studies and real data analysis, respectively. 
			Concluding remarks are provided in Section \ref{sec:disc}, and proofs of the main results can be found in the supplementary material.

			\subsection{Notation}\label{subsec:notation}
			
			For any given constants $a$ and $b$, we use the notation $a\vee b$ and $a\wedge b$ to represent the maximum and minimum between the two, respectively. When considering a vector $x=(x_1,\ldots, x_p)^T$ and a positive integer $q$, the vector $\ell_q$-norm is denoted as $\|x\|_q = (\sum_{j=1}^p x_j^q )^{1/q}$. 
			When $q=0$, it means the vector $\ell_0$-norm, $\|x\|_0 = \sum_{j=1}^p I(x_j \neq 0)$.
			For positive sequences $a_n$ and $b_n$, the notation $a_n\ll b_n$ or equivalently $a_n = o(b_n)$ indicates that $a_n /b_n \lra 0$ as $n\to\infty$.  
			$a_n= O(b_n)$ means there exists a constant $C>0$ such that $a_n/b_n \le C$ for all large $n$. 
			The notation $a_n\asymp b_n$ implies both $a_n=O(b_n)$ and $b_n = O(a_n)$. 
			For a given matrix $A\in \bbR^{p\times p}$, the Frobenius norm is denoted as $\|A\|_F = ( \sum_{i=1}^p \sum_{j=1}^p a_{ij}^2 )^{1/2}$, the matrix $\ell_1$-norm as $\|A\|_1 = \sup_{x\in \bbR^p , \|x\|_1=1 } \|Ax\|_1$, the spectral norm as $\|A\| = \sup_{x\in \bbR^p , \|x\|_2=1 }\|Ax\|_2$, and the matrix maximum norm as $\|A\|_{\max} = \max_{1\le i\le j\le p}|a_{ij}|$. 
			For given $I_1 \subseteq \{1,\ldots, n\}$ and $I_2\subseteq \{1,\ldots, p\}$, denote $\bfX_{I_1, I_2}$ as a sub-matrix of $\bfX_n$ consisting of $I_1$th rows and $I_2$th columns.
			For simplicity, let $\bfX_{I_1,\cdot}:= \bfX_{I_1, 1:p}$ for any $I_1 \subseteq \{1,\ldots, n\}$.

			\section{Change point detection in mean structure}\label{sec:change_mean}

			Suppose  we observe the data $X_1,\ldots, X_n$ independently from $p$-dimensional Gaussian distribution with a common covariance. Specifically,  $X_i \overset{ind.}{\sim} N_p(\mu_{i}, \sg)$ for $i=1,\ldots,n$.
			We are interested in testing the following hypotheses: 
			\begin{eqnarray}
				\begin{split}\label{H01_change_mean}
					H_{0}^{\mu}: \mu_i = \mu_{i+1} \text{ for all } 1\le i \le n-1  \quad\text{ versus }\quad
					H_{1}^{\mu}: \text{ not } H_{0}^{\mu}.
				\end{split}
			\end{eqnarray}
			In this section, the main goals are (i) to develop a consistent Bayesian test for  change point detection problem  described in \eqref{H01_change_mean},  and (ii) to estimate the number and locations of change points, should they exist.

			\subsection{Testing the existence of change points in mean structure}\label{subsec:cpoint_H0_mean}
			
%
%
			Rather than directly testing the hypothesis for change point detection in \eqref{H01_change_mean}, we decompose the problem into a series of smaller hypothesis tests. Specifically, we begin by testing whether there is a change in the mean vector when moving from  $X_{l-1}$ to $X_{l}$, for each given $l$. For each individual hypothesis test, only the observations within an appropriate window size are considered, rather than using all observations on both sides. Finally, we address the overall change point detection problem in \eqref{H01_change_mean} by aggregating the Bayes factors from these smaller hypothesis tests.

			Let $n_w>1$ be a window size.
			For a positive integer $n_w+1\le l \le n-n_w+1$, consider the model
			\bean\label{model_mean_l}
			\begin{split}
				X_{l-n_w}, \ldots, X_{l-1} \mid \mu_{l-1} ,\sg \,\,&\overset{i.i.d.}{\sim}\,\, N_p ( \mu_{l-1} , \sg), \\
				X_{l}, \ldots, X_{l+n_w-1} \mid \mu_{l}, \sg \,\,&\overset{i.i.d.}{\sim}\,\, N_p ( \mu_{l} , \sg) ,
			\end{split}
			\eean
			where $\mu_{l-1}=(\mu_{l-1,1},\ldots,\mu_{l-1,p})^T$ and $\mu_l =(\mu_{l,1}, \ldots, \mu_{l,p})^T$,
			and the testing problem 
			\bean\label{hypo_mean_l}
			H_0^{\mu,l}: \mu_{l-1}=\mu_{l} &\text{ versus }& H_1^{\mu, l}: \mu_{l-1} \neq \mu_{l} .
			\eean
			That is, we focus on $n_w$ observations to the left of $X_{l-1}$, say $\bfX_{ (l-n_w):(l-1),\cdot}$, and $n_w$ observations to the right of $X_{l}$, say $\bfX_{l:(l+n_w-1), \cdot}$, to detect a change point.
			We call $n_w$ the  {\it window size}.
			If $H_0^{\mu,l}$ is true, we say that $l$ is a change point.
			Note that $H_0^{\mu,l}$ is true for all $n_w +1 \le l \le n-n_w+1$ if and only if $H_{0}^{\mu}$ is true.
			Therefore, we will construct Bayes factors for \eqref{hypo_mean_l} and then combine them to form the final Bayes factor for \eqref{H01_change_mean}.

			To construct the Bayes factor for hypothesis testing problem \eqref{hypo_mean_l}, we apply the maximum pairwise Bayes factor (mxPBF)  framework \citep{lee2021maximum}.
			Specifically, the marginal models for the $j$th variable in model \eqref{model_mean_l} are
			\bean\label{model_mean_l_j}
			\begin{split}
				X_{l-n_w, j}, \ldots, X_{l-1,j} \mid \mu_{l-1,j} ,\sigma_{jj} \,\,&\overset{i.i.d.}{\sim}\,\, N ( \mu_{l-1,j} , \sigma_{jj}), \\
				X_{l,j}, \ldots, X_{l+n_w-1,j} \mid \mu_{l,j}, \sigma_{jj} \,\,&\overset{i.i.d.}{\sim}\,\, N ( \mu_{l,j} , \sigma_{jj}) ,
			\end{split}
			\eean
			for $j=1,\ldots, p$, 
			where $\sigma_{jj}$ is the $(j,j)$  entry of $\sg$. 
			For given window size $n_w>1$ and $n_w+1\le l \le n-n_w+1$, the hypothesis testing problem \eqref{hypo_mean_l} can be decomposed into $p$ small testing problems 
			\bea
			H_{0,j}^{\mu, l} : \mu_{l-1,j} = \mu_{l,j} &\text{ versus }& H_{1,j}^{\mu, l} : \mu_{l-1,j} \neq \mu_{l,j} ,
			\eea
			for $j=1,\ldots ,p$, in the sense that $H_{0}^{\mu,l}$ is true if and only if $H_{0,j}^{\mu, l} $ is true for all $j=1,\ldots ,p$.
			Thus, we first calculate the {\it pairwise Bayes factors} (PBFs) for each hypothesis testing problem $H_{0,j}^{\mu, l} $ versus $H_{1,j}^{\mu, l}$, and then integrate them.
			Under $H_{0,j}^{\mu, l}: \mu_{l-1,j}= \mu_{l,j} \equiv \mu_j$, we impose the following prior $\pi_{0,j}^{\mu,l}(\mu_{j}, \sigma_{jj}) $,
			\bea
			\mu_j \mid \sigma_{jj} \,\,\sim\,\, N \Big( \bar{\bfX}_{ (l-n_w):(l+n_w-1), j}, \frac{\sigma_{jj}}{2n_w \gamma_{n_w}}  \Big), \quad
			\pi (\sigma_{jj}) \,\,\propto\,\, \sigma_{jj}^{-1},
			\eea 
			and the following prior  $\pi_{1,j}^{\mu,l}(\mu_{l-1,j},\mu_{l,j},\sigma_{jj})$ under $H_{1,j}^{\mu, l}$, 
			\bea
			\mu_{l-1,j} \mid \sigma_{jj} &\sim& N \Big( \bar{\bfX}_{(l-n_w):(l-1), j}, \frac{\sigma_{jj}}{n_w \gamma_{n_w}}  \Big), \\
			\mu_{l,j} \mid \sigma_{jj} &\sim& N \Big( \bar{\bfX}_{l:(l+n_w-1),j}, \frac{\sigma_{jj}}{n_w \gamma_{n_w}}  \Big), 
			\quad \pi (\sigma_{jj}) \,\,\propto\,\, \sigma_{jj}^{-1},
			\eea
			where $\bar{\bfX}_{a:b, j} = (b-a+1)^{-1} \sum_{i= a}^{b} X_{ij}$, $\gamma_{n_w} = (n_w \vee p)^{-\alpha}$ and $\alpha >0$.
			The above priors, originally proposed in \cite{lee2024bayesian}, have been tailored to fit our setting.
			Then, the resulting log PBF is
			\bea
			\log B_{10}^{\mu}( \bfX_{ (l-n_w):(l+n_w-1), j}  )  
			&=& \log \frac{ p(\bfX_{ (l-n_w):(l+n_w-1), j} \mid H_{1,j}^{\mu,l}) }{ p(\bfX_{ (l-n_w):(l+n_w-1), j} \mid H_{0,j}^{\mu,l}) }  \\
			&=& \frac{1}{2} \log \Big( \frac{\gamma_{n_w} }{1+\gamma_{n_w} }\Big) + n_w \log \Big( \frac{ 2n_w \what{\sigma}_{n_w,l}^2 }{ n_w \what{\sigma}_{n_w,l1}^2 + n_w \what{\sigma}_{n_w, l2}^2 } \Big)  ,
			\eea
			where 
			\begin{align*}
				p(\bfX_{ (l-n_w):(l+n_w-1), j} \mid H_{0,j}^{\mu,l}) &= \iint p(\bfX_{ (l-n_w):(l+n_w-1), j} \mid \mu_{j}, \sigma_{jj}) \pi_{0,j}^{\mu,l}(\mu_{j}, \sigma_{jj}) d\mu_j d\sigma_{jj} \\
				p(\bfX_{ (l-n_w):(l+n_w-1), j} \mid H_{1,j}^{\mu,l}) &=  \iiint p(\bfX_{ (l-n_w):(l+n_w-1), j} \mid \mu_{l-1,j}, \mu_{l,j} , \sigma_{jj}) \\ 
				&\quad\quad\quad\quad\quad \times \pi_{1,j}^{\mu,l}(\mu_{l-1,j},\mu_{l,j},\sigma_{jj}) d \mu_{l-1,j} d\mu_{l,j} d\sigma_{jj} , \\
				\what{\sigma}_{n_w,l}^2 &= (2n_w)^{-1} \bfX_{ (l-n_w):(l+n_w-1), j}^T (I_{2n_w} - H_{1_{2n_w}}) \bfX_{ (l-n_w):(l+n_w-1), j}, \\
				\what{\sigma}_{n_w,1l}^2 &= (n_w)^{-1} \bfX_{ (l-n_w):(l-1), j}^T (I_{n_w} - H_{1_{n_w}}) \bfX_{ (l-n_w):(l-1), j} , \\
				\what{\sigma}_{n_w,2l}^2 &= (n_w)^{-1} \bfX_{ l:(l+n_w-1), j}^T (I_{n_w} - H_{1_{n_w}}) \bfX_{ l:(l+n_w-1), j} .
			\end{align*}
			Here, $p(\bfX_{ (l-n_w):(l+n_w-1), j} \mid \mu_{j}, \sigma_{jj}) $ and $p(\bfX_{ (l-n_w):(l+n_w-1), j} \mid \mu_{l,j}, \mu_{l+1,j} , \sigma_{jj})$ are likelihood functions of \eqref{model_mean_l_j} under $H_{0,j}^{\mu,l}$ and $H_{1,j}^{\mu,l}$, respectively.
			Refer to \cite{lee2024bayesian} for a detailed derivation of the resulting PBF.
			Now, the mxPBF idea can be adopted to construct the Bayes factor for \eqref{hypo_mean_l}:
			\bean\label{mxPBF_change_mean_l}
			B_{\max, 10}^{\mu, l, n_w}(\bfX_n )  
			&:=& \max_{1\le j \le p} B_{10}^{\mu}( \bfX_{ (l-n_w):(l+n_w-1), j}  )  .
			\eean

			To obtain the final mxPBF for testing the existence of change points, \eqref{H01_change_mean}, we take the maximum once again over $l$ and define
			\bean\label{mxPBF_change_mean}
			B_{\max,10}^{\mu, n_w}(\bfX_n) &:=&  \max_{n_w+1 \le l \le n-n_w+1 } B_{\max, 10}^{\mu, l, n_w}(\bfX_n ) .
			\eean
			We call \eqref{mxPBF_change_mean} the mxPBF for testing the existence of change points in mean structure.
			For a given threshold $C_{\rm cp}>0$, we conclude that $H_1^{\mu}$ is true if $B_{\max,10}^{\mu, n_w}(\bfX_n) > C_{\rm cp}$, or otherwise, we conclude that $H_0^{\mu}$ is true.
			Note that the mxPBF $B_{\max,10}^{\mu, n_w}(\bfX_n)$ exceeds the specified threshold $C_{\rm cp}$ if and only if at least one PBF, $B_{10}^{\mu}(\bfX_{(l-n_w):(l+n_w-1), j})$, exceeds $C_{\rm cp}$.
			Therefore, this Bayesian test based on the mxPBF detects a change point if any component of the mean vector undergoes a significant change at some point.

			To justify the proposed Bayesian test based on the mxPBF, we establish its consistency. 
			The consistency of mxPBF, denoted $B_{\max,10}^{\mu, n_w}(\bfX_n)$,  means that  under the null, we have $H_0^{\mu}$  $B_{\max,10}^{\mu, n_w}(\bfX_n) \overset{p}{\lra} 0$, and under  the alternative $H_1^{\mu}$,  we have $\{B_{\max,10}^{\mu, n_w}(\bfX_n)\}^{-1} \overset{p}{\lra} 0$ as $n\to\infty$.
			Throughout the paper, we assume the following condition for the window size $n_w$: 
			\begin{itemize}
				\item[(A1)] $n_w \le n/2$ and $\epsilon_{n_w} := \log (n_w\vee p)/n_w = o(1)$ as $n\to\infty$.
			\end{itemize}
			Theorem \ref{thm:mean_changepoint} states that the resulting mxPBF defined in \eqref{mxPBF_change_mean} is consistent for testing the existence of change points.

			\begin{theorem}\label{thm:mean_changepoint}
				Suppose that we observe $X_i \overset{ind.}{\sim} N_p(\mu_{0i}, \sg_{0}), i=1,\ldots,n$, and consider a hypothesis testing  \eqref{H01_change_mean}.
				Assume that a window size $n_w$ satisfies condition (A1).
				Let $C_1$ and $C_2$ be constants arbitrarily close to but slightly larger than $1$ and $2$, respectively.
				\begin{itemize}
					\item[(i)] Suppose $H_{0}^{\rm \mu}$ is true, i.e., $\mu_{0i} \equiv \mu_0$ for all $1\le i \le n$.
					If $ p (n-2n_w +1) (n_w \vee p)^{-C_{2, \rm low}}  = o(1)$ for some constants $0<C_{2,\rm low} < C_2$ and
					\bean\label{alpha_mean_cond}
					\alpha &>& \frac{2 [1 +2\{C_2  \log (n_w\vee p) \}^{-1/2}] }{1-3\sqrt{C_2 \epsilon_{n_w}}} , 
					\eean
					then we have, for some constant $c>0$,
					\bea
					B_{\max,10}^{\mu, n_w}(\bfX_n) &=&O_p\big\{(n_w\vee p)^{-c} \big\}  .
					\eea
					\item[(ii)] Suppose $H_{1}^{\mu}$ is true and $\{i_k \}_{k=1}^{K_0}$ is the set of change points.
					If there exist $k$ and $j$ such that   $ (i_{k} - i_{k-1}) \wedge (i_{k+1} - i_{k}) \ge n_w$ and 
					\bean
					\frac{(\mu_{0 i_{k},j}-\mu_{0 i_{k+1},j})^2}{\sigma_{0,jj}} \,\,\ge\,\,  4\Big[ \sqrt{2C_1} + \sqrt{2C_1 + \alpha C_1 \{ 1+ (1+ 8 C_1)\epsilon_0 \} } \Big]^2 \frac{\log(n_w\vee p)}{n_w}  \,\,=:\,\, \xi^2_{n_w,p} \label{mean_change2}
					\eean
					for some constant $\epsilon_0>0$, then we have, for some constant $c'>0$,
					\bea
					\{B_{\max,10}^{\mu, n_w}(\bfX_n)\}^{-1} &=&O_p\big\{(n_w\vee p)^{-c'}\big\}  .
					\eea
				\end{itemize}
			\end{theorem}

			When $p\ge n$, the condition $p(n-2n_w +1) (n_w \vee p)^{-C_{2, \rm low}}  = p^{1- C_{2, \rm low}}(n-2n_w +1)  = o(1)$ is satisfied if $C_{2, \rm low}>2$.
			In this case, we assume that $C_{2, \rm low}$ is arbitrary close to $2$, i.e., $C_{2, \rm low} = 2+\epsilon$ for some small constant $\epsilon>0$, and $C_2 =2 +2\epsilon$.

			Recently, \cite{enikeeva2019high} proposed a test for the change point detection in mean under sparse alternatives.
			The sparse alternative means that the changes occur only in $s_0 = p^{1-\beta}$ variables among $p$ variables, for some constant $\beta\in[0,1)$.
			Note that this method does not permit changes to occur in only a fixed number of components; instead, the number of components undergoing changes must grow with the dimension of the data. 
			Furthermore, they considered an identity covariance matrix, $\sg_0 = I_p$, and assumed there is at most one change point.
			For theoretical results, they assumed $\log(n s_0)/\log\binom{p}{s_0} \lra 0$ and $\log n /\{s_0\log (p/s_0) \} \lra 0$ as $\min(n,p) \to\infty$.
			This condition implies that the number of observations, $n$, cannot be too large, which may not be a natural assumption. 
			In Theorem \ref{thm:mean_changepoint}, we do not impose such a requirement.
			Their test can detect a change point if (1) $\|\mu_{0 i_1}- \mu_{0 i_2}\|_2^2 \ge C \sqrt{p \log (p \log n)}/ \{i_1 (1- i_1/n)\}$ in {\it moderate sparsity regime} where $\beta\in[0,1/2)$ or (2) $\|\mu_{0 i_1}- \mu_{0 i_2}\|_2^2 \ge C' s_0\log (p/s_0) / \{i_1 (1- i_1/n)\}$ in {\it high sparsity regime} where $\beta\in(1/2,1)$ for some positive constants $C$ and $C'$, which turned out to be minimax or nearly minimax separation rate under the $\ell_2$-norm.
			On the other hand, we consider a general unknown covariance matrix $\Sigma_0$, allow changes to occur only at fixed locations, and assume that there can be multiple change points.
			However, our condition \eqref{mean_change2} can be seen as $\max_{1\le k \le K_0} \|\mu_{0 i_k}- \mu_{0 i_k} \|_{\max}^2 \ge C \log (n_w \vee p)/n_w$ for some constant $C>0$ when $\sg_0 = I_p$, thus it uses a different type of criterion, the maximum of the elementwise maximum norm, compared to the conditions in \cite{enikeeva2019high}. 
			Therefore, our test and the test in \cite{enikeeva2019high} have different characteristics which are not directly comparable.

			\subsection{Estimation of change points  in mean structure}\label{subsec:cpoint_est_mean}
			
			
			The Bayesian test proposed in  Section \ref{subsec:cpoint_H0_mean}  can identify the presence of change points but does not provide estimates of their locations. In this section, we propose a procedure for estimating both the number and locations of the change points based on the maximum Bayes factor (mxPBF).
			For given window size $n_w$ and threshold $C_{\rm cp}>0$, define an initial estimate of the first change point
			\bea
			\tilde{i}_1 &=&  \min \Big\{ n_w+1 \le l \le n-n_w+1 :   B_{\max,10}^{\mu, l,n_w}(\bfX_n) > C_{\rm cp}  \Big\}  , 
			\eea
			where $B_{\max,10}^{\mu, l,n_w}(\bfX_n) $ is defined at \eqref{mxPBF_change_mean_l}. 
			Suppose that we have found $\tilde{i}_1$ and assume that $i_1 \ge n_w$.
			By Theorem \ref{thm:mean_changepoint}, with probability tending to one, the mxPBF $B_{\max,10}^{\mu, l,n_w}(\bfX_n)$ is (i) not greater than a fixed threshold $C_{\rm cp}$  if $l \le i_1 - n_w$ and (ii) greater than $C_{\rm cp}$ if $l = i_1 $. 
			When $ l \in [i_1 - n_w+1 , i_1 -1]$, we cannot ensure the probabilistic behavior of mxPBF.
			What we can expect is that $\tilde{i}_1 \in [i_1 - n_w +1 , i_1]$ with high probability tending to one, which implies that $\bbP_0(\tilde{i}_1 \le i_1 \le \tilde{i}_1+n_w-1) \lra 1$ as $n\to\infty$.
			Therefore, we aim to obtain a refined estimate for the first change point, say $\what{i}_1$, by focusing on the interval $[\tilde{i}_1, \tilde{i}_1+n_w-1]$ and identifying the point with the strongest evidence for being a change point based on the mxPBF criterion:
			\bea
			\what{i}_1 &=& \argmax_{\tilde{i}_1 \le l \le \tilde{i}_1 + n_w-1}   B_{\max,10}^{\mu,l,n_w}(\bfX_n) .
			\eea
			
			After obtaining the refined estimate $\what{i}_1$, the process of detecting subsequent change points is similar to the described process for detecting the first change point.
			Specifically, the proposed process for detecting change points can be described as follows, where  $\what{i}_k$ is the refined estimate for the $k$th change point and $\what{i}_0\equiv 1$:
			\bea
			\tilde{i}_k &=&  \min \Big\{ \what{i}_{k-1} + n_w \le l \le n-n_w+1 :   B_{\max,10}^{\mu, l,n_w}(\bfX_n) > C_{\rm cp}  \Big\}  , \\
			\what{i}_k &=& \argmax_{\tilde{i}_k \le l \le \tilde{i}_k + n_w-1}   B_{\max,10}^{\mu,l,n_w}(\bfX_n) \quad \text{ for } k \ge 1 .
			\eea
			Let $\{\what{i}_k \}_{k=1}^{\what{K}}$ be the estimated change points.
			By the definition, we require $\what{i}_k - \what{i}_{k-1} \ge n_w$ for any $1\le k \le \what{K}+1$, where $\what{i}_{\what{K} +1}\equiv n$.
			This is a reasonable assumption since we need enough data between the change points to consistently detect them.

			Let $\calP^*(K_0, \delta_n, \psi_{\min}^2, \sg_0)$ be the class of distributions of $\bfX_n$ with $X_i \overset{ind.}{\sim} N_p(\mu_{0i}, \sg_0)$, $i=1,\ldots,n$, where $K_0$ change points $\{i_1,\ldots, i_{K_0}\}$ exist with $i_0=1$, $i_{K_0+1}=n$, 
			$\delta_n = \min_{0 \le k\le K_0}(i_{k+1}- i_k)$
			and
			$\psi_{\min}^2 = \min_{0 \le k\le K_0} \max_{1\le j\le p} ( \mu_{0i_k, j} - \mu_{0i_{k+1},j} )^2 / \sigma_{0,jj}$.
			Therefore, $\delta_n$ represents the minimum distance between change points, and $\psi_{\min}^2$ denotes the minimum scaled-mean difference at the change point.
			The following theorem shows that the final estimate $\{\what{i}_k \}_{k=1}^{\what{K}}$ consistently detects (i)  the number and (ii) the locations of change points with the error bound $n_w$.

			\begin{theorem}\label{thm:changepoint_est_mean}
				Suppose that $\bfX_n \sim P \in \calP^*(K_0, \delta_n, \psi_{\min}^2, \sg_0)$.
				Let  $C_{1}$ and $C_{2}$ be constants arbitrarily close to but slightly larger than 1 and 2, respectively.
				Assume that $n_w \le \delta_n$, $\psi_{\min}^2 \ge \xi^2_{n_w,p}$, $K_0 (n_w \vee p)^{-C_{1, \rm low}} = o(1)$, $p (n-2n_w +1) (n_w \vee p)^{-C_{2,\rm low}} = o(1)$, (A1) and condition \eqref{alpha_mean_cond} in Theorem \ref{thm:mean_changepoint} hold, for some constants $1<C_{1, \rm low}<C_1$ and $0<C_{2, \rm low}<C_2$, where $\xi^2_{n_w,p}$ is defined in \eqref{mean_change2}.
				Then we have, as $n\to\infty$,
				\bea
				\bbP_0 \Big( \what{K} = K_0 , \,\,  \max_{1\le k \le K_0} |i_k - \what{i}_k| \le n_w  \Big) &\lra& 1 .
				\eea
			\end{theorem}
			
			\cite{wang2018high} proposed a change point estimation method for high-dimensional mean vectors based on a sparse projection.
			In their settings, they assumed the data are independently generated from a $p$-dimensional Gaussian distribution with mean vectors $\{\mu_{0i}\}_{i=1}^n$ and a common diagonal covariance matrix $\sg_0 = \sigma^2 I_p$ for some known $\sigma^2>0$. 
			Let $\max_{1\le k \le K_0}\| \mu_{0 i_{k}} - \mu_{0 i_{k+1}}\|_0   \le s_0 $ and  $\tau \le n^{-1} \min_{0 \le k\le K_0} (i_{k+1} - i_k )$.
			To obtain the localization rate for change point detection (in their Theorem 2), they assumed 
			\bean\label{wang_dense}
			\min_{1 \le k\le K_0} \| \mu_{0 i_{k}} - \mu_{0 i_{k+1}} \|_2^2 
			&\ge& C\sigma^2  \frac{ s_0 \log ( np  )}{n \tau^2 }
			\eean
			for some constant $C>0$.
			Note that $n\tau$ in their notation corresponds to $\delta_n$ in our notation, so the rate of the lower bound in \eqref{wang_dense} can be written as $ \sigma^2 s_0 \log(np)/ ( \delta_n \tau )$.
			Furthermore, \eqref{wang_dense} implies $\min_{1 \le k\le K_0} \| \mu_{0 i_{k}} - \mu_{0 i_{k+1}} \|_{\max}^2 \ge C   \sigma^2 \log (n p)/(\delta_n \tau)$.
			On the other hand, when $n_w\le \delta_n$ and $\sigma_{0,jj} \equiv \sigma^2$ for all $j$, our condition $\psi_{\min}^2 \ge \xi^2_{n_w,p}$ implies that $\min_{1 \le k\le K_0} \| \mu_{0 i_{k}} - \mu_{0 i_{k+1}} \|_{\max}^2 \ge C_\star   \sigma^2 \log (n_w \vee p)/\delta_n$ for some constant $C_\star>0$.
			Thus, condition $\psi_{\min}^2 \ge \xi^2_{n_w,p}$ is much weaker than condition \eqref{wang_dense} used in \cite{wang2018high}.
			The difference between those conditions increases as $\tau$ approaches $0$.

			We say that the {\it localization rate} for change point detection is $\epsilon_n^\star$ if 
			\bea 
			\bbP_0 \Big( \what{K} = K_0 , \,\,  \max_{1\le k \le K_0} \frac{|i_k - \what{i}_k|}{n} \le \epsilon_n^\star  \Big)  &\lra& 1 .
			\eea 
			Thus, Theorem \ref{thm:changepoint_est_mean} implies that the localization rate is equal to  $n_w/n$.
			The next theorem says that, under certain conditions, this is indeed a nearly minimax rate up to a factor of $\log(n_w \vee p)$.
			
			\begin{theorem}\label{thm:changepoint_est_mean_lb}
				Let $\calP^*_1 =\calP^*_1(K_0=1, \delta_n, \psi_{\min}^2, \sg_0)$ be the class of distributions of $\bfX_n$ with $X_i\overset{ind.}{\sim} N_p(\mu_{0i} , \sg_0)$, where only one change point $i_1$ exists and $\|\mu_{0i_1 } - \mu_{0i_0}\|_0 = 1$.
				Suppose 
				$1/\delta_n \le \psi_{\min}^2 \le 1$. 
				Then, we have 
				\bea
				\inf_{\what{i}_1} \sup_{P \in \calP_1^*}  \frac{1}{n} \bbE_P |\what{i}_1- i_1| &\ge& \frac{1}{16 n \psi_{\min}^2 }  .
				\eea
			\end{theorem}
			
			If condition $\psi_{\min}^2 \ge \xi^2_{n_w,p}$ in Theorem \ref{thm:changepoint_est_mean} holds, particularly when equality holds, it implies that $\psi_{\min}^2 \equiv C_\star \log(n_w \vee p)/n_w$ for some constant $C_\star >0$, and  the rate of the lower bound in Theorem \ref{thm:changepoint_est_mean_lb} becomes $n_w / \{n\log(n_w\vee p)\}$.
			Thus, in this case, the obtained rate in Theorem \ref{thm:changepoint_est_mean} is nearly minimax-optimal  up to a factor of $\log (n_w\vee p)$.

			

			\section{Change point detection in covariance structure}\label{sec:change_cov}
			
			Suppose that we observe the data $X_1,\ldots,X_n$ independently from $p$-dimensional Gaussian distributions with mean zero, that is, $X_i \overset{ind.}{\sim} N_p(0, \sg_i)$ for $i=1,\ldots,n$.
			Consider a hypothesis testing 
			\begin{eqnarray}
				\begin{split}\label{H01_change_cov}
					H_{0}^{\sg}: \sg_i = \sg_{i+1} \text{ for all } 1\le i \le n-1  \quad\text{ versus }\quad
					H_{1}^{\sg}: \text{ not } H_{0}^{\sg} .
				\end{split}
			\end{eqnarray}
			The main goals in this section are (i) to develop a consistent Bayesian test for the change point detection problem \eqref{H01_change_cov} and (ii) to estimate the number and locations of change points if they exist.

			\subsection{Testing the existence of change points in covariance structure}\label{subsec:cpoint_H0}

			Similar to the mean vector case in Section \ref{subsec:cpoint_H0_mean}, we approach the hypothesis testing by decomposing it into multiple small testing problems.
			For a given window size $n_w>1$ and a positive integer $n_w+1\le l \le n-n_w+1$, consider the following model
			\bean\label{model_cov_l}
			\begin{split}
				X_{l-n_w}, \ldots, X_{l-1} \mid \sg_{l-1} \,\,&\overset{i.i.d.}{\sim}\,\, N_p ( 0 , \sg_l), \\
				X_{l}, \ldots, X_{l+n_w-1} \mid \sg_{l} \,\,&\overset{i.i.d.}{\sim}\,\, N_p ( 0 , \sg_{l+1})
			\end{split}
			\eean
			and the testing problem 
			\bean\label{hypo_cov_l}
			H_0^{\sg, l}: \sg_{l-1}=\sg_{l} &\text{versus}& H_1^{\sg, l}: \sg_{l-1} \neq \sg_{l} .
			\eean 
			Thus, if $H_0^{\sg, l}$ is true, then $l$ is a change point.
			Note that $H_0^{\sg, l}$ is true for all $n_w+1 \le l \le n-n_w+1$ if and only if $H_0^{\sg}$ is true.
			Hence, we will construct Bayes factors for \eqref{hypo_cov_l} and then combine them to form the final Bayes factor for \eqref{H01_change_cov}.

			To construct the Bayes factor for \eqref{hypo_cov_l}, we again apply the mxPBF idea \citep{lee2021maximum}. 
			Let $\sg_l = (\sigma_{l,ij})$, $a_{l,ij} = \sigma_{l,ij}/\sigma_{l,jj}$ and $\tau_{l,ij}= \sigma_{l,ii}(1- \rho_{l,ij}^2)$ for any $n_w+1\le l \le n-n_w+1$, $1\le i\neq j \le p$, where $R_{l} = (\rho_{l,ij})$ is the correlation matrix based on $\sg_l$.
			For a given pair $(i,j)$ with $1\le i \neq j \le p$, the conditional models from model \eqref{model_cov_l} can be written as 
			\bean\label{model_cov_l_ij}
			\begin{split}
				\bfX_{(l-n_w):(l-1), i} \mid \bfX_{(l-n_w):(l-1), j}, a_{l-1, ij} , \tau_{l-1, ij}  \,\,&\sim\,\, N_{n_w}( a_{l-1,ij} \bfX_{(l-n_w):(l-1), j} ,  \tau_{l-1, ij} I_{n_w} )    \\
				\bfX_{l:(l +n_w-1), i} \mid \bfX_{l:(l +n_w-1), j}, a_{l, ij} , \tau_{l, ij}    \,\,&\sim\,\, N_{n_w}( a_{l,ij} \bfX_{l:(l +n_w-1), j} ,  \tau_{l, ij} I_{n_w} )  .
			\end{split}
			\eean
			For a given integer $n_w+1\le l \le n-n_w+1$, the hypothesis testing problem \eqref{hypo_cov_l} can be decomposed into the following small testing problems, 
			\bea
			H_{0,ij}^{\sg, l} : a_{l-1,ij} = a_{l,ij} \text{ and } \tau_{l-1,ij}=\tau_{l,ij}  &\text{ versus }& H_{1,ij}^{\sg, l} : \text{ not } H_{0,ij}^{\sg, l} ,
			\eea
			for $1\le i \neq j \le p$, in the sense that $H_0^{\sg, l}$ is true if and only if $H_{0,ij}^{\sg, l}$ is true for all pairs $(i,j)$ such that $1\le i\neq j\le p$.
			Let $IG(a,b)$ represent the inverse-gamma distribution with shape parameter $a$ and rate parameter $b$.
			We will calculate the PBFs for each hypothesis $H_{0,ij}^{\sg, l}$ versus $H_{1,ij}^{\sg, l}$.
			Let $\gamma_{n_w} = (n_w\vee p)^{-\alpha}$ for some $\alpha>0$ as before.
			Under $H_{0,ij}^{\sg, l}$, we use the following prior $\pi_{0,ij}^{\sg, l}(a_{ij}, \tau_{ij})$, 
			\bea
			a_{ij} \mid \tau_{ij} \,\,\sim\,\, N\Big( \what{a}_{l, ij}, \frac{\tau_{ij}}{\gamma_{n_w} \|\bfX_{ (l-n_w):(l+n_w-1),j}\|_2^2   } \Big), &&
			\tau_{ij} \,\,\sim\,\, IG(a_0, b_{0,ij}) ,
			\eea
			where $a_{ij} = a_{l,ij}=a_{l+1,ij}$ and $\tau_{ij} = \tau_{l-1,ij} = \tau_{l,ij}$, and the following prior $\pi_{1,ij}^{\sg, l}(a_{l-1,ij} , a_{l,ij}, \tau_{l-1,ij}, \tau_{l,ij})$ under $H_{1,ij}^{\sg,l}$,
			\begin{align*}
				a_{l-1,ij}\mid \tau_{l-1,ij} \,\,\sim\,\, N\Big( \what{a}_{1 l-1,ij}, \frac{\tau_{l-1,ij}}{\gamma_{n_w}\|\bfX_{ (l-n_w):(l-1),j}\|_2^2} \Big), &\quad a_{l,ij}\mid \tau_{l,ij} \,\,\sim\,\, N\Big( \what{a}_{2l,ij}, \frac{\tau_{l,ij}}{\gamma_{n_w}\|\bfX_{l:(l+n_w-1), j}\|_2^2} \Big), \\
				\tau_{l-1,ij} \,\,\sim\,\, IG(a_0, b_{01 ,ij}), &\quad \tau_{l,ij} \,\,\sim\,\, IG(a_0, b_{02 ,ij})
			\end{align*}
			where  
			\begin{align*}
				\what{a}_{l,ij} &= \bfX_{ (l-n_w):(l+n_w-1),i}^T \bfX_{ (l-n_w):(l+n_w-1),j} / \| \bfX_{ (l-n_w):(l+n_w-1),j} \|_2^2, \\ 
				\what{a}_{1l-1,ij} &= \bfX_{ (l-n_w):(l-1),i}^T \bfX_{ (l-n_w):(l-1),j}/ \| \bfX_{ (l-n_w):(l-1),j}\|_2^2 , \\
				\what{a}_{2l,ij} &= \bfX_{ l:(l+n_w-1),i}^T \bfX_{ l:(l+n_w-1),j} / \| \bfX_{ l:(l+n_w-1),j}\|_2^2 .
			\end{align*}
			Then, the resulting log PBF is 
			\begin{align*}\label{logPBF_cov}
				&\log B_{10}^{\sg}( \bfX_{(l-n_w):(l+n_w-1),i }, \bfX_{(l-n_w):(l+n_w-1),j } )  \\
				&= \log \frac{p ( \bfX_{(l-n_w):(l+n_w-1),i }\mid \bfX_{(l-n_w):(l+n_w-1),j } , H_{0,ij}^{\sg, l}) }{p(\bfX_{(l-n_w):(l+n_w-1),i }\mid \bfX_{(l-n_w):(l+n_w-1),j } , H_{1,ij}^{\sg, l} ) }  \\
				&= \frac{1}{2} \log \Big( \frac{\gamma_{n_w} }{1+\gamma_{n_w} }\Big) + 2\log \Gamma \Big( \frac{n_w}{2}+a_0 \Big)  - \log \Big\{ \Gamma  ( n_w +a_0 ) \Gamma(a_0) \Big\} + a_0\log \Big( \frac{b_{01,ij} b_{02,ij} }{b_{0,ij} } \Big) \\
				&- \Big(\frac{n_w}{2}+a_0 \Big) \log \big( b_{01,ij} + \frac{n_w}{2} \what{\tau}_{1l-1,ij} \big) 
				- \Big(\frac{n_w}{2}+a_0 \Big) \log \big( b_{02,ij} + \frac{n_w}{2} \what{\tau}_{2l,ij} \big) \\
				&+ (n_w + a_0) \log \big( b_{0,ij} + \frac{n_w}{2} \what{\tau}_{l,ij} \big) ,
			\end{align*}
			where $\Gamma$ is the gamma function, 
			\begin{align*}
				\what{\tau}_{l,ij} &= \frac{1}{2n_w} \bfX_{ (l-n_w):(l+n_w-1),i}^T (I_{2n_w} - H_{\bfX_{ (l-n_w):(l+n_w-1),j}} )\bfX_{ (l-n_w):(l+n_w-1),i} , \\ 
				\what{\tau}_{1l-1,ij} &= \frac{1}{n_w} \bfX_{ (l-n_w):(l-1),i}^T (I_{n_w} - H_{\bfX_{ (l-n_w):(l-1),j}} )\bfX_{ (l-n_w):(l-1),i} ,\\
				\what{\tau}_{2l,ij} &= \frac{1}{n_w} \bfX_{ l:(l+n_w-1),i}^T (I_{n_w} - H_{\bfX_{ l:(l+n_w-1),j}} )\bfX_{ l:(l+n_w-1),i} ,
			\end{align*}
			and 
			\begin{align*}
				& p \big( \bfX_{(l-n_w):(l+n_w-1),i }\mid \bfX_{(l-n_w):(l+n_w-1),j } , H_{0,ij}^{\sg, l}\big) \\
				&=   \iint  p \big( \bfX_{(l-n_w):(l+n_w-1),i }\mid \bfX_{(l-n_w):(l+n_w-1),j } , a_{ij}, \tau_{ij}\big) \pi_{0,ij}^{\sg,l}(a_{ij},\tau_{ij}) da_{ij} d\tau_{ij} ,  \\
				& p \big(  \bfX_{(l-n_w):(l+n_w-1),i }\mid \bfX_{(l-n_w):(l+n_w-1),j } , H_{1,ij}^{\sg, l}\big) \\
				&= \iiiint p \big( \bfX_{(l-n_w):(l+n_w-1),i }\mid \bfX_{(l-n_w):(l+n_w-1),j } , a_{l-1,ij}, a_{l,ij}, \tau_{l-1,ij}, \tau_{l,ij}\big)  \\
				&\quad\quad \times \,\, \pi_{1,ij}^{\sg,l}(a_{l-1,ij}, a_{l,ij}, \tau_{l-1,ij}, \tau_{l,ij}) d a_{l-1,ij} da_{l,ij} d\tau_{l-1,ij} d\tau_{l,ij} .
			\end{align*}
			Note that $p \big( \bfX_{(l-n_w):(l+n_w-1),i }\mid \bfX_{(l-n_w):(l+n_w-1),j } , a_{ij}, \tau_{ij}\big)$ and 
			$$p \big( \bfX_{(l-n_w):(l+n_w-1),i }\mid \bfX_{(l-n_w):(l+n_w-1),j } , a_{l-1,ij}, a_{l,ij}, \tau_{l-1,ij}, \tau_{l,ij}\big) $$ 
			are likelihood functions of \eqref{model_cov_l_ij} under $H_{0,ij}^{\sg,l}$ and $H_{1,ij}^{\sg,l}$, respectively.
			Refer to \cite{lee2024bayesian} for a detailed derivation of the resulting PBF.
			Then the mxPBF for \eqref{hypo_cov_l} is defined as
			\bea
			B_{\max,10}^{\sg, l,n_w}(\bfX_n) &:=& \max_{1\le i \neq j \le p} B_{10}^{\sg}( \bfX_{ (l-n_w):(l+n_w-1), i},  \bfX_{ (l-n_w):(l+n_w-1), j} )  .
			\eea
			Similarly to the case of the mean vector, the final mxPBF for testing the existence of change points in the covariance structure is defined as follows:
			\bean\label{Bayes_test_cpoint}
			B_{\max,10}^{\sg, n_w}(\bfX_n) \,\,:=\,\, \max_{n_w+1 \le l \le n-n_w+1 } B_{\max,10}^{\sg, l,n_w}(\bfX_n).
			\eean
			For given threshold $C_{\rm cp}>0$, we conclude that $H_{1}^{\sg}$ in \eqref{H01_change_cov} is true if $B_{\max,10}^{\sg, n_w}(\bfX_n) > C_{\rm cp}$, or otherwise, we conclude that $H_{0}^{\sg}$ is true.

			We now introduce the following sufficient conditions to guarantee consistency of the mxPBF in \eqref{Bayes_test_cpoint}.
			
			\begin{itemize}
				
				\item[(A2)] $\min_{i\neq j}\tau_{0 ,ij} \gg \{\log (n_w\vee p)\}^{-1}$.
				
				\item[(A3)] 
				There exist an index $k$ and a pair $(i,j)$ with $i\neq j$ satisfying 
				\bea
				\{\log (n_w\vee p)\}^{-1} \ll \tau_{0i_{k},ij}\wedge \tau_{0i_{k+1},ij} &\le& \tau_{0i_{k},ij}\vee \tau_{0i_{k+1},ij} \ll (n_w \vee p)
				\eea
				and  
				\bea
				\frac{\tau_{0i_{k},ij}}{\tau_{0i_{k+1},ij}} \vee \frac{\tau_{0i_{k+1},ij}}{\tau_{0i_k,ij}} &>& \frac{1+ C_{\rm bm}\sqrt{\epsilon_{n_w}}}{1- 4 \sqrt{C_1\epsilon_{n_w}}}  
				\eea
				for the constant $C_1>1$ and some constant $C_{\rm bm}^2 > 8 (\alpha+1)$.
				
				\item[(A3$^\star$)] 
				There exist an index $k$ and a pair $(i,j)$ with $i\neq j$ such that $\sigma_{0i_{k},ii} \vee \sigma_{0i_{k+1},ii} \ll (n_w\vee p)$,
				\bea
				(a_{0i_{k},ij} - a_{0i_{k+1},ij})^2 &\ge& \frac{25}{2} C_1 \sum_{u=k}^{k+1} \Big\{\frac{\tau_{0i_u,ij}\epsilon_{n_w} }{\sigma_{0i_u,jj}(1-2\sqrt{C_1 \epsilon_{n_w}}) }  \Big\} ,  \\
				(a_{0i_{k},ij} - a_{0i_{k+1},ij})^2 &\ge& \frac{10n_w}{n_w+a_0} \sum_{u=k}^{k+1} \Big\{ \frac{\epsilon_{n_w}}{ \sigma_{0i_u,jj}  (1-2\sqrt{C_1\epsilon_{n_w}}) }  \Big\}  \\
				&& \times\, \Big[ \frac{b_{0 i_k,ij}}{\log (2n_w\vee p)} + \Big(\sum_{u=k}^{k+1} \sigma_{0i_u,ii}(1+4\sqrt{C_1\epsilon_{n_w}}) + \frac{b_{0l,ij}}{n_w} \Big) C_{{\rm bm}, a}   \Big] 
				\eea
				for some constant $C_{{\rm bm}, a} > a_0+\alpha +1$ and $C_1>1$.
			\end{itemize}
			
			Condition (A2) is a sufficient condition for consistency of the mxPBF in \eqref{Bayes_test_cpoint} under the null $H_0^{\sg}$.
			When there is no change point, it allows the cases where possibly $\sigma_{0,ii}\to 0$ and $\rho_{0,ij}^2 \to 1$ as $p\to\infty$ at certain rates.
			When $H_1^{\sg}$ is true, suppose there exist change points $\{ i_1,\ldots, i_{K_0}\}$.
			Conditions (A3) and (A3$^\star$) represent the minimum lower bound condition of the signal under the alternative $H_1^{\sg}$.
			In fact, they can be roughly understood as the squared maximum standardized-difference condition between $\sg_{0i_k}$ and $\sg_{0i_{k+1}}$, say 
			\bean\label{mx_std_diff_cpoint}
			\min_{1\le k \le K_0} \max_{1\le i\le j\le p}\frac{ (\sigma_{0i_{k},ij} - \sigma_{0i_{k+1},ij})^2 }{ \sigma_{0i_{k},ii}\sigma_{0i_{k},jj} + \sigma_{0i_{k+1},ii}\sigma_{0i_{k+1},jj} } &\ge& C_\star  \, \frac{\log (n_w \vee p)}{n_w} 
			\eean
			for some constant $C_\star>0$.
			Specifically, let 
			\bean\label{cond_Ln}
			\begin{split}
				\max_{1\le k \le K_0}\max_{1\le i\neq j \le p}\rho_{0i_k,ij}^2 &\le 1 - c_0 \\
				\{ \log (n_w \vee p)\}^{-1} \le \min_{1\le k \le K_0}\min_{1\le i \le p} \sigma_{0i_k,ii} &\le \max_{1\le k \le K_0}\max_{1\le i \le p} \sigma_{0i_k,ii} \le (n_w \vee p)
			\end{split}
			\eean
			for some small constant $c_0>$.
			If \eqref{mx_std_diff_cpoint} and \eqref{cond_Ln} hold for $(\sg_{0i_k},\sg_{0 i_{k+1}})$ and some $1\le k\le K_0$, then condition (A3) or (A3$^\star$) is met.
			For a detailed statement, see \cite{lee2024bayesian} (Lemma 3.1 in Supplementary material).

			Theorem \ref{thm:cov_changepoint} says that we can consistently detect the existence of change points in the covariance structure based on the mxPBF in \eqref{Bayes_test_cpoint}.

			\begin{theorem}\label{thm:cov_changepoint}
				Suppose that we observe $X_i \overset{ind.}{\sim} N_p(0, \sg_{0i}), i=1,\ldots,n$, and consider a hypothesis testing problem \eqref{H01_change_cov}.
				Assume that a window size $n_w$ satisfies condition (A1).
				Let $C_3$ be constant arbitrarily close to but slightly larger than 3.
				\vspace{-.2cm}
				\begin{itemize}
					\item[(i)] Suppose $H_{0}^{\sg}$ is true, i.e., $\sg_{0i} \equiv \sg_0$ for all $1\le i \le n$.
					If $\alpha > 6C_3 $, $(n-2n_w +1) p^2 (n_w \vee p)^{-C_{3,\rm low}} = o(1)$ for some positive constants $C_{3,\rm low} < C_3$ and condition (A2) holds, then for some constant $c>0$,
					\bea
					B_{\max,10}^{\sg, n_w}(\bfX_n) &=&O_p\{(n_w\vee p)^{-c}\}  .
					\eea
					\item[(ii)] Suppose $H_{1}^{\sg}$ is true and $\{i_k \}_{k=1}^{K_0}$ is the set of change points.
					If there exist $k$ and a pair $(i,j)$ such that  $(i_{k} - i_{k-1}) \vee (i_{k+1} - i_{k}) \ge n_w$ and satisfy condition (A3) or (A3$^\star$), then for some constant $c'>0$,
					\bea
					\{B_{\max,10}^{\sg, n_w}(\bfX_n)\}^{-1} &=&O_p\{(n_w\vee p)^{-c'}\}  .
					\eea
				\end{itemize}
			\end{theorem}

			When $p \ge n$, the condition $(n-2n_w +1) p^2 (n_w \vee p)^{-C_{3,\rm low}}=(n-2n_w +1) p^{2-C_{3,\rm low}}  = o(1)$ is satisfied if $C_{3,\rm low}>3$.
			In this case, we  assume that $C_{3,\rm low}$ is arbitrarily close to $3$, i.e., $C_{3,\rm low} = 3+\epsilon$ for some small constant $\epsilon>0$, and $C_3 = 3+2\epsilon$.

			Theorem \ref{thm:cov_changepoint} assumes much weaker conditions compared with the existing literature.
			\cite{avanesov2018change} suggested a testing procedure based on bootstrap and multiscale idea for a change point detection.
			A set of window sizes ${\mathcal{N}}$ was used to construct the test statistic.
			They investigated consistency of the test in high-dimensional settings under the sparsity assumption for precision matrices, where $d= \max_{1\le i\le n} \max_{1\le k\le p} \sum_{j=1}^p I( (\Sigma_{0i}^{-1})_{jk} \neq 0 )$.
			Let $s \le n$ be the number of subsamples for constructing bootstrap statistics, and let $n_{-}$ and $n_{+}$ be the minimum and maximum window sizes belonging to $\mathcal{N}$, respectively.
			To prove consistency of the proposed test, \cite{avanesov2018change} assumed $d |{\mathcal{N}}|^3 \{\log(np) \}^{10} = o( \{s \wedge n_{-}^2 \}^{1/4} )$ and $d|{\mathcal{N}}| \{\log(np)\}^7 = o(\{s \vee n_{-} \}^{1/2})$, which require the true precision matrices to be sufficiently sparse.
			Furthermore, irrepresentability conditions of the precision matrix were also assumed for the existence of a consistent estimator for the precision matrices.
			Hence, our test based on the mxPBF requires much weaker conditions on $(n,p)$ for consistency, while allowing multiple change points and not assuming any sparsity condition on covariance (or precision) matrices.

			\subsection{Estimation of change points  in covariance structure}\label{subsec:cpoint_est}
			
			In this section, we suggest a procedure for estimating the number and locations of change points based on the mxPBF in Section \ref{subsec:cpoint_H0}.
			For given window size $n_w$ and threshold $C_{\rm cp}>0$, let $\what{i}_0 \equiv 1$ and
			\bea
			\tilde{i}_k &=& \min \Big\{ \what{i}_{k-1} + n_w \le l \le n-n_w +1:  B_{\max,10}^{\sg, l,n_w}(\bfX_n) > C_{\rm cp}  \Big\}   ,  \\
			\what{i}_k &=& \argmax_{\tilde{i}_k \le l\le \tilde{i}_k +n_w -1}   B_{\max,10}^{\sg, l,n_w}(\bfX_n)   \,\, \text{ for }  k \ge 1 .
			\eea
			Similar to Section \ref{subsec:cpoint_est_mean}, let $\{\what{i}_k \}_{k=1}^{\what{K}}$ be the estimated change points.

			Let $\calP^*_{\rm cov}(K_0, \delta_n)$ be the class of distributions of $\bfX_n$ with $X_i \overset{ind.}{\sim} N_p(0, \sg_{0i})$, $i=1,\ldots,n$, where $K_0$ change points $\{i_1,\ldots, i_{K_0}\}$ exist with $i_0=1$, $i_{K_0+1}=n$, 
			$\delta_n = \min_{0 \le k\le K_0}(i_{k+1}- i_k)$
			and
			conditions (A3) or (A3$^\star$) hold for all $k=1,\ldots, K_0$.
			The following theorem shows that $\{\what{i}_k \}_{k=1}^{\what{K}}$ consistently detect (i)  the number and (ii) the locations of change points with the error bound $n_w$.

			\begin{theorem}\label{thm:changepoint_est}
				Suppose that $\bfX_n \sim P \in \calP^*_{\rm cov}(K_0, \delta_n)$.
				Let $C_1$ and $C_3$ be constants arbitrarily close to but slightly larger than 1 and 3, respectively.
				Assume that $n_w \le \delta_n$, $K_0 (n_w \vee p)^{-C_{1, \rm low}} = o(1)$, $ p^2 (n-2n_w +1)(n_w \vee p)^{-C_{3,\rm low}} = o(1)$, $\alpha > 6C_3 $ and condition (A1) holds for some constants $1<C_{1, \rm low}<C_1$ and $0<C_{3,\rm low} < C_3$.
				Then we have 
				\bea
				\bbP_0 \Big( \what{K} = K_0 , \,\,  \max_{1\le k \le K_0} |i_k - \what{i}_k| \le n_w  \Big) &\lra& 1 .
				\eea
			\end{theorem}
			
			We compare Theorem \ref{thm:changepoint_est} with existing results in the literature.
			\cite{wang2021optimal} proposed two procedures to detect change points in covariance structures for sub-Gaussian random vectors. 
			Their second procedure, called WBSIP, has a nearly minimax localization rate, $\epsilon_n =C B^4 \kappa^{-2} \log n$, where $\kappa = \min_k \|\sg_{0 i_k} - \sg_{0 i_{k+1}}\| $ and $\max_k \|\sg_{0 i_k} \|\le 2B^2$.
			To obtain theoretical results, they assumed that for any $\xi>0$, there exists a constant $C>0$ such that $\min_k \|\sg_{0 i_k} - \sg_{0 i_{k+1}}\|^2 \ge C p (\log n )^{1+\xi} B^4 /\delta_n$.
			If we assume that $n_w =  \delta_n$ and $B$ is a fixed constant, it becomes $\min_k \|\sg_{0 i_k} - \sg_{0 i_{k+1}}\|^2 \ge C' p (\log n )^{1+\xi} /n_w$ for some constant $C'>0$. 
			On the other hand, under the same conditions, conditions (A3) and (A3$^\star$) are satisfied if $\min_k \| \sg_{0 i_{k}} - \sg_{0 i_{k+1}} \|_{\max}^2 \ge C \log(n_w \vee p)/ n_w$ for some constant $C>0$.
			Thus their condition is stronger than ours if $\|\sg_{0 i_k} - \sg_{0 i_{k+1}}\| \le \sqrt{p} \|\sg_{0 i_k} - \sg_{0 i_{k+1}}\|_{\max}$ and $(\log n)^{1+\xi}  \ge \log (n_w \vee p)$, but a direct comparison is difficult due to the different criteria used to measure the magnitude of the change.

			Recently, \cite{dette2022estimating} developed a two-step approach to detect a change point in covariance structures when there is only one change point and each component of an observation is a sub-Gaussian random variable.
			Let $i_1$ be the true change point and $\lambda$ be a lower bound for the smallest nonzero entry in $|\sg_{0 i_0} - \sg_{0 i_1}|$.
			They obtained the localization rate $\Delta /n$ (Corollary 3.2 in \cite{dette2022estimating}) for some $\Delta >0$ and $c'>0$ such that $p^2 n = o( e^{ c' \lambda \Delta} )$.
			To obtain this result, they assumed $i_1 > n^e$, $\lambda \ge C(\tau/n)^{1/2} \max \{ n^2 /(n-i_1)^2, n^2 /i_1^2 \}$, $p = o(n^{M_1})$ and $\max ( \|\sg_{0 i_0}\|_{\max},  \|\sg_{0 i_1}\|_{\max}) \le M_2$ for some constants $0<e<1$, $C>0$, $M_1>0$ and $M_2>0$, where $\tau$ is a threshold for the dimension reduction in their method.
			If we further assume that $c' \lambda >3$, $\tau = C\log (n \vee p)$ and $i_1 \asymp n$, then their localization rate becomes $\log (n\vee p)/n$.
			Note that if we  set $n_w \asymp \log (n\vee p)$, the localization rate in Theorem \ref{thm:changepoint_est} coincides with $\log (n\vee p)/n$. 
			Furthermore, $c'\lambda >3$ implies that the {\it smallest} nonzero entry in $|\sg_{0 i_0} - \sg_{0 i_1}|$ is larger than a fixed positive constant, which is stronger than our condition assuming the {\it maximum} nonzero entry in $|\sg_{0 i_0} - \sg_{0 i_1}|$ is larger than $C \log (n_w\vee p)/n_w$.
			Therefore, our result establishes the localization rate of the proposed change point detection method under relatively weaker assumptions compared to existing literature.

			Theorem \ref{thm:changepoint_est} implies that the localization rate is equal to  $n_w/n$.
			We will show that, in the following two theorems, this rate is minimax-optimal up to a factor of $\log (n_w\vee p)$ under certain conditions.

			\begin{theorem}\label{thm:changepoint_est_lowerbound_diverge}
				For given constant $C_d>0$ and $0<\tilde{\sigma}_{{\rm diff}}<3/4$, define a class of joint distributions of $(X_1,\ldots,X_n)$,
				\bea
				\calG(\tilde{\sigma}_{{\rm diff}}, C_d)   &:=&  \Big\{ P : X_i \overset{ind.}{\sim} N_p(0, \sg_{0i}),\, i=1,\ldots,n, \\
				&& \quad \text{there is only one change point } i_1 \text{ and it satisfies }\, \frac{C_d \log p}{\tilde{\sigma}_{{\rm diff}}^2} \le \delta_n \le \frac{n}{3} , \\
				&& \quad \max_{1\le i\le j\le p}\frac{ (\sigma_{0i_{0},ij} - \sigma_{0i_{1},ij})^2 }{ \sigma_{0i_{0},ii}\sigma_{0i_{0},jj} + \sigma_{0i_{1},ii}\sigma_{0i_{1},jj} } \ge \tilde{\sigma}_{{\rm diff}}^2 \, \Big\} .
				\eea
				If $0<C_d < 1/2$, then we have 
				\bea
				\inf_{\what{i} } \sup_{P \in \calG(\tilde{\sigma}_{{\rm diff}},\, C_d)} \bbE_{P} |\what{i} - i_1| &\ge& c  \, n
				\eea
				for some constant $c>0$ and all sufficiently large $p$.
			\end{theorem}

			Theorem \ref{thm:changepoint_est_lowerbound_diverge} says that if $\delta_n \tilde{\sigma}_{\rm diff}^2/ \log p$ is smaller than certain constant, consistent estimation in terms of the localization rate is impossible.
			Specifically, this theorem implies that if 
			$\tilde{\sigma}_{\rm diff}^2 < \log p /( 2 \delta_n)$, 
			the localization rate  is larger than some constant $c>0$, at least in the worst case.
			Theorem \ref{thm:changepoint_est_lowerbound} shows that if $C_d >2$, the lower bound of localization rate becomes $\tilde{\sigma}_{{\rm diff}}^{-2} n^{-1}$ instead of $c>0$.


			\begin{theorem}\label{thm:changepoint_est_lowerbound}
				If $C_{d}>2$, we have 
				\bea
				\inf_{\what{i} } \sup_{P \in \calG(\tilde{\sigma}_{{\rm diff}},\, C_d)} \bbE_{P} |\what{i} - i_1| &\ge& c \,\tilde{\sigma}_{{\rm diff}}^{-2}
				\eea
				for some constant $c>0$ and all sufficiently large $p$.
			\end{theorem}

			This results means that if conditions \eqref{mx_std_diff_cpoint} and \eqref{cond_Ln} hold, the localization rate given at Theorem \ref{thm:changepoint_est} coincides with the obtained minimax lower bound up to $\log(n_w \vee p)$.
			Since $\tilde{\sigma}_{{\rm diff}}^{2} \asymp \log(n_w \vee p)/ n_w$ in our setting by \eqref{mx_std_diff_cpoint} and \eqref{cond_Ln},  $\tilde{\sigma}_{{\rm diff}}^{-2} n^{-1}$ has the same rate with $n_w/\{n \log (n_w \vee p) \}$.
			Thus, the localization rate in Theorem \ref{thm:changepoint_est}, $n_w/n$, is nearly optimal in terms of the minimax.
			Furthermore, Theorems \ref{thm:changepoint_est_lowerbound_diverge} and \ref{thm:changepoint_est_lowerbound} show that $\delta_n \tilde{\sigma}_{{\rm diff}}^{2}$ should be larger than $2 \log p$ to have the minimax lower bound of the localization rate, $\tilde{\sigma}_{{\rm diff}}^{-2} n^{-1}$.
			In our settings, $\delta_n \tilde{\sigma}_{{\rm diff}}^{2} \ge 2C_\star \log(n_w \vee p)$ for some constant $C_\star>0$ by \eqref{mx_std_diff_cpoint} and the condition $n_w\le \delta_n$.
			Thus, if $\log(n_w\vee p) \asymp \log p$, our condition is also rate-optimal.


			\section{Simulation study}\label{sec:simulation}
			In this section, we compare the proposed change point detection methods with several state-of-the-art methods. The comparison spans a wide range of scenarios, including both single and multiple change points, varying dimensions, different numbers and magnitudes of signals, and various covariance matrix structures. We consider situations where changes occur either in the mean or in the covariance, but not both.
			As competing methods,  we choose the change point detection methods proposed by \cite{dette2022estimating}, \cite{grundy2020high}, \cite{wang2018high} and \cite{matteson2014nonparametric}.
			For simplicity, we refer to these competing methods as Dette, Geomcp, Inspect and Edivisive, respectively.
		\textbf{Dette} is designed to detect a single change in the covariance structure and will be included only in relevant scenarios.
\textbf{Inspect } is intended for detecting changes in the mean structure and will be compared only in applicable settings.
\textbf{Geomcp} can detect change points in both the mean and diagonal covariance structures.
\textbf{Edivisive} is a nonparametric change point detection method.


			\subsection{Multiscale method}\label{subsec:multiscale}
	Selecting an appropriate window size is critical for the performance of the proposed methods. However, the optimal window size is case-dependent, which makes it challenging to choose the right size when there is limited prior knowledge about the data.
			To alleviate this issue and improve stability across various scenarios, we propose a multiscale approach inspired by \cite{avanesov2018change}.
			Let \(\mathcal{N} = \{n_w^{(r)} : r=1\ldots, |\mathcal{N}| \} \) denote the set of window sizes, where 
			$n_w^{(1)}< \cdots < n_w^{(|\mathcal{N}|)}$ and $|\mathcal{N}|$ is an odd number.
			For $r = 1, \ldots, |\mathcal{N}|$, let \( \mathcal{I}_r = \{\widehat{i}_k^{(r)} : k = 1, \ldots, \widehat{K}_r\} \) denote the set of detected change points from the mxPBF with window size \( n_w^{(r)} \).
			We sequentially use the window sizes $n_w^{(r)}$ to group the change points obtained from each mxPBF.
			Specifically, for each \( \widehat{i}_k^{(1)} \in \mathcal{I}_1 \), we first construct intervals based on the narrowest window $n_w^{(1)}$, defined as \( [\widehat{i}_k^{(1)} - n_{w}^{(1)} + 1, \widehat{i}_k^{(1)} + n_{w}^{(1)} - 1] \).
			We then count how many points are included in each interval among the estimated change points $\{\mathcal{I}_r : r = 1, \ldots, |\mathcal{N}| \}$.
			If an interval contains at least $(|\mathcal{N}|+1)/2$ points, the change points within that interval are grouped together. 
			Once no interval contains $(|\mathcal{N}|+1)/2$ or more points, we proceed to the next window size, $n_w^{(2)}$, and use the change points in $\mathcal{I}_2$ to construct the intervals $ [\widehat{i}_k^{(2)} - n_w^{(2)} + 1, \widehat{i}_k^{(2)} + n_w^{(2)} - 1] $. 
			From these intervals, we again group change points if at least $(|\mathcal{N}|+1)/2$ change points, excluding those that have already been grouped, fall within a single interval.
			If overlapping intervals contain exactly the same number of points, we prioritize grouping the interval with the smaller sample variance of the points within it.
			This process is repeated until groups are identified using the final window size $n_w^{(|\mathcal{N}|)}$. 
			Each of the resulting groups corresponds to a single change point, and the final estimate of the change point is computed as the average of $\widehat{i}_k^{(r)}$ within each group.

			Since the final decision is based on a majority rule, we refer to this multiscale approach as $\text{mxPBF\_major}$ throughout.
			In the numerical studies, we found that $\text{mxPBF\_major}$ achieves reasonable performance across a variety of scenarios, as detailed in the supplementary material.
			This multiscale method can also be theoretically justified. Specifically, if at least the window sizes $n_w^{(1)} < \cdots < n_w^{((|\mathcal{N}|+1)/2)}$ in $\mathcal{N}$ satisfy the conditions of Theorems \ref{thm:changepoint_est_mean} and \ref{thm:changepoint_est}, and if $n_w^{(r)} \le n_w^{(r+1)}/2$ holds for $r=1, \ldots, (|\mathcal{N}|-1)/2$, then it can be shown that the localization rate of the proposed multiscale approach is at most $n_w^{((|\mathcal{N}|+1)/2)} /n$.

			\subsection{Choice of hyperparameters}\label{subsec:hyper_param}
			The proposed mxPBFs for detecting changes in mean vectors and covariance matrices require the selection of hyperparameters. 
			The mxPBF in \eqref{mxPBF_change_mean}, designed for mean structures, includes the hyperparameter \( \alpha \). 
			Similarly, the mxPBF in \eqref{Bayes_test_cpoint}, for covariance structures, involves the hyperparameters \( a_0 \), \( b_{0,ij} \), \( b_{01,ij} \), \( b_{02,ij} \), and \( \alpha \). 
			We recommend selecting \( a_0 = b_{0,ij} = b_{01,ij} = b_{02,ij} = 0.01 \) for all \( 1 \leq i \neq j \leq p \) to minimize their impact on mxPBF.
			The choice of the hyperparameter \( \alpha \) in the mxPBFs \eqref{mxPBF_change_mean} and \eqref{Bayes_test_cpoint} is more crucial for performance.
			Although Theorems \ref{thm:mean_changepoint} and \ref{thm:cov_changepoint} provide theoretical conditions for selecting \( \alpha \), they may be overly conservative in practice. 
			Therefore, we extend and adapt the method proposed in \cite{lee2024bayesian} to select $\alpha$, which controls the empirical false positive rate (FPR).
			We select $\alpha$ for each window size $n_w$ using the proposed FPR-based method, with detailed explanations provided in the supplementary material.
			By default, we set the prespecified FPR to 0.05, the threshold  \(C_{\rm cp}\)  to 10, and the number of simulated datasets $N$ to $300$.

			\subsection{Evaluation metrics}
			To illustrate the performance of each method, we use the F1 score and the Hausdorff distance.
			For a given integer $M > 0$, let true positive (TP) be the number of true change points whose differences from the estimated change points are less than $M$, i.e.,
			${\rm TP}(\mathcal{I}, \hat{\mathcal{I}}) := \{ i \in \mathcal{I} \mid \exists \hat{i} \in \hat{\mathcal{I}} \text{ such that } |\hat{i} - i| < M \}$, 
			where \( \mathcal{I} \) and \( \hat{\mathcal{I}} \) are sets of true and detected change points, respectively.
			The precision and recall are then defined as ${\rm Prec}(\mathcal{I}, \hat{\mathcal{I}}) := | {\rm TP}(\mathcal{I}, \hat{\mathcal{I}})| / |\hat{\mathcal{I}}|$ and ${\rm Rec}(\mathcal{I}, \hat{\mathcal{I}}) := | {\rm TP}(\mathcal{I}, \hat{\mathcal{I}}) |/ |\mathcal{I}|$, respectively. 
			The $F1$ score is the harmonic mean of precision  and recall:
			\bea
			F1(\mathcal{I}, \hat{\mathcal{I}}) &:=&   \frac{2 \text{Prec}(\mathcal{I}, \hat{\mathcal{I}})   \text{Rec}(\mathcal{I}, \hat{\mathcal{I}})}{\text{Prec}(\mathcal{I}, \hat{\mathcal{I}}) + \text{Rec}(\mathcal{I}, \hat{\mathcal{I}})}.
			\eea
			The F1 score always falls between 0 and 1, with values closer to 1 indicating better performance.
			
			The Hausdorff distance measures the robustness of detection methods by quantifying the largest discrepancy between a true change point and its corresponding estimated change point:
			\bea
			{\rm Haus}(\mathcal{I}, \hat{\mathcal{I}}) &:=& \max\left(\max_{\hat{i} \in \hat{\mathcal{I}}} \min_{i \in \mathcal{I}} \vert i - \hat{i} \vert, \max_{i \in \mathcal{I}} \min_{\hat{i} \in \hat{\mathcal{I}}} \vert \hat{i} - i \vert\right).
			\eea
			The Hausdorff distance is non-negative, with values closer to 0 indicating better performance.
			
			Note that the F1 score is not defined if \( \vert {\rm TP}(\mathcal{I}, \hat{\mathcal{I}}) \vert = 0 \), and the Hausdorff distance is undefined if \( \vert \mathcal{I} \vert = 0 \) or \( \vert \hat{\mathcal{I}} \vert = 0 \). To address this, we include \( i=1 \) and \( i=n \) as trivial change points both in \( \mathcal{I} \) and \( \hat{\mathcal{I}} \). 
			This approach is suggested by \cite{van2020evaluation}.
			This does not affect the interpretation of the detection results, as change points are considered the start of a new segment.


			\subsection{Simulation study: change point detection in mean structure}\label{subsec:simul_mean}
			
			First, we compare the performance of mxPBF\_major with Inspect, Geomcp and Edivisive in detecting change points in the mean structure through simulation studies. 
			We set \( n = 500 \) and \( p \in \{200, 500, 800\} \) as the dimensions of the dataset. 
			In ``single change'' scenarios, the data is generated as follows: \( X_{1}, \ldots, X_{i-1} \overset{i.i.d.}{\sim} N_p(\mu_1, \Sigma) \) and \( X_{i}, \ldots, X_{n} \overset{i.i.d.}{\sim} N_p(\mu_2, \Sigma) \), with the true change point \( \mathcal{I} = \{ i_1 = 250 \} \). 
			Under the null hypothesis, \( H_{0}^{\mu}: \mu_1 = \mu_2 \), we set \( \mu_1 = \mu_2 = 0 \in \mathbb{R}^p \). 
			Under the alternative hypothesis, \( H_{1}^{\mu}: \mu_1 \neq \mu_2\), we set \( \mu_1 = 0 \in \mathbb{R}^p \) and randomly choose \( n_0 \) entries in \( \mu_2 \), setting them to \( \mu > 0 \), while the remaining entries are set to 0. 
			In ``multiple change'' scenarios, the data is generated as follows: \( X_{1}, \ldots, X_{i_1-1} \overset{i.i.d.}{\sim} N_p(\mu_1, \Sigma) \), \( X_{i_1}, \ldots, X_{i_2-1} \overset{i.i.d.}{\sim} N_p(\mu_2, \Sigma) \), \( X_{i_2}, \ldots, X_{i_3-1} \overset{i.i.d.}{\sim} N_p(\mu_3, \Sigma) \) and \( X_{i_3}, \ldots, X_{n} \overset{i.i.d.}{\sim} N_p(\mu_4, \Sigma) \), with the true change point set  \(\mathcal{I} = \{ i_1 = 150, i_2 = 300, i_3 = 350 \}\).
			Under the null hypothesis, we set \( \mu_1=\mu_2=\mu_3=\mu_4= 0\in \bbR^p \).
			Under the alternative hypothesis, we set \( \mu_1 = \mu_3 = 0 \in \mathbb{R}^p \) and randomly select \( n_0 \) entries in \( \mu_2 \) and \( \mu_4 \), setting them to \( \mu > 0 \), while the remaining entries are set to 0. 
			Here, \( n_0 \) represents the number of signals involved in a change, and \( \mu \) is the magnitude of these signals. 
			
			We set $\mu\in \{0.1, 0.3, 0.5, 0.7, 1, 1.2, 1.5, 2\}$, and the following scenarios for the number of signals \( n_0 \) are considered:
			\begin{itemize}
				\item[1.] ($H_{1,R}$: Rare signals) 
				To consider a scenario with only a few signals, we set $n_0 = 5$.
				
				\item[2.] ($H_{1,M}$: Many signals) 
				To consider a scenario with a lot of signals, we set $n_0 = p/2$.
			\end{itemize}
			Furthermore, we consider the following two structures for the true precision matrix $\Omega$:
			\begin{itemize}
				\item[1.] (Sparse $\Omega=\Sigma^{-1}$) We randomly select $1\%$ of entries in $\Omega= (\omega_{ij})$ and set their value to $0.3$. The remaining entries in $\Omega$ are set to 0. When the resulting $\Omega$ is not positive definite, we make it positive definite by adding $\{-\lambda_{\min}(\Omega)+0.1^3\}I_p$ to $\Omega$. 
				
				\item[2.] (Dense $\Omega=\Sigma^{-1}$) We randomly choose $40\%$ of entries in $\Omega= (\omega_{ij})$ and set their value to 0.3. The rest of the steps for constructing $\Sigma$ is the same as above.
			\end{itemize}
			Note that the terms “rare” and “many” refer to the number of nonzero entries in $\mu_{k+1} - \mu_k \in \mathbb{R}^p$ under the alternative hypothesis, indicating the number of signals in a change. 
			On the other hand, the terms “sparse” and “dense” refer to the number of nonzero entries in the precision matrix $\Omega$, irrespective of the number of signals.

			When applying mxPBF\_major, we use a set of window sizes \( \mathcal{N} = \{25,\ 60,\ 100\} \), where the other hyperparameters are selected as described in Section \ref{subsec:hyper_param}.
			The R package \texttt{hdbcp}, which implements mxPBF\_major, and the reproduction code for the simulation study are available on GitHub.\footnote{GitHub repository: \url{https://github.com/JaeHoonKim98/hdbcp}}
			To apply Geomcp, we use the R package \texttt{changepoint.geo} \citep{grundy2020cpgeo} and \texttt{changepoint.np} \citep{haynes2016cpnp}. 
			Since the covariance matrices in the scenarios are not diagonal, we employ the empirical cost function with the number of quantiles set to $4\log(n)$, as recommended by \cite{grundy2020high} and \cite{haynes2016nonparametric}. 
			We use the default MBIC penalty \citep{zhang2007modified} and set both the minimum segmentation parameter and the threshold for the reconciling method to \(\xi = n_w^1\). 
			For the implementation of Edivisive method, we use the R package \texttt{ecp} \citep{matteson2013ecp}. 
			Following the recommendations of \cite{matteson2014nonparametric}, we set \( \alpha = 1 \), a significance level of 0.05 and \( R = 499 \). 
			Additionally, we set the minimum segment size to \( n_w^1 \) in line with our assumption. 
			To implement the Inspect method, we use the R package \texttt{InspectChangepoint} \citep{wang2016inspect} with the default setting of \( Q = 0 \). 
			Setting \( Q = 0 \) implements a Binary Segmentation approach \citep{scott1974cluster, vostrikova1981detecting} for detecting multiple change points.
			When setting \( Q = 1000 \), as suggested in \cite{wang2018high}, a Wild Binary Segmentation approach \citep{fryzlewicz2014wild} is implemented. 
			However, it detected more false positives compared to the default setting.
			The threshold for identifying change points is determined using the data-driven approach proposed by \cite{wang2018high}. 
			Since \texttt{InspectChangepoint} does not provide a parameter for setting a minimum segment length during binary segmentation, we include a post-processing step that removes change points with lower CUSUM statistics if the distance between detected points is less than  \(n_w^1\).
			This process yields the same results as directly setting a minimum segment length parameter. 
			Also, it significantly improves performance by reducing false positives through the elimination of redundant detections.

			\begin{figure}[!bt]
				\centering
				\includegraphics[width=0.97\textwidth]{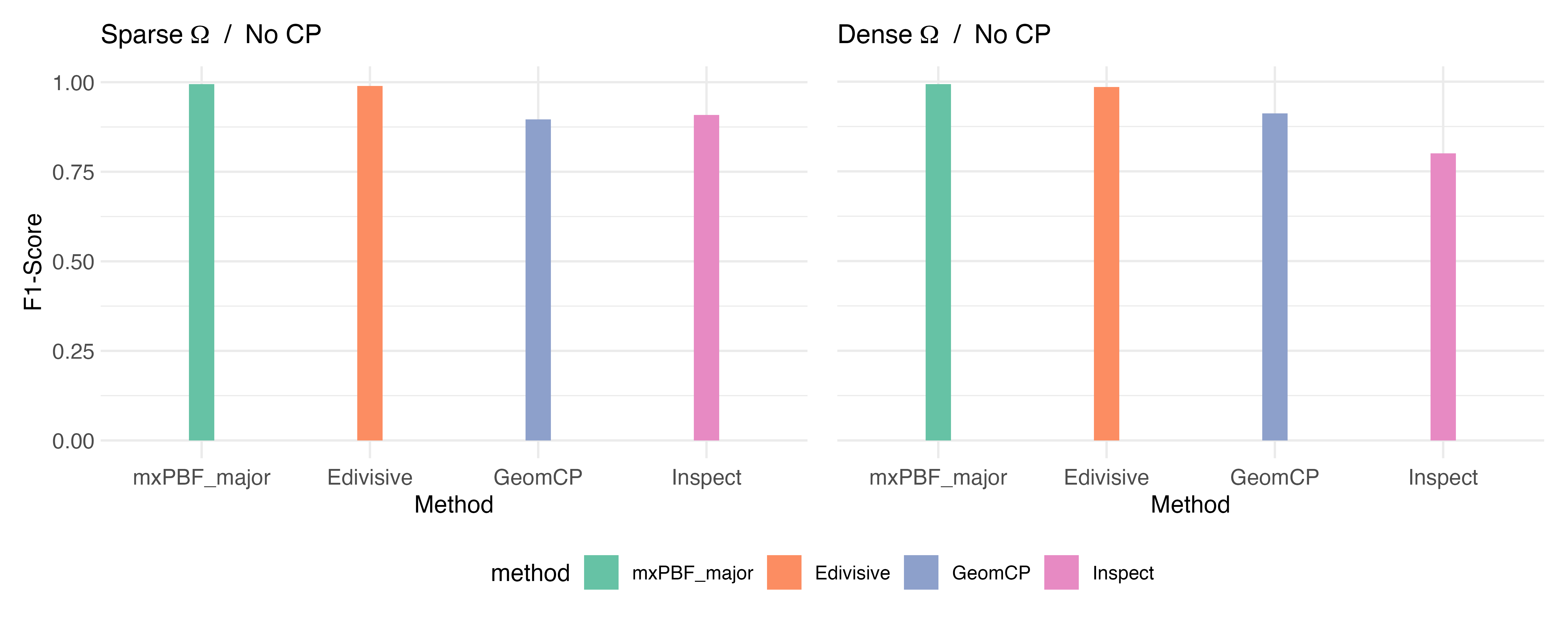}
				\caption{F1 scores for change point detection in mean structure based on 50 simulated datasets for $H_0$ with $p = 200$. The methods \texttt{mxPBF\_major}, \texttt{Edivisive}, \texttt{Inspect} and \texttt{GeomCP} correspond to the multiscale method proposed in this paper, and to those by \cite{matteson2014nonparametric}, \cite{wang2018high}, and \cite{grundy2020high}, respectively.}
				\label{fig:Fscore_null_mean}
			\end{figure}
			
			Figure \ref{fig:Fscore_null_mean} shows the F1 scores for each method, based on 50 simulated datasets under $H_0$ when $p = 200$. 
			In this scenario, where no true change points exist, a higher F1 score indicates better performance in reducing false positives. 
			Among the methods, mxPBF\_major and Edivisive exhibit superior performance in this regard.

			\begin{figure}[!tb]
				\centering
				\includegraphics[width=0.97\textwidth]{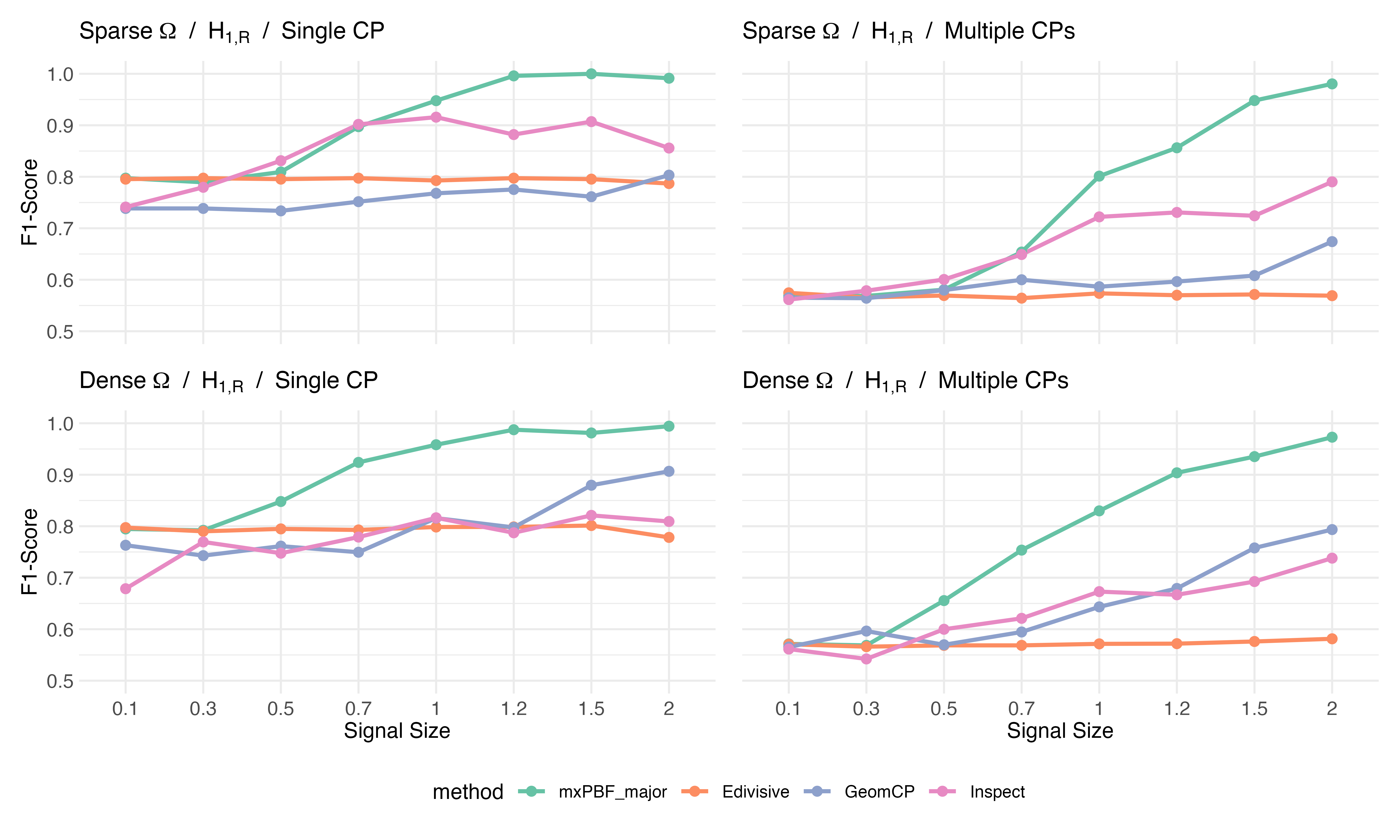}
				\includegraphics[width=0.97\textwidth]{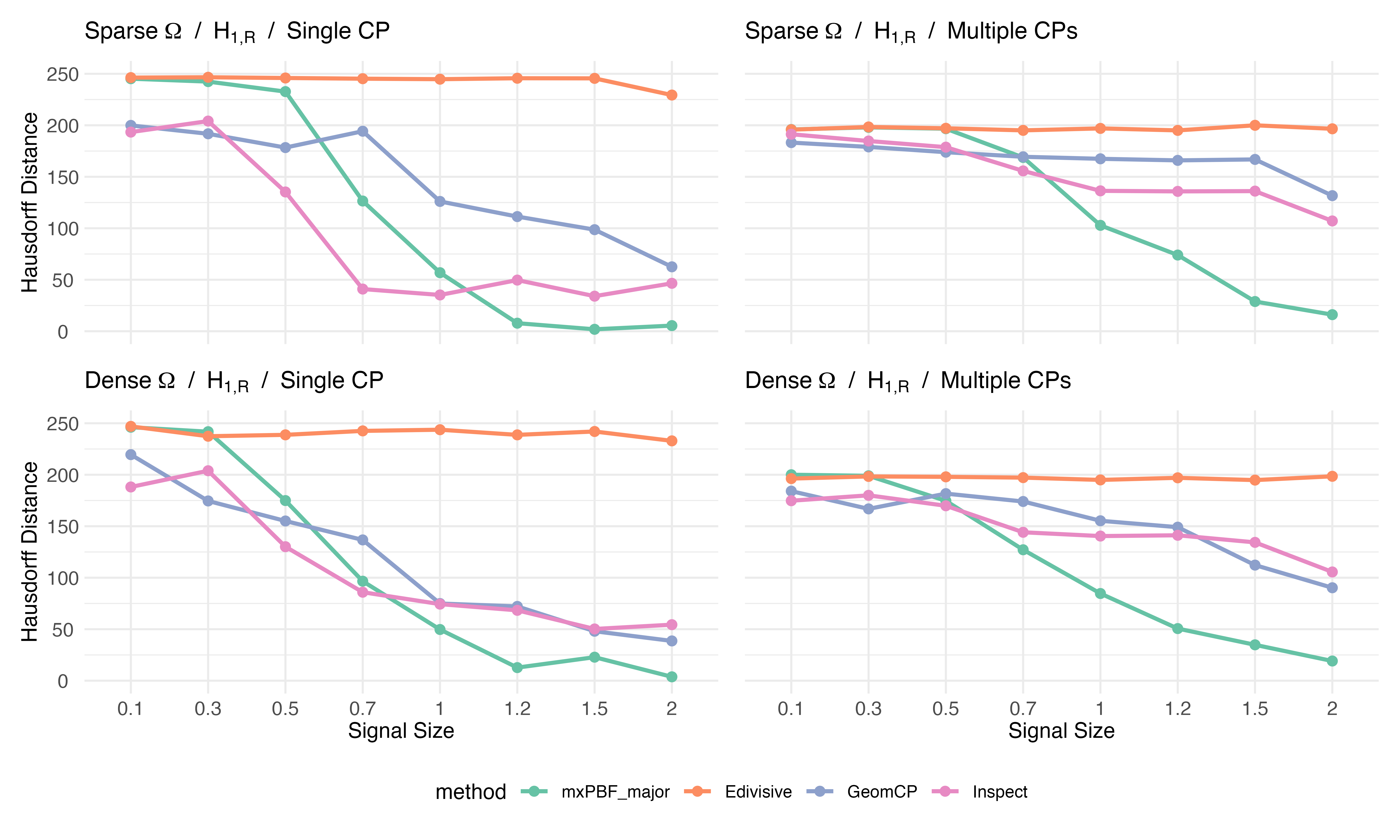}
				\caption{F1 scores and Hausdorff distances for change point detection in mean structure based on 50 simulated datasets under \( H_{1,R} \) with \( p = 200 \).}
				\label{fig:FH_H1R_mean}
			\end{figure}
			
			Figure \ref{fig:FH_H1R_mean} presents the F1 score and Hausdorff distance based on 50 simulated datasets under ``rare signals'' scenario $H_{1,R}$, with $p = 200$. 
			Under $H_{1,R}$, the difference vector \( \mu_{k+1} - \mu_k \in \mathbb{R}^p \) contains only five nonzero elements, each with a signal size of \( \mu \). 
			Note that mxPBF\_major demonstrates significant improvement as the signal size increases and outperforms the other methods in most scenarios.
			Inspect and Geomcp show improved performance as the signal size increases but tends to produce a large number of false positives. 
			Additionally, Inspect performs better in scenarios where $\Omega$ is sparse, likely because this method assumes a diagonal structure for $\Omega$.
			Edivisive appears to struggle in accurately detecting changes, which may stem from the method being more suited to scenarios involving numerous signals.

			\begin{figure}[!tb]
				\centering
				\includegraphics[width=0.97\textwidth]{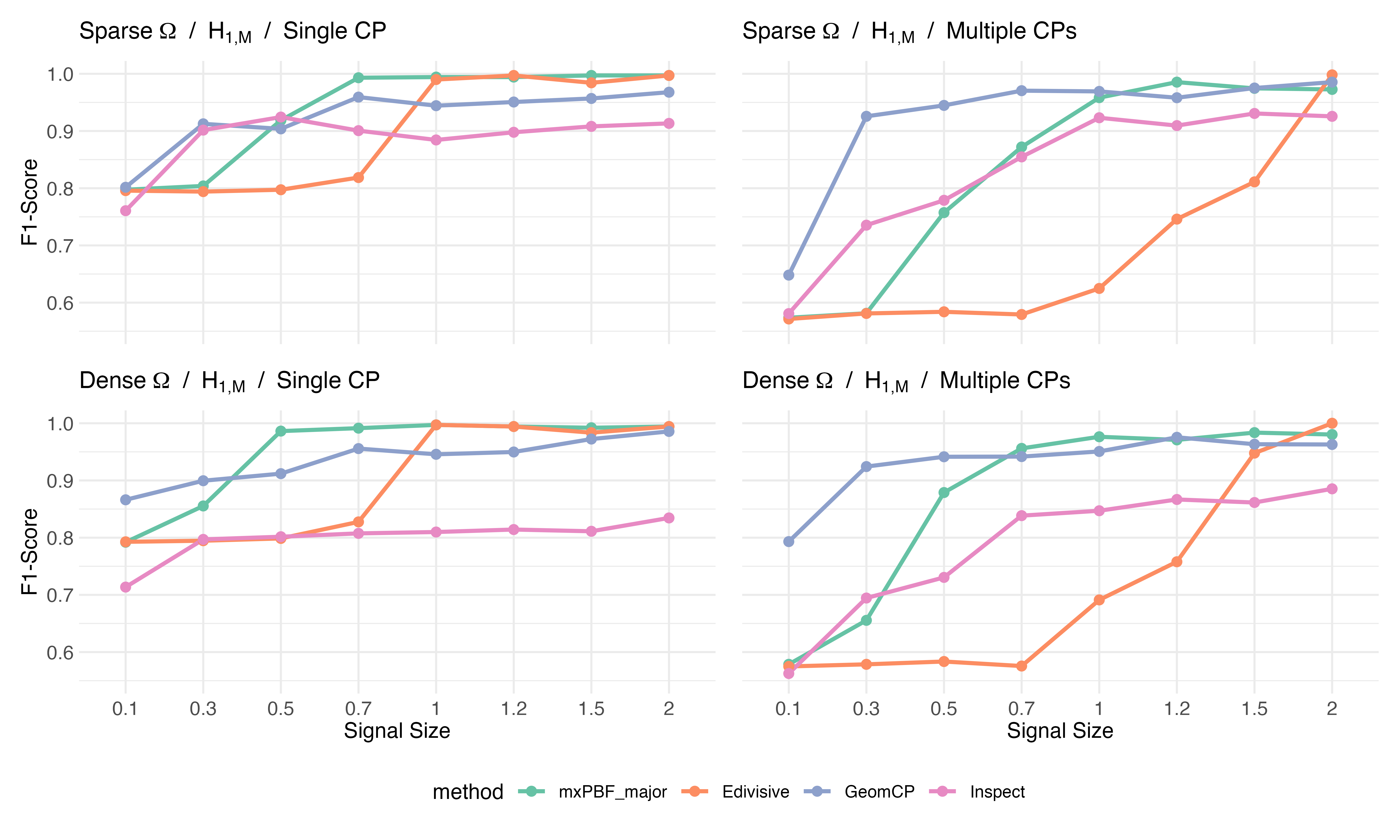}
				\includegraphics[width=0.97\textwidth]{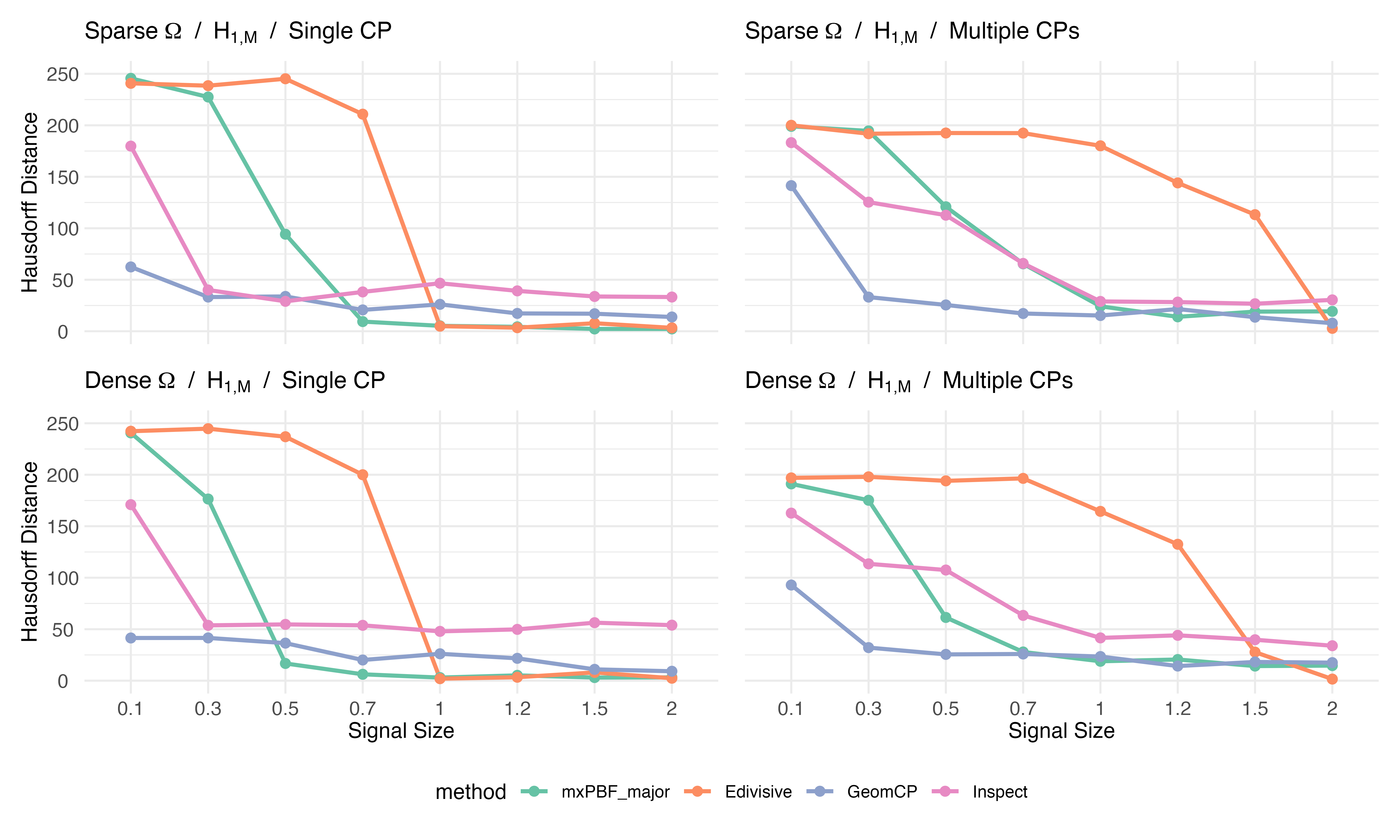}
				\caption{F1 scores and Hausdorff distances for change point detection in mean structures based on 50 simulated datasets under \( H_{1,M} \) with \( p = 200 \).}
				\label{fig:FH_H1M_mean}
			\end{figure}
			
			Figure \ref{fig:FH_H1M_mean} presents the F1 score and Hausdorff distance based on 50 simulated datasets under ``many signals'' scenario $H_{1,M}$, with $p = 200$. 
			Under $H_{1,M}$, the difference vector \(\mu_{k+1} - \mu_k \in \mathbb{R}^p\) contains $p/2$ nonzero elements, with each element having a signal size of \(\mu\).
			When signals are small, Inspect and Geomcp show decent performance. 
			However, similar to the rare signals scenario, Inspect appears to detect numerous false positives.
			When signals are large, mxPBF\_major and Edivisive also demonstrate improved performance, achieving results comparable to Geomcp. 
			Notably, the performance of mxPBF\_major improves rapidly as the signal size increases.

			Since similar results are observed for \(p = 500\) and \(p = 800\), we omit these results for brevity. 
			The detailed results are provided in the supplementary material.

			\subsection{Simulation study: change point detection in covariance structure}\label{subsec:simul_cov}
			
			Next, we demonstrate the performance of mxPBF\_major, along with Dette, Geomcp and Edivisive, for detecting change points in the covariance structure.
			Similar to the settings in Section \ref{subsec:simul_mean}, we set \( n = 500 \) and \( p \in \{200, 500, 800\} \) as the dimensions of the dataset.
			In ``single change'' scenarios, the data is generated as follows: \( X_{1}, \ldots, X_{i-1} \overset{i.i.d.}{\sim} N_p(0, \Sigma_1) \) and \( X_i, \ldots, X_{n} \overset{i.i.d.}{\sim} N_p(0, \Sigma_2) \), with the true change point \( \mathcal{I} = \{ i_1 = 250 \} \). 
			Under the null hypothesis, \( H_{0}^{\Sigma}: \Sigma_1 = \Sigma_2 \), we set \( \Sigma_1 = \Sigma_2 \equiv \Sigma \in \mathbb{R}^{p \times p} \). 
			Under the alternative hypothesis, \( H_{1}^{\Sigma}: \Sigma_1 \neq \Sigma_2 \), we set \( \Sigma_1 \equiv \Sigma \) and \( \Sigma_2 = \Sigma + U \), where \( U \in \mathbb{R}^{p \times p} \) is a symmetric matrix containing the signals. 
			If \( \Sigma_1 \) or \( \Sigma_2 \) is not positive definite, we add $[-\min\{\lambda_{\min}(\Sigma_1), \lambda_{\min}(\Sigma_2)\} + 0.05]I_p$ to both matrices to make them positive definite. 
			In ``multiple change'' scenarios, the data is structured as follows: \( X_{1}, \ldots, X_{i_1-1} \overset{i.i.d.}{\sim} N_p(0, \Sigma_1) \), \( X_{i_1}, \ldots, X_{i_2-1} \overset{i.i.d.}{\sim} N_p(0, \Sigma_2) \), \( X_{i_2}, \ldots, X_{i_3-1} \overset{i.i.d.}{\sim} N_p(0, \Sigma_3) \), and \( X_{i_3}, \ldots, X_{n} \overset{i.i.d.}{\sim} N_p(0, \Sigma_4) \), with the true change point set \( \mathcal{I} = \{ i_1 = 150, i_2 = 300, i_3 = 350 \} \). 
			Under the null hypothesis, we set \( \Sigma_k \equiv \Sigma \) for all \( k \). 
			Under the alternative hypothesis, we set \( \Sigma_1 = \Sigma_3 \equiv \Sigma \) and set \( \Sigma_2 \) and \( \Sigma_4 \) to \( \Sigma + U \), with \( U \) generated independently.
			Note that the symmetric matrix \( U \) determines both the number and magnitude of signals under the alternative hypothesis. 
			We set $\psi\in \{0.5, 1, 1.3, 2, 4, 6, 8, 10\}$ and consider the following two scenarios for generating \( U \):
			\begin{itemize}
				\item[1.] ($H_{1,R}$: Rare signals) 
				To consider a scenario with only a few signals, we randomly select five entries in the lower triangular part of $U$ and generate their values from ${\rm Unif}(0, \psi)$.
				To ensure that $U$ is symmetric, the corresponding upper triangular part is assigned the same nonzero values, while the remaining elements are set to 0.
				
				\item[2.] ($H_{1,M}$: Many signals) To consider a scenario with a lot of signals, we set $U=u u^T$, where  $u=(u_1,\ldots,u_p)^T$ and $u_j\overset{i.i.d.}{\sim} {\rm Unif}(0, \psi)$.
				It leads to $p(p+1)/2$ signals in $U$, except upper triangular part.
				
			\end{itemize}
			Note that $\psi$ is closely related to the magnitude of the signals. 
			Furthermore, we consider the following two settings for the common covariance matrix $\sg$:
			\begin{itemize}
				\item[1.] (Sparse $\Sigma$) We randomly select $5\%$ of the entries in the lower triangular part of $\Delta_1 = (\delta_{1,jk})$ and set their values to $\delta_{1,jk} = 0.5$. 
				To ensure that $\Delta_1$ is symmetric, the corresponding upper triangular part is assigned the same nonzero values, while the remaining elements are set to 0.
				We define $\Delta = \Delta_1 + \delta I_p$, where where $\delta = |\lambda_{\min}(\Delta_1)| + 0.05$, and set $\Sigma = D^{1/2} \Delta D^{1/2}$, where  $D = \text{diag}(d_j)$  and  $d_j \overset{i.i.d.}{\sim}  {\rm Unif}(0.5, 2.5)$.
				This corresponds to Model 3 in \cite{cai2013two}.
				
				\item[2.] (Dense $\Sigma$) We set $\Sigma = O \Delta O$, where $O = \text{diag}(o_j)$ with $o_j \overset{i.i.d.}{\sim} \text{Unif}(1, 5)$, $\Delta = (\delta_{ij})$ where $\delta_{ij} = (-1)^{i+j}0.4^{\vert i-j\vert^{1/10}}$.
				This corresponds to Model 4 in \cite{cai2013two}.
			\end{itemize}
			Similar to the previous case, the terms ``rare” and ``many” describe the number of nonzero entries in $\Sigma_2 - \Sigma_1 \in \mathbb{R}^{p \times p}$ under the alternative hypothesis and, thus, relate to the number of signals. 
			On the other hand, the terms ``sparse” and “dense” refer to the number of nonzero entries in the common covariance matrix, $\Sigma$, regardless of the number of signals .

			Since Dette is a method designed to detect a single change point, it was applied only in the ``single change'' scenario.
			When applying Dette, the threshold for dimension reduction is selected based on the resampling approach proposed by \cite{dette2022estimating}.
			For the other methods, we use the same settings as described in Section \ref{subsec:simul_mean}.

			\begin{figure}[!bt]
				\centering
				\includegraphics[width=0.97\textwidth]{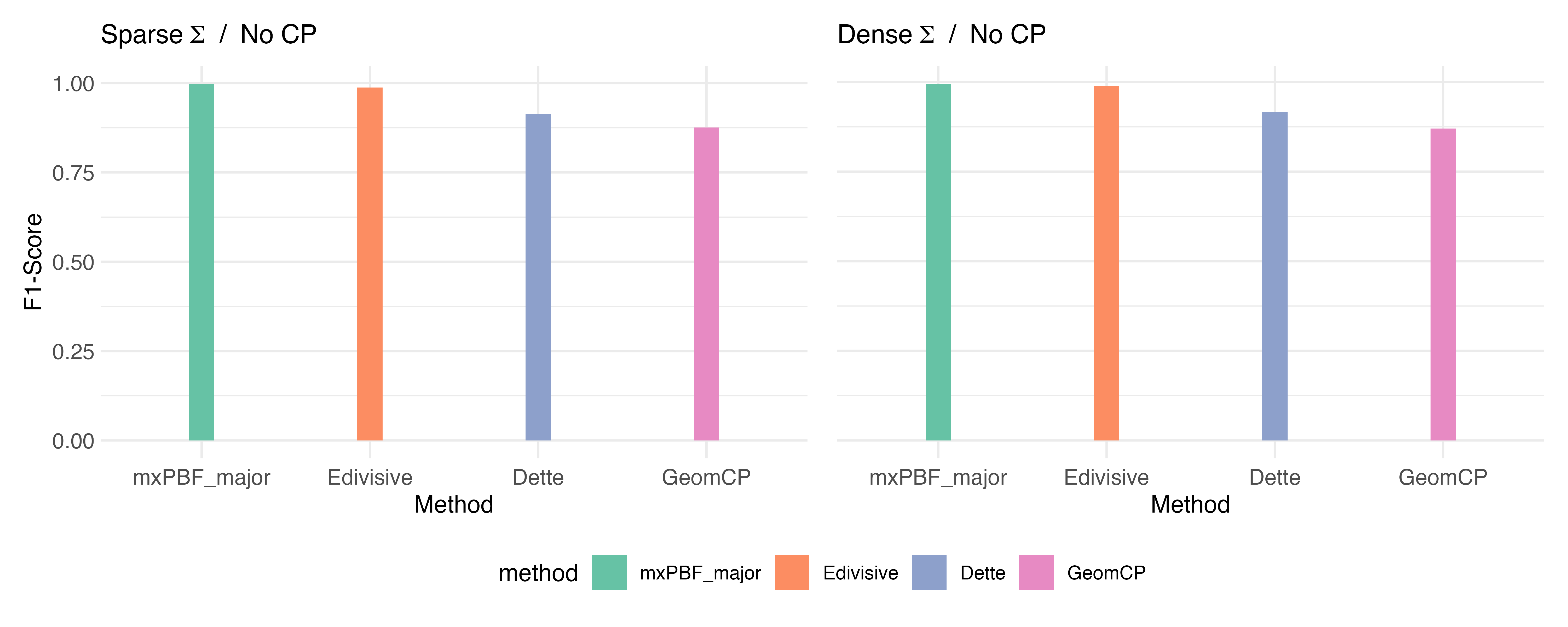}
				\caption{F1 scores for change point detection in covariance structure based on 50 simulated datasets for $H_0$ with $p = 200$.}
				\label{fig:Fscore_null_cov}
			\end{figure}
			
			Figure \ref{fig:Fscore_null_cov} shows the F1 scores for each method, based on 50 simulated datasets under $H_0$ with $p = 200$. 
			Similar to the change point detection results for the mean structure,  mxPBF\_major and Edivisive outperform other methods,

			\begin{figure}[!bt]
				\centering
				\includegraphics[width=0.97\textwidth]{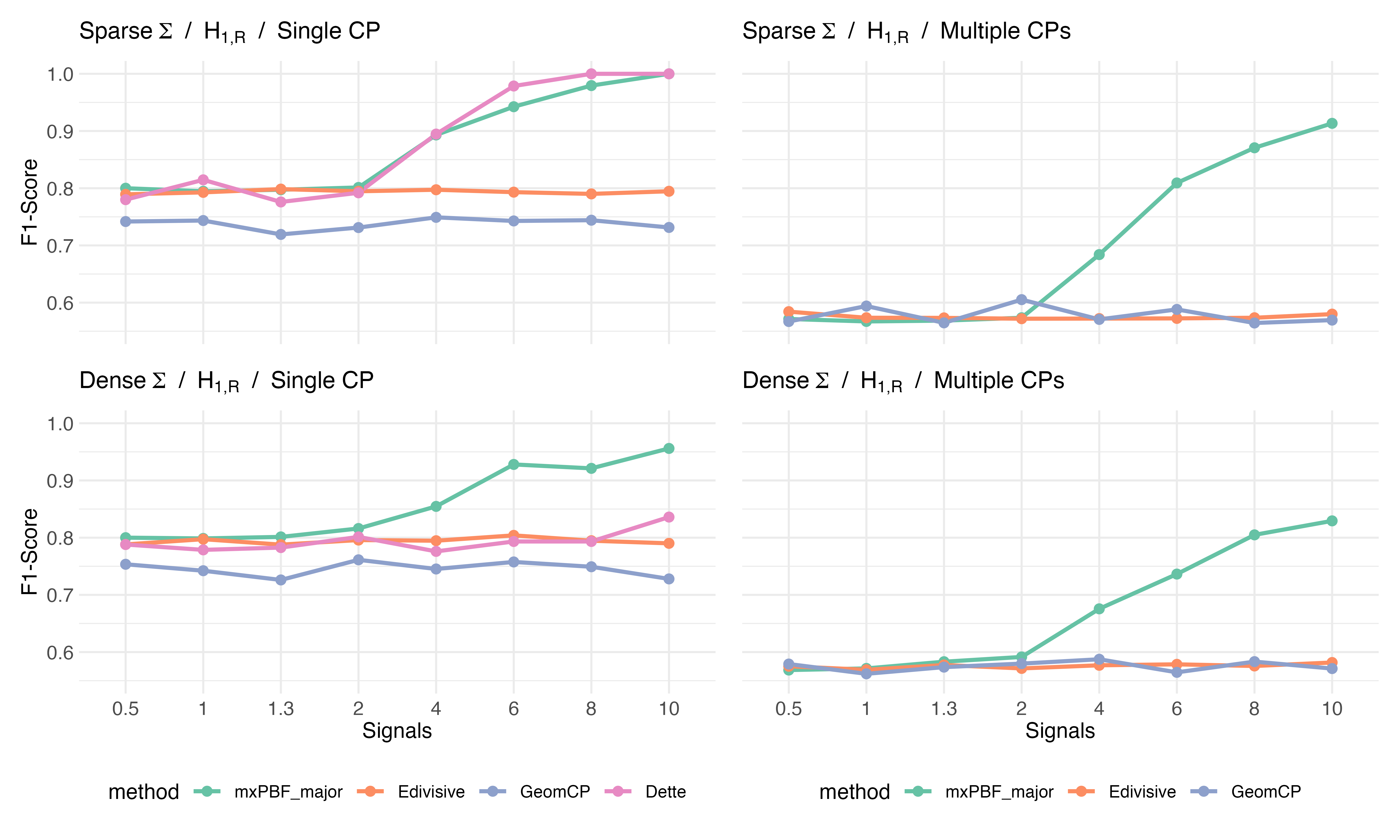}
				\includegraphics[width=0.97\textwidth]{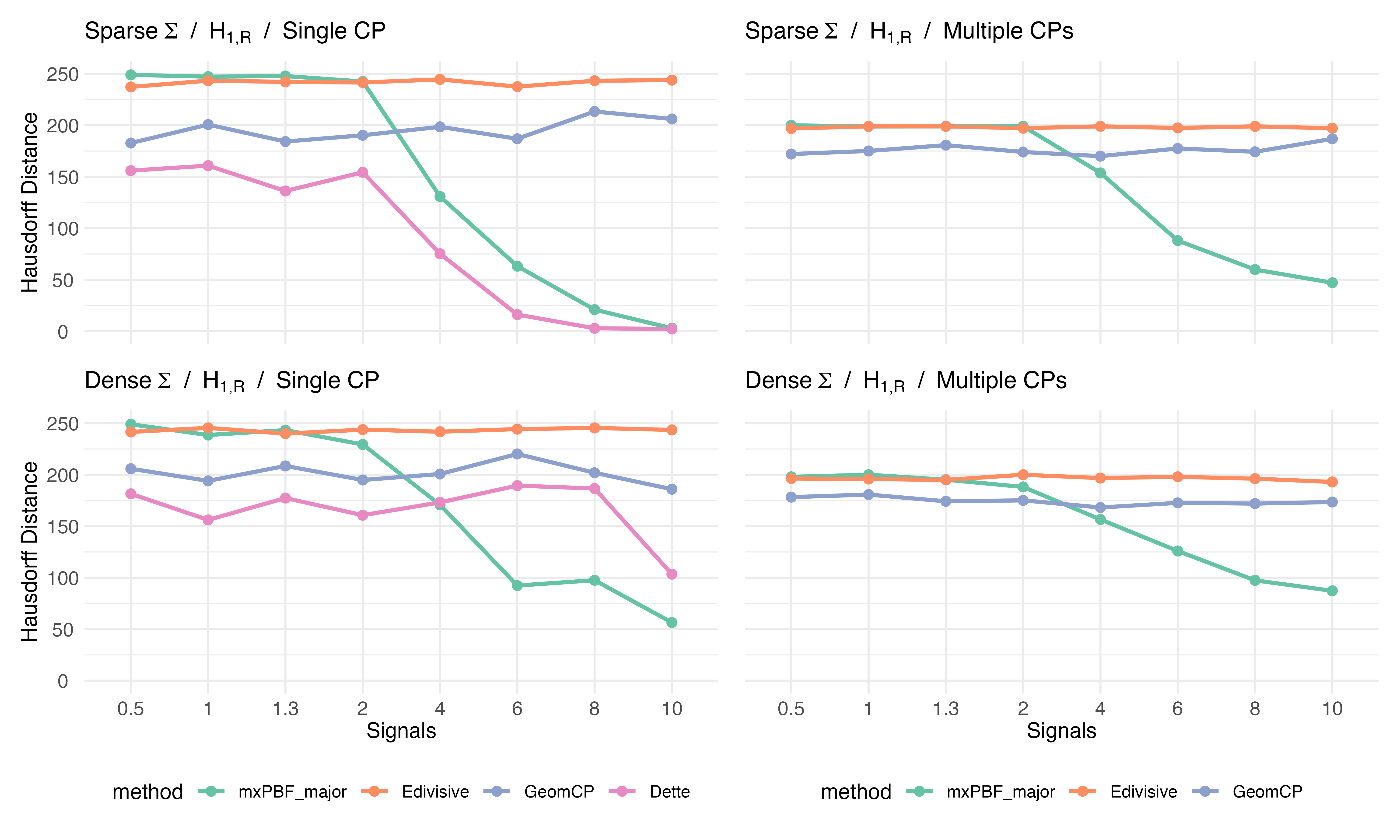}
				\caption{F1 scores and Hausdorff distances for change point detection in covariance structure based on 50 simulated datasets under \( H_{1,R} \) with \( p = 200 \).}
				\label{fig:FH_H1R_cov}
			\end{figure}
			
			Figure \ref{fig:FH_H1R_cov} shows F1 score and Hausdorff distance based on 50 simulated datasets under ``rare signals'' scenario $H_{1,R}$, with $p = 200$. 
			Note that mxPBF\_major outperforms the other methods in most scenarios, with its performance steadily improving as the signal size increases.
			Dette performs well in the sparse covariance scenario but struggles to detect changes when the covariance is dense.
			Edivisive and Geomcp struggle to detect change points, with their performance showing no improvement as the signal size increases. 
			This may again be due to the fact that they were not designed to detect rare signals.

			\begin{figure}[!bt]
				\centering
				\includegraphics[width=0.97\textwidth]{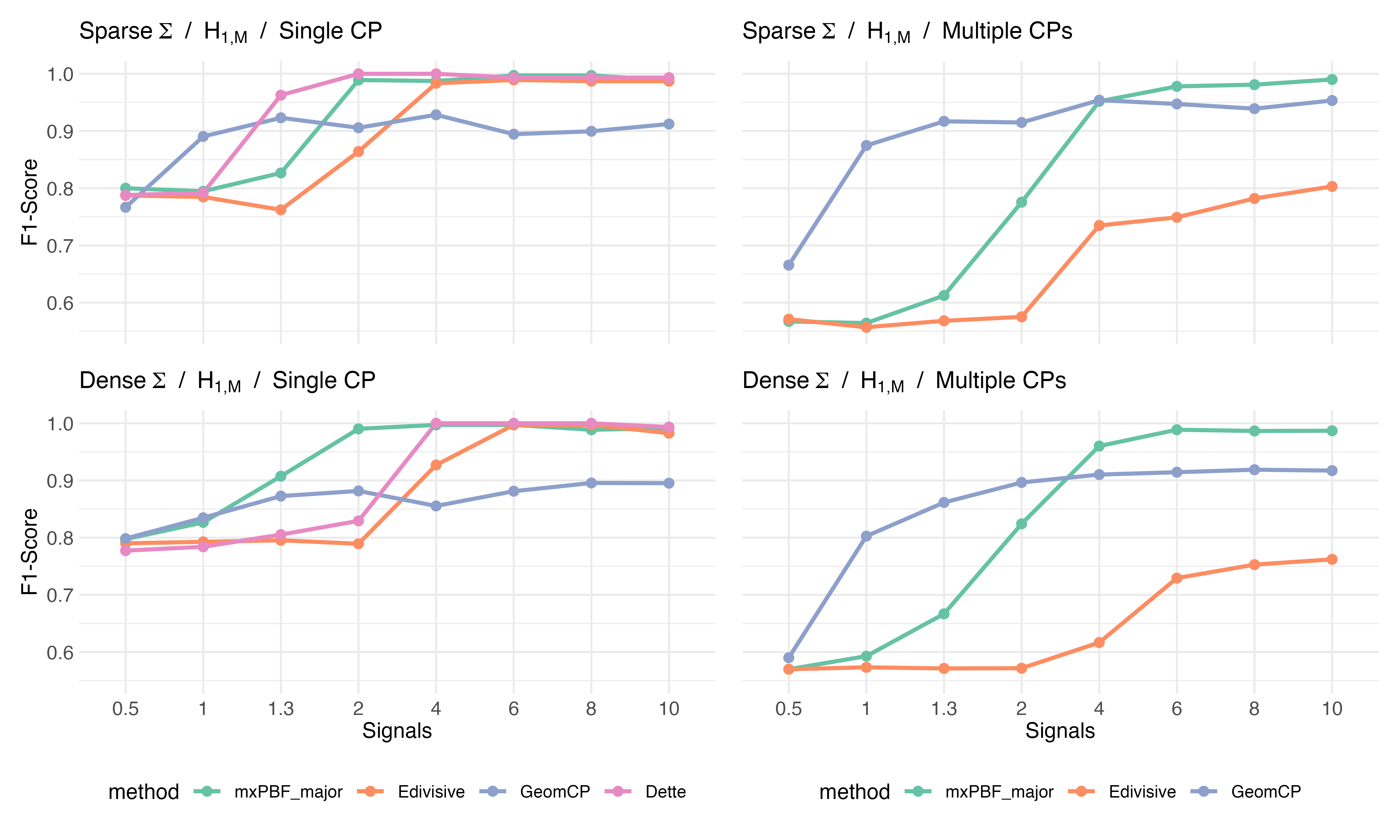}
				\includegraphics[width=0.97\textwidth]{Plots/haus_cov_200_h1r.png}
				\caption{F1 scores and Hausdorff distances for change point detection in covariance structure based on 50 simulated datasets under \( H_{1,M} \) with \( p = 200 \).}
				\label{fig:FH_H1M_cov}
			\end{figure}
			

			Figure \ref{fig:FH_H1M_cov} presents the F1 score and Hausdorff distance based on 50 simulated datasets under ``many signals'' scenario $H_{1,M}$, with $p = 200$. 
			When signals are small, Geomcp outperforms other methods. However, the improvement in its performance metrics is limited due to numerous false positives.
			In most settings, mxPBF\_major outperforms other methods when signals are large, say \(\psi \geq 4\).
			Dette and Edivisive effectively detect the change point in the single change scenario.
			However, in the multiple changes scenario, Edivisive struggles to accurately detect changes, even when signals are large.

			We observe that the experiments for \(p = 500\) and \(p = 800\) show similar patterns. 
			Thus, for brevity, the results are provided in the supplementary material.

			\section{Real data analysis}\label{sec:real}
			In this section, we apply the proposed change point detection methods to real datasets.
			In practice, it is unclear whether changes occur in the mean, covariance or both.
			To address this issue, we apply both of the proposed change point detection methods sequentially.
			We first apply the method for detecting change points in the covariance structure.
			To satisfy the assumption that the mean is a zero vector, we roughly center the observations using a given window size $n_w$.
			Specifically, given \(X_{ij}\) where \(1 \leq i \leq n\) and \(1 \leq j \leq p\), and a window size \(n_w \in \mathcal{N}\), the centered data is defined as
			\(\tilde{X}_{ij} = X_{ij} - \bar{X}_{\max(1, i - \lfloor n_w/2 \rfloor):\min(i + \lfloor n_w/2 \rfloor, n), j}\), where
			\(\bar{X}_{a:b, j} = (b-a+1)^{-1} \sum_{k=a}^b X_{kj}\).
			Then we apply the change point detection method for covariance matrices.
			If no changes are detected, we then apply the proposed change point detection method for mean vectors to the original, uncentered dataset. 
			If one or more change points are detected, we divide the data based on the estimated change points. 
			Since each segment is expected to share a common covariance matrix, we apply the proposed change point detection method for mean vectors within each segment.
			In this case, if some segments are smaller than twice the largest window size we used, we only use the available windows.
			Finally, we combine the change points detected in both the mean and covariance structures.
			Note that this process prioritizes changes in covariance matrices over changes in mean vectors.

			\subsection{Comparative Genomic Hybridization}\label{subsec:realgenomic}
			We first examine the comparative genomic hybridization microarray dataset from \cite{bleakley2011group}, which examines DNA copy number variations in individuals with bladder tumors. 
			This technique involves comparing the fluorescence intensity levels of DNA fragments between a test sample and a normal reference sample to detect chromosomal copy number abnormalities. 
			The dataset consists of log-intensity ratio measurements from 43 individuals $(p=43)$ with bladder tumors, collected at 2,215 positions $(n=2215)$ across the genome. 
			This dataset is available in the R package \texttt{ecp} \citep{matteson2014nonparametric}.
			Following the method described in \cite{grundy2020high}, we scale each series using the median absolute deviation prior to applying the detection methods. 
			For Geomcp, we use the CROPS algorithm \citep{haynes2017computationally} as suggested in \cite{grundy2020high}.
			For the other parameters, we use the same settings as described before.

			\begin{figure}[!bt]
				\centering
				\includegraphics[width = \textwidth]{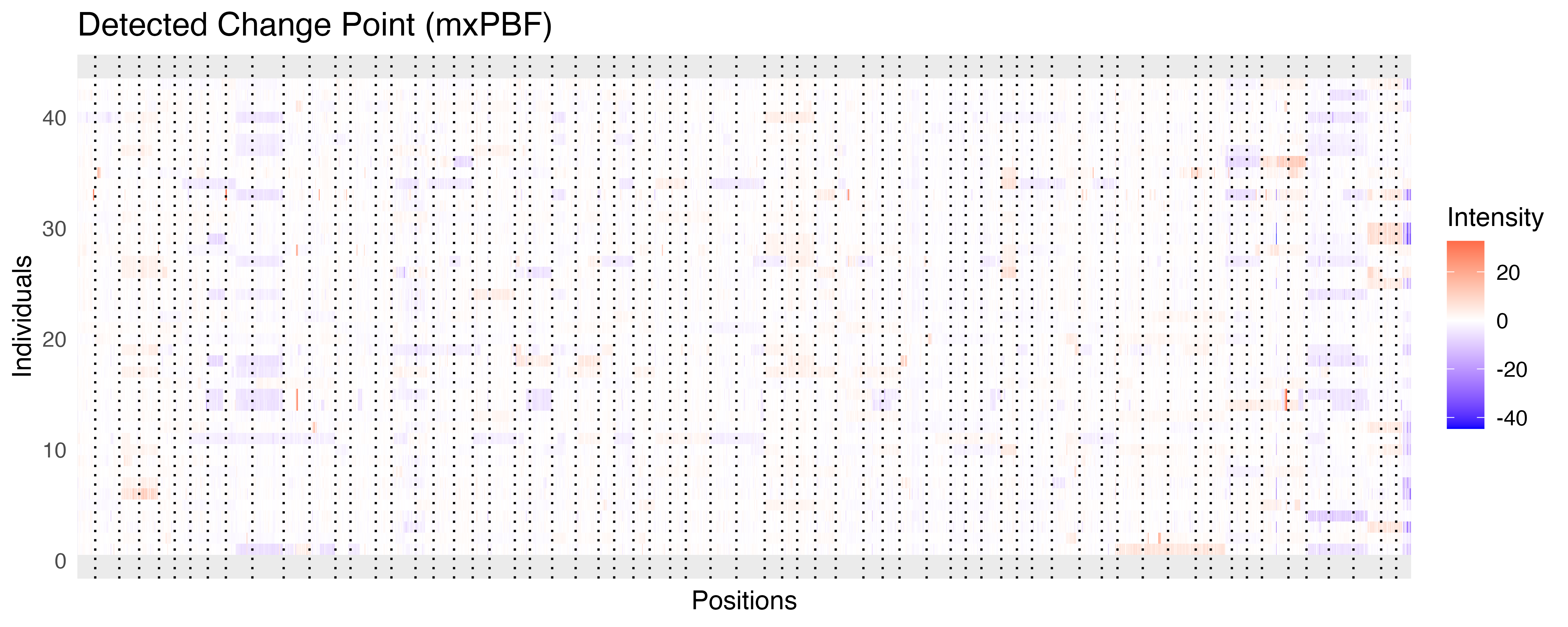}
				\caption{Log-intensity ratio measurements from microarray data of 43 individuals, with vertical dotted lines marking the identified change points.}
				\label{fig:realgenom}
			\end{figure}
			
			As illustrated in Figure \ref{fig:realgenom}, mxPBF\_major identifies 64 change points, while Inspect and Edivisive detect 59 and 64, respectively.
			There is a strong overlap among the change points detected by these methods.
			Within a margin of 15 points, approximately 83\% of the change points identified by mxPBF\_major match those detected by Inspect, and 86\% overlap with those detected by Edivisive.
			Similarly, about 80\% of the change points detected by Inspect and 88\% of those detected by Edivisive correspond to those identified by mxPBF\_major.
			Meanwhile, Geomcp detects 48 change points when the number of changes in distance and angle mappings are set to 46 and 42, respectively, based on diagnostic plots.
			The relatively small number of change points detected by Geomcp can be attributed to its design, which identifies only those changes that appear common across multiple individuals \citep{grundy2020high}.
			Approximately 85\% of the change points detected by Geomcp align with those identified by mxPBF\_major, suggesting that mxPBF\_major is also capable of detecting these changes.
			Thus, mxPBF\_major effectively detects not only individually specific change points but also those common across multiple individuals.

			\subsection{S\&P500 Stock Prices}\label{subsec:realstock}
			We now consider the daily log returns of the closing stock prices for companies in the S\&P 500 from January 2015 to December 2016. 
			The dataset was loaded using the R package \texttt{SP500R} \citep{foretsp500r}.
			The initial dataset is a matrix with \(n = 380\) rows, corresponding to dates, and 500 columns representing companies.
			First, we remove 21 columns containing missing values, which results in \(p = 479\).
			We then scale each series using the median absolute deviation, following the approach outlined in \cite{grundy2020high}.
			For Geomcp, we use the normal cost function and the CROPS algorithm \citep{haynes2017computationally}, as suggested in \cite{grundy2020high}.
			For the other parameters, we use the same settings as described previously.

			\begin{figure}[!bt]
				\centering
				\includegraphics[width = \textwidth]{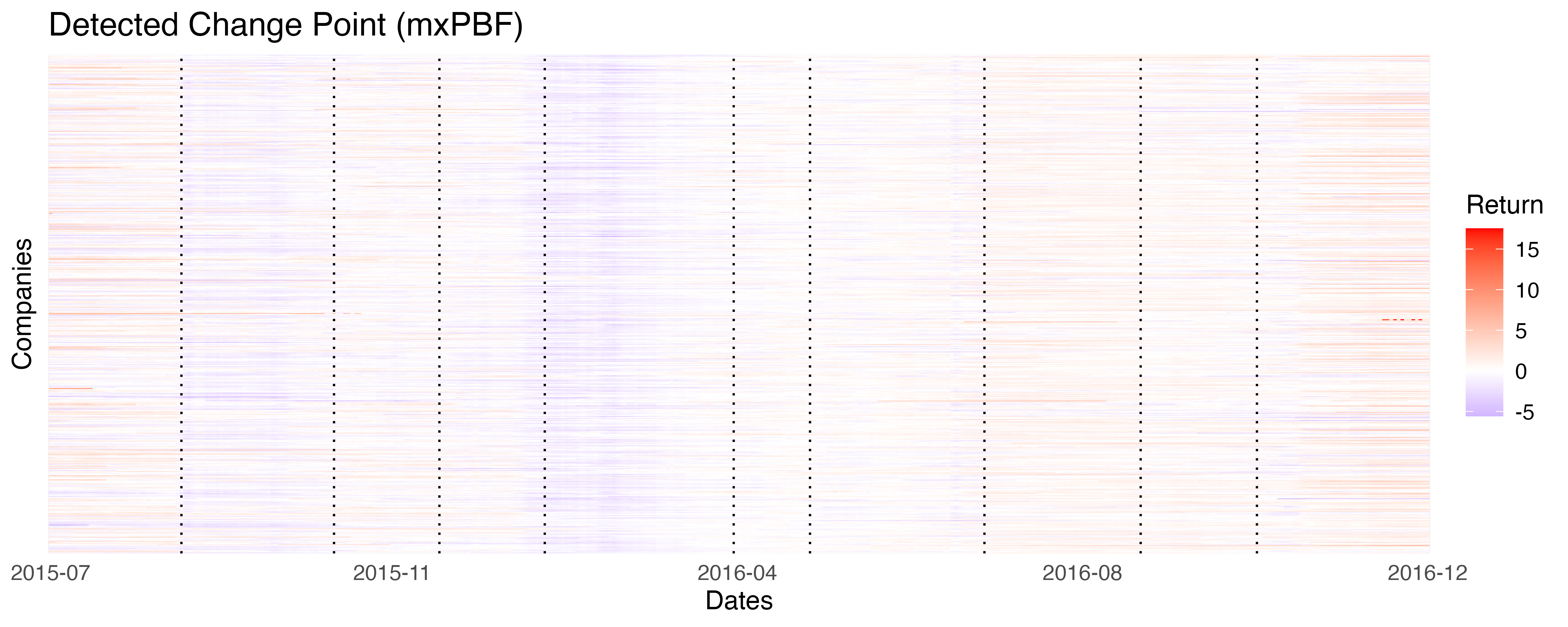}
				\caption{Log returns of 479 companies within the S\&P 500 are displayed, with vertical dotted lines indicating the identified change points.}
				\label{fig:realstock}
			\end{figure}
			
			As shown in the figure \ref{fig:realstock}, mxPBF\_major identifies 9 change points. 
			These detected change points correspond to events that broadly impacted many companies.
			For example, the change points in August 2015 and January 2016 align with significant declines in the Chinese stock markets, which affected global markets, including the S\&P 500 index. 
			The changepoint in early July 2016 likely corresponds to the announcement and outcome of the British referendum to leave the European Union, while the one in late October reflects the U.S. presidential election.
			Inspect and Edivisive methods detect 11 and 10 change points, respectively. 
			For Geomcp, we select 13 and 5 as the number of changes in distance and angle mappings, respectively, resulting in a total of 8 final change points.
			Similar to the result in \eqref{subsec:realgenomic}, approximately 88\% change points detected by Geomcp align with those identified by mxPBF\_major within 15 time points.
			Furthermore, mxPBF\_major and Edivisive demonstrate a strong relationship, as all change points detected by both methods overlap.

			\section{Discussion}\label{sec:disc}

			In this paper, we proposed Bayesian change point detection methods for high-dimensional mean and covariance structures, building upon the mxPBF framework  \citep{lee2024bayesian}.  We established the consistency of mxPBF and derived localization rates for the proposed methods under conditions that are either weaker or comparable to those required by existing approaches. Through extensive evaluations, the proposed methods demonstrated superior performance across a broad range of scenarios, outperforming current state-of-the-art techniques.

Several promising directions for future research emerge from this work. The first avenue involves enhancing the computational efficiency of the proposed methods. The primary bottleneck lies in selecting the hyperparameter  $\alpha$  which controls the empirical false positive rate. This process necessitates generating a large number of datasets and calculating mxPBF for each, significantly increasing the computational cost. To address this challenge, a possible solution is to calculate all pairwise Bayes factors (PBFs) for a given $\alpha$, retaining only those variables (or pairs of variables, in the case of covariance matrices) with “relatively large” PBF values. For example, the top 10\% quantile could serve as the threshold for identifying these values. Since this approach relies on relative magnitudes of the PBFs,  $\alpha$ can initially be fixed arbitrarily. Subsequently, the false positive rate-based method can be applied to this reduced dataset to fine-tune  $\alpha$, significantly alleviating the computational burden.

A second direction for future work is to develop adaptive change point detection methods capable of detecting both large changes in a few entries and small changes distributed across many entries. The current methods rely on mxPBF, a maximum-type Bayes factor, which may encounter difficulties in identifying change points involving numerous small-magnitude signals. Designing a Bayes factor capable of aggregating small changes across multiple components could address this limitation. By integrating such an aggregation-based approach with mxPBF, a more robust change point detection method could be developed, effectively detecting both prominent changes in a few components and subtle changes spread across many components.

		\section*{Acknowledgements}
		We would like to thank Dr. Qing Yang for providing the MATLAB code to implement their method.

			\newpage
			\appendix

			\section{Additional numerical results}\label{sec:numresults}
			\subsection{Change point detection in mean structure}
			In this section, we present additional simulation results for change point detection in the mean structure. Recall that mxPBF\_major, Geomcp, Inspect, and Edivisive refer to our multiscale method and those from \cite{grundy2020high}, \cite{wang2018high}, and \cite{matteson2014nonparametric}, respectively.
			
\begin{figure}[!bt]
				\centering
				\includegraphics[width=0.95\textwidth]{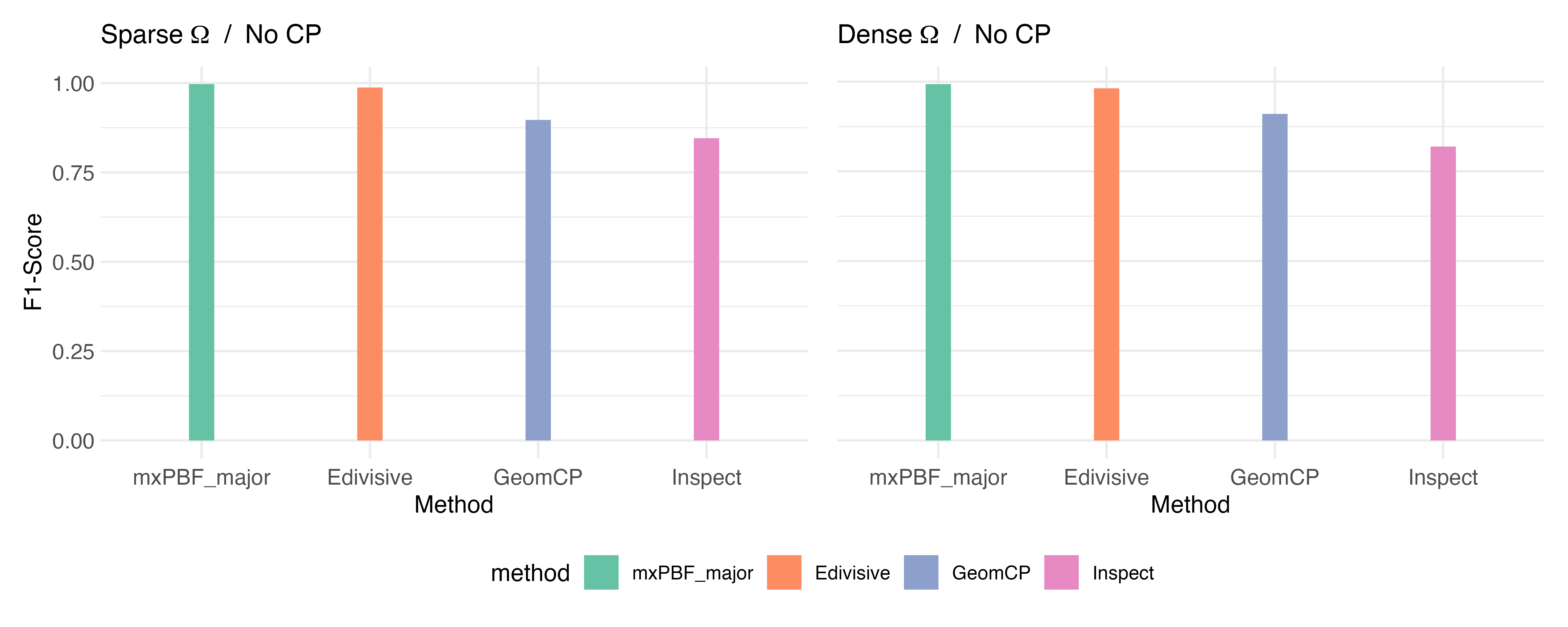}
				\includegraphics[width=0.95\textwidth]{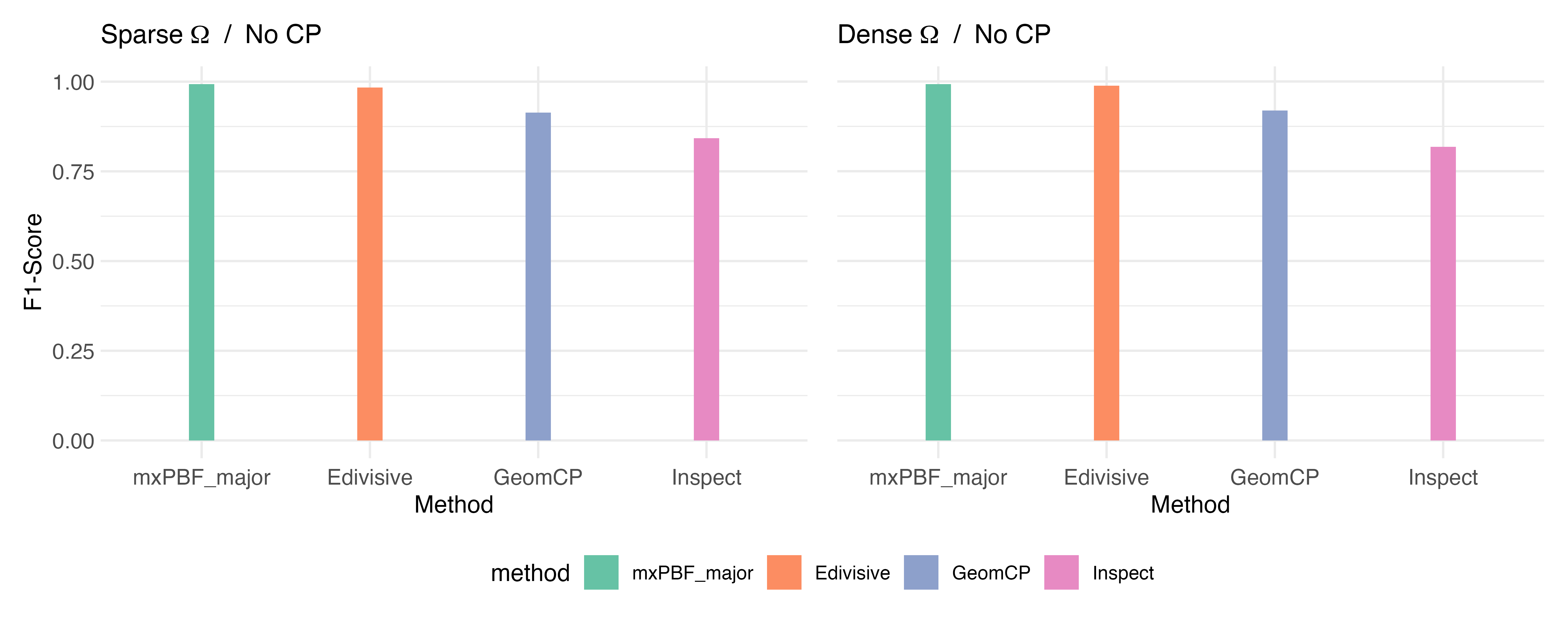}
				\caption{F1 scores for change point detection in mean structure based on 50 simulated datasets for $H_0$ with $p = 500$ (top) and $p = 800$ (bottom).}
				\label{fig:Fscore_null_mean_500}
			\end{figure}

			Figure \ref{fig:Fscore_null_mean_500} shows the F1 scores for each method, based on 50 simulated datasets under $H_0$ with $p = 500$ and $p = 800$. 
			Among the methods, mxPBF\_major and Edivisive demonstrate superior performance, consistent with the results observed when $p = 200$.

			\begin{figure}[!tb]
				\centering
				\includegraphics[width=0.97\textwidth]{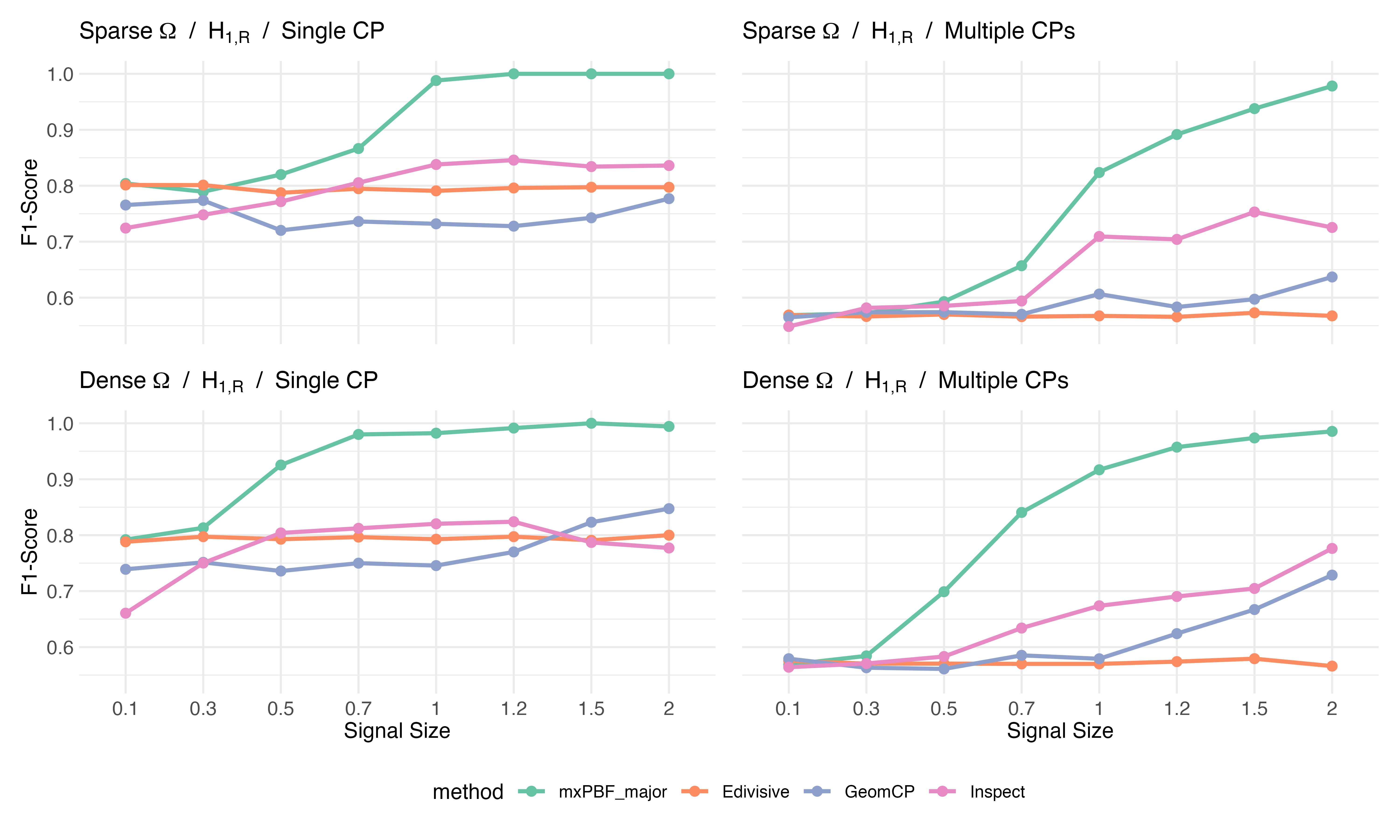}
				\includegraphics[width=0.97\textwidth]{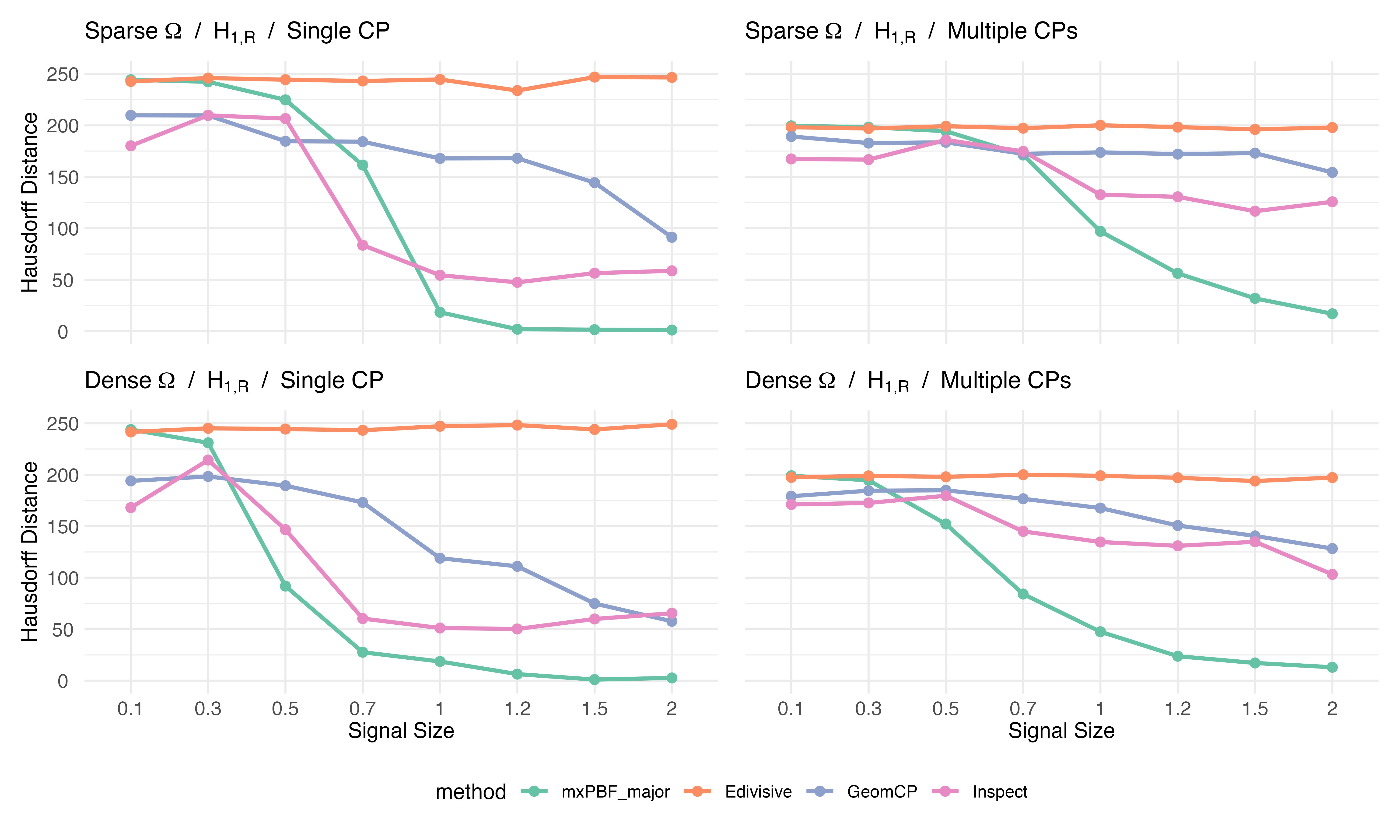}
				\caption{F1 scores and Hausdorff distances for change point detection in mean structure based on 50 simulated datasets under \( H_{1,R} \) with \( p = 500 \).}
				\label{fig:FH_H1R_mean_500}
			\end{figure} 
			\begin{figure}[!tb]
				\centering
				\includegraphics[width=0.97\textwidth]{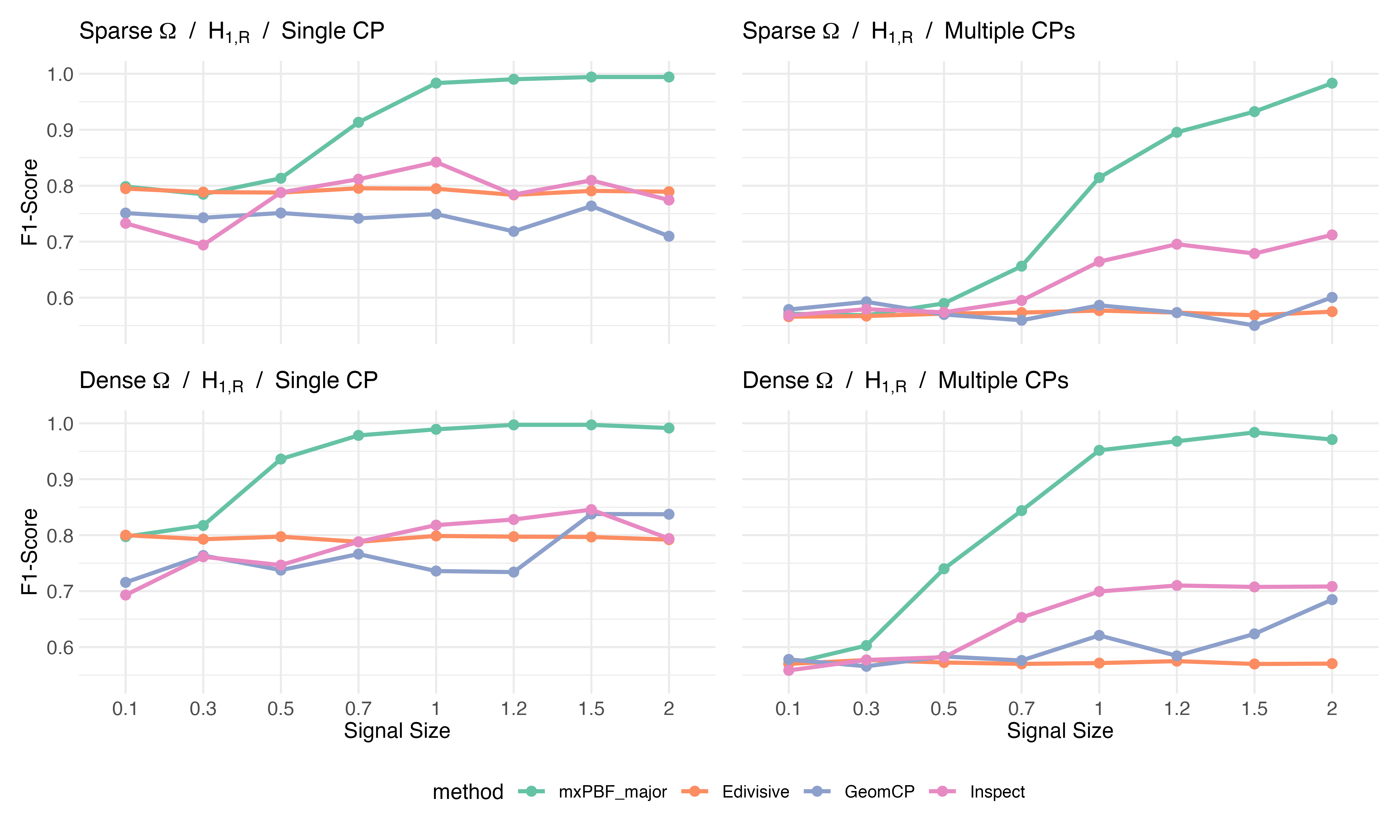}
				\includegraphics[width=0.97\textwidth]{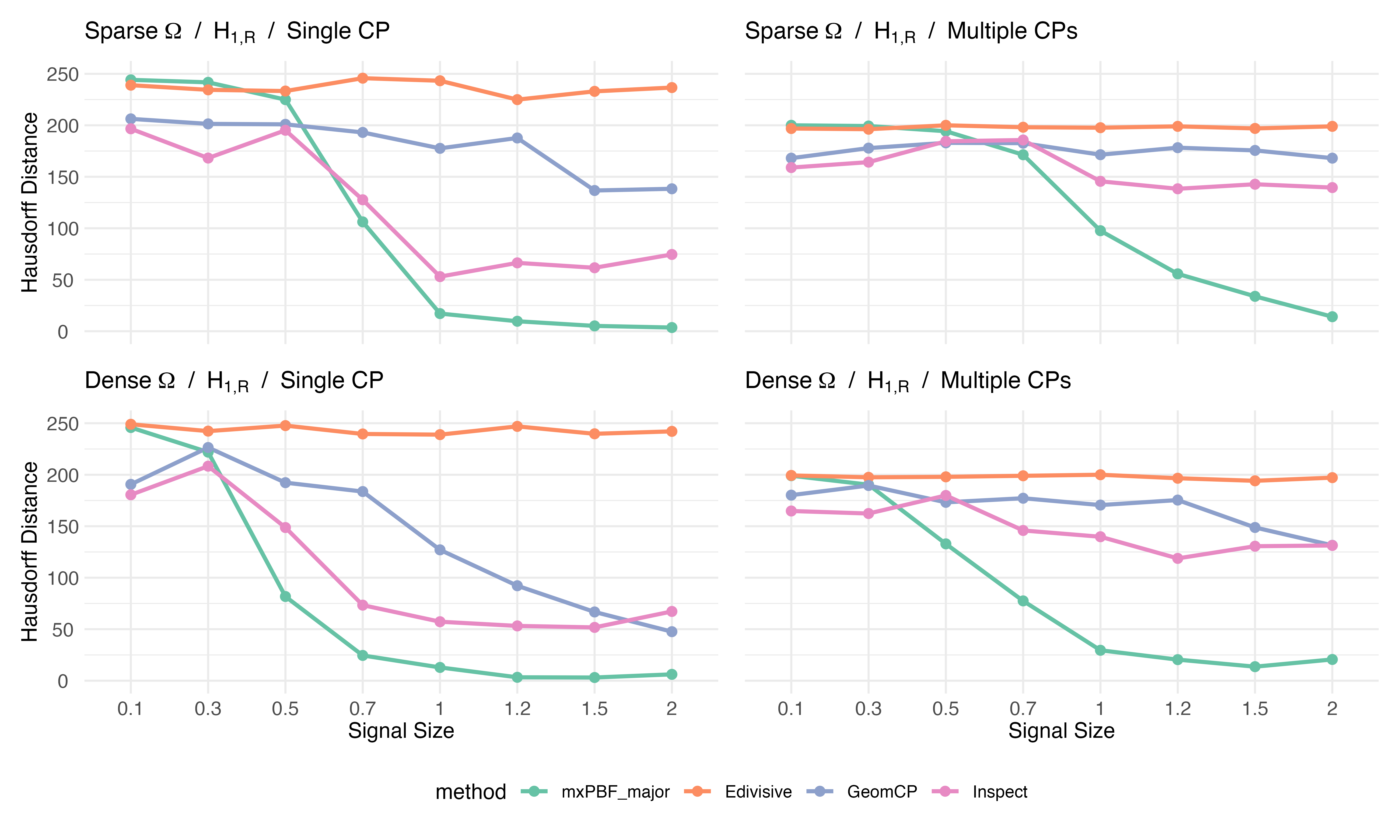}
				\caption{F1 scores and Hausdorff distances for change point detection in mean structure based on 50 simulated datasets under \( H_{1,R} \) with \( p = 800 \).}
				\label{fig:FH_H1R_mean_800}
			\end{figure}
			Figures \ref{fig:FH_H1R_mean_500} and \ref{fig:FH_H1R_mean_800} describe the suggested metrics (F1 score and Hausdorff distance) based on 50 simulated datasets under ``rare signals'' scenario  $H_{1,R}$, with $p = 500$ and $p = 800$, respectively. 
			The results show similar trends to those observed when $p = 200$.
			mxPBF\_major demonstrates significant improvement as the signal size increases and outperforms the other methods in most scenarios.
			Inspect and Geomcp show improved performance as the signal size increases but tends to produce a large number of false positives, and Edivisive appears to struggle in accurately detecting changes.
			

			\begin{figure}[!tb]
				\centering
				\includegraphics[width=0.97\textwidth]{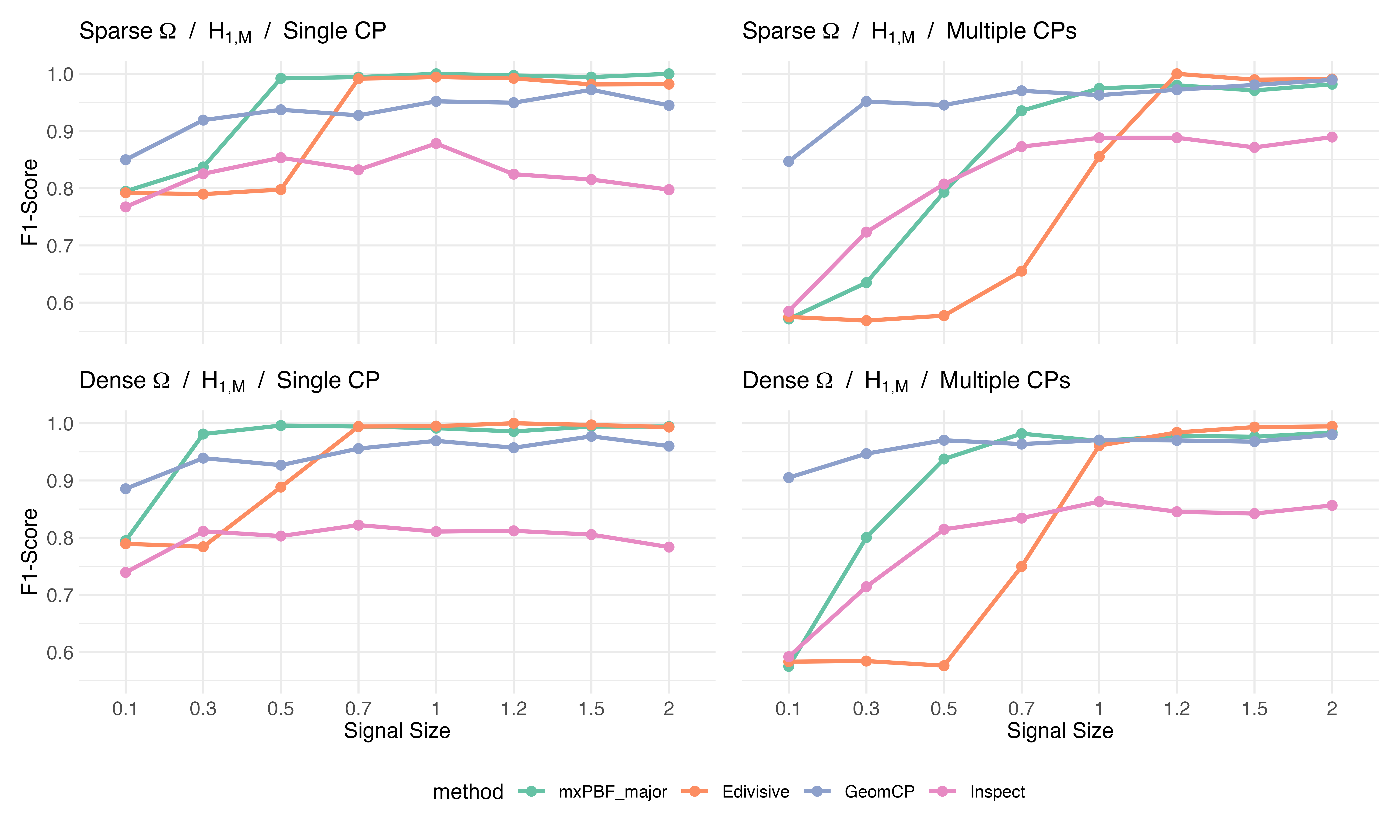}
				\includegraphics[width=0.97\textwidth]{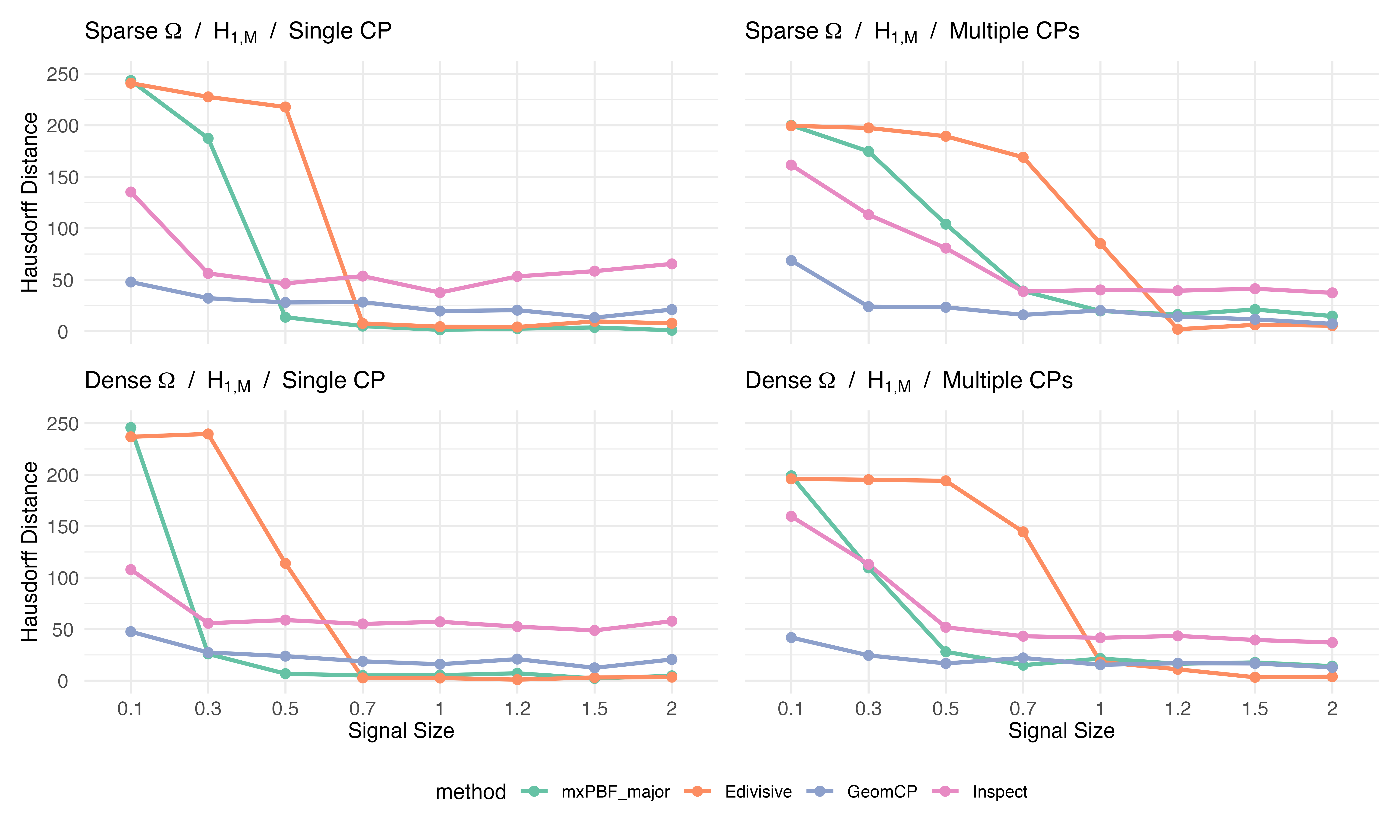}
				\caption{F1 scores and Hausdorff distances based for change point detection in mean structure on 50 simulated datasets under \( H_{1,M} \) with \( p = 500 \).}
				\label{fig:FH_H1M_mean_500}
			\end{figure} 
			\begin{figure}[!bt]
				\centering
				\includegraphics[width=0.97\textwidth]{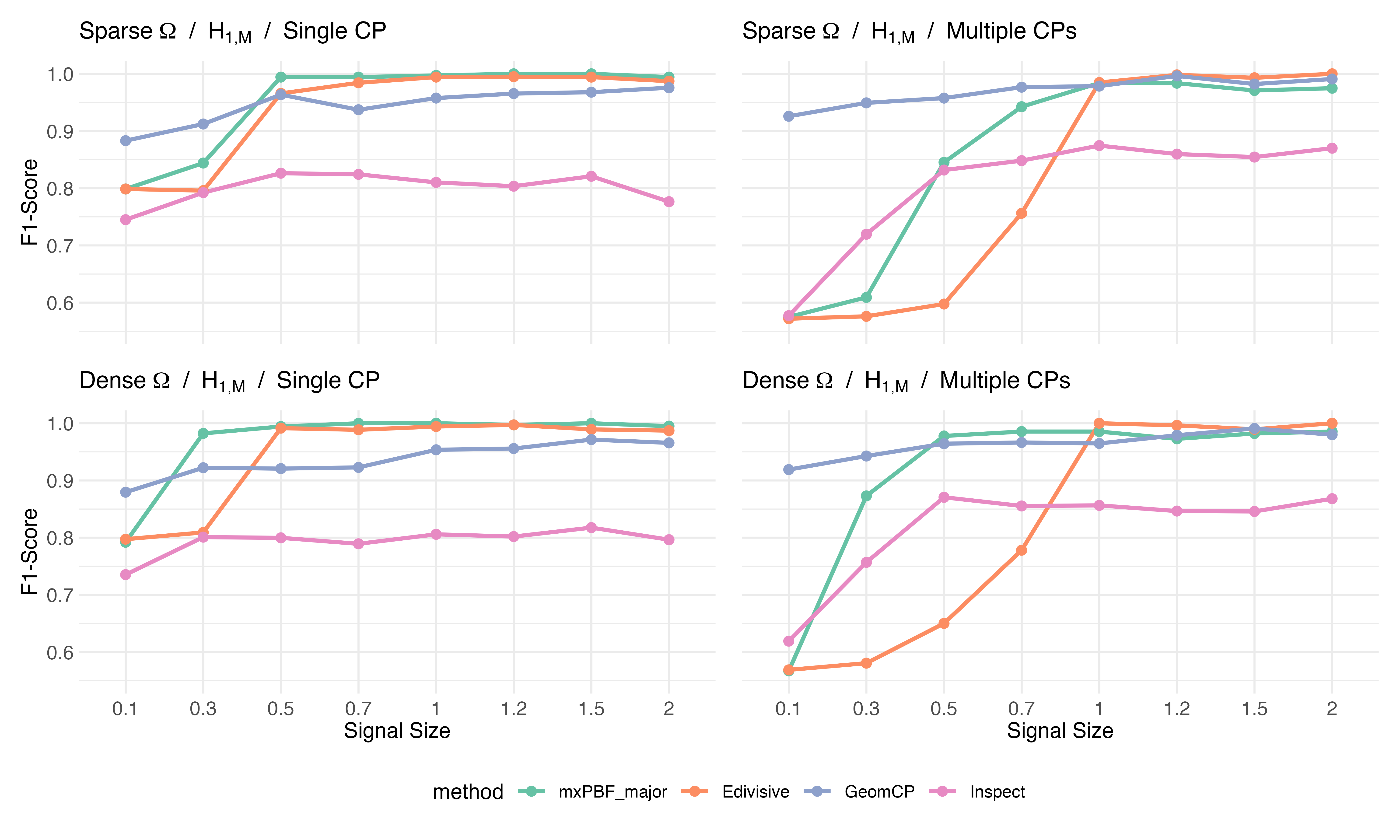}
				\includegraphics[width=0.97\textwidth]{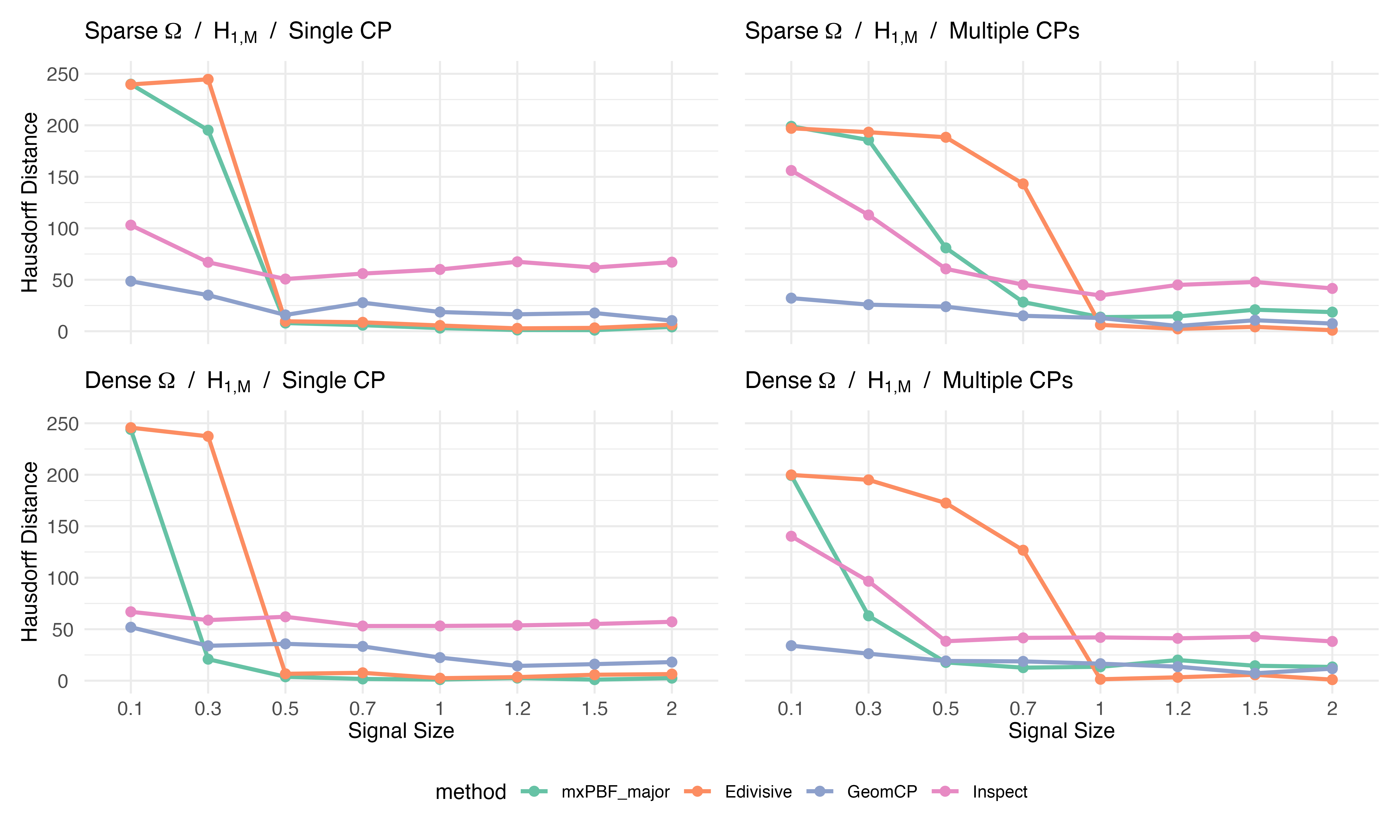}
				\caption{F1 scores and Hausdorff distances for change point detection in mean structure based on 50 simulated datasets under \( H_{1,M} \) with \( p = 500 \).}
				\label{fig:FH_H1M_mean_800}
			\end{figure}
			
			Figures \ref{fig:FH_H1M_mean_500} and \ref{fig:FH_H1M_mean_800} present the suggested metrics based on 50 simulated datasets under ``many signals'' scenario $H_{1,M}$, with $p = 500$ and $p = 800$, respectively.
			The results show similar trends to those observed when $p = 200$.
			When signals are small, Inspect and Geomcp show decent performance, but mxPBF\_major and Edivisive also demonstrate improved performance when signals are large.
			Notably, the performance of mxPBF\_major improves rapidly as the signal size increases.
			

			\subsection{Change point detection in covariance structure}
			Next, we present additional numerical results for change point detection in the covariance structure. Recall that mxPBF\_major, Dette, Geomcp, and Edivisive refer to our multiscale method and those from \cite{dette2022estimating}, \cite{grundy2020high}, and \cite{matteson2014nonparametric}, respectively.

			\begin{figure}[!bt]
				\centering
				\includegraphics[width=0.97\textwidth]{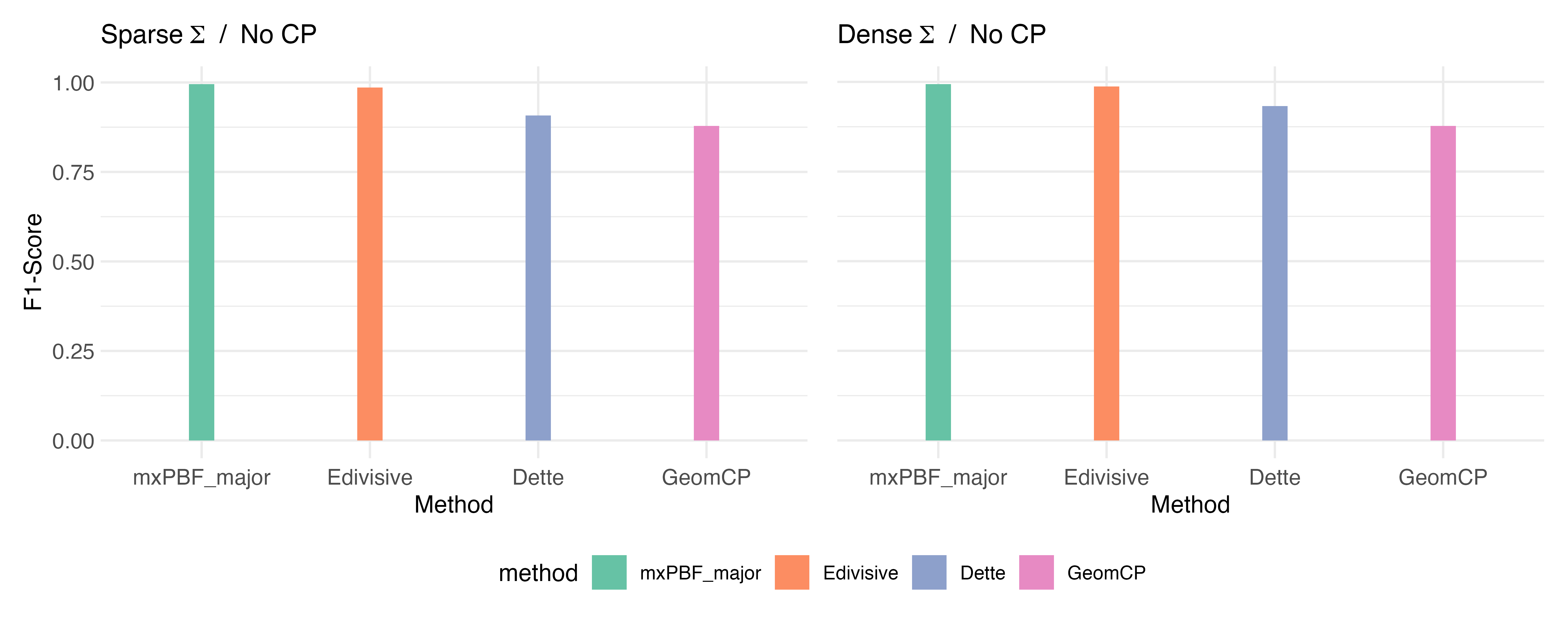}
				\includegraphics[width=0.97\textwidth]{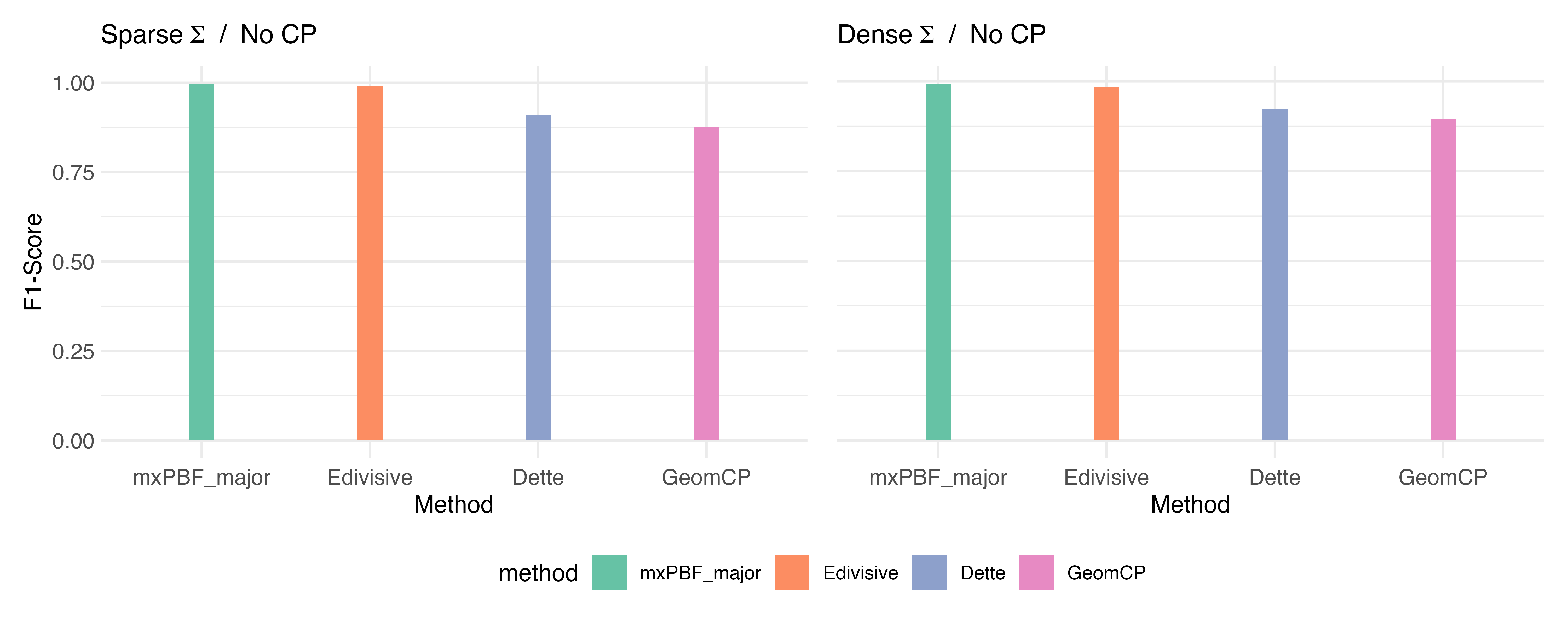}
				\caption{F1 scores for change point detection in covariance structure based on 50 simulated datasets for $H_0$ with $p = 500$ (top) and $p=800$ (bottom).}
				\label{fig:Fscore_null_cov_500}
			\end{figure}
			Figure \ref{fig:Fscore_null_cov_500} shows the F1 scores for each method, based on 50 simulated datasets under $H_0$, with $p = 500$ and $p = 800$.
			Similar to the result observed when $p = 200$, mxPBF\_major and Edivisive demonstrate superior performance in avoiding false positives.

			\begin{figure}[!tb]
				\centering
				\includegraphics[width=0.97\textwidth]{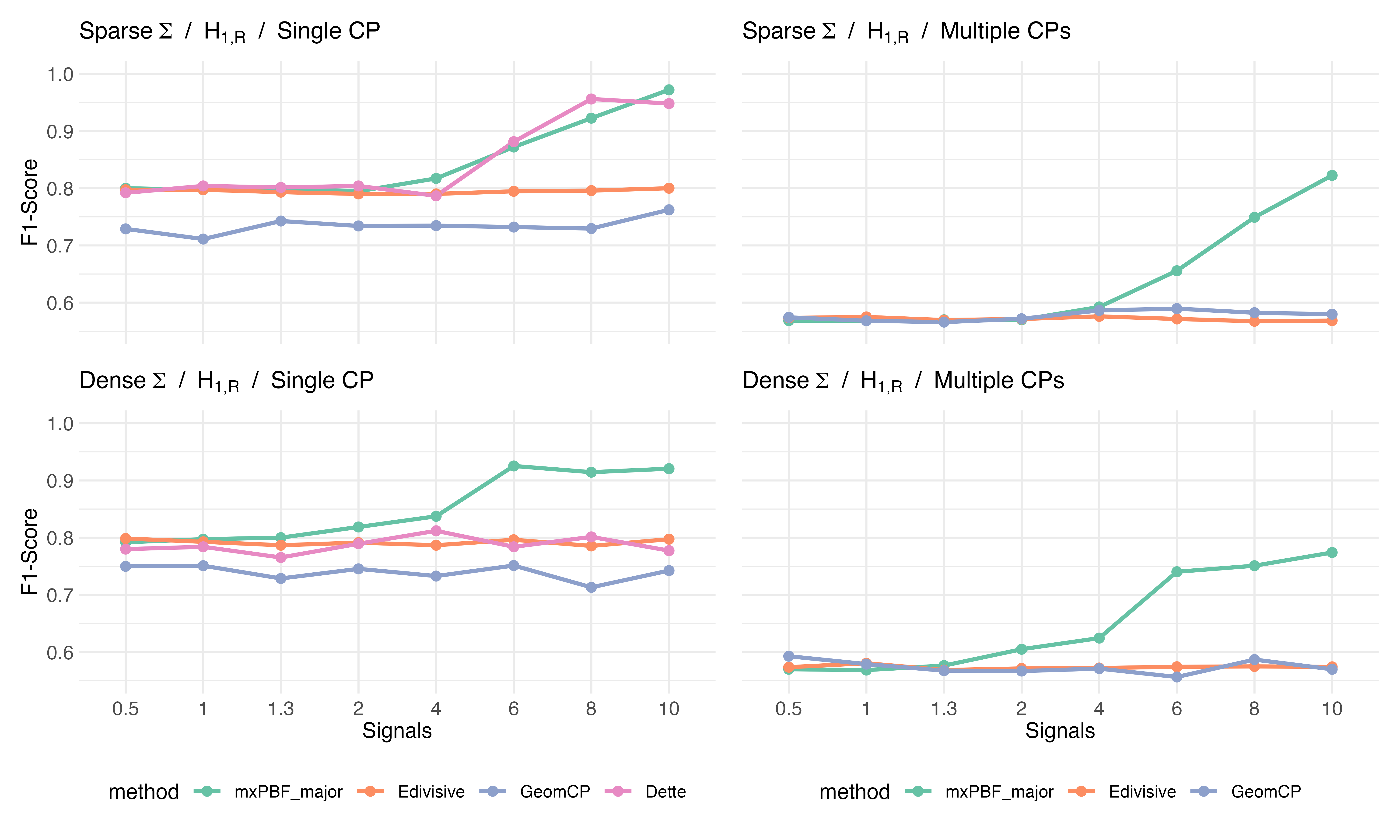}
				\includegraphics[width=0.97\textwidth]{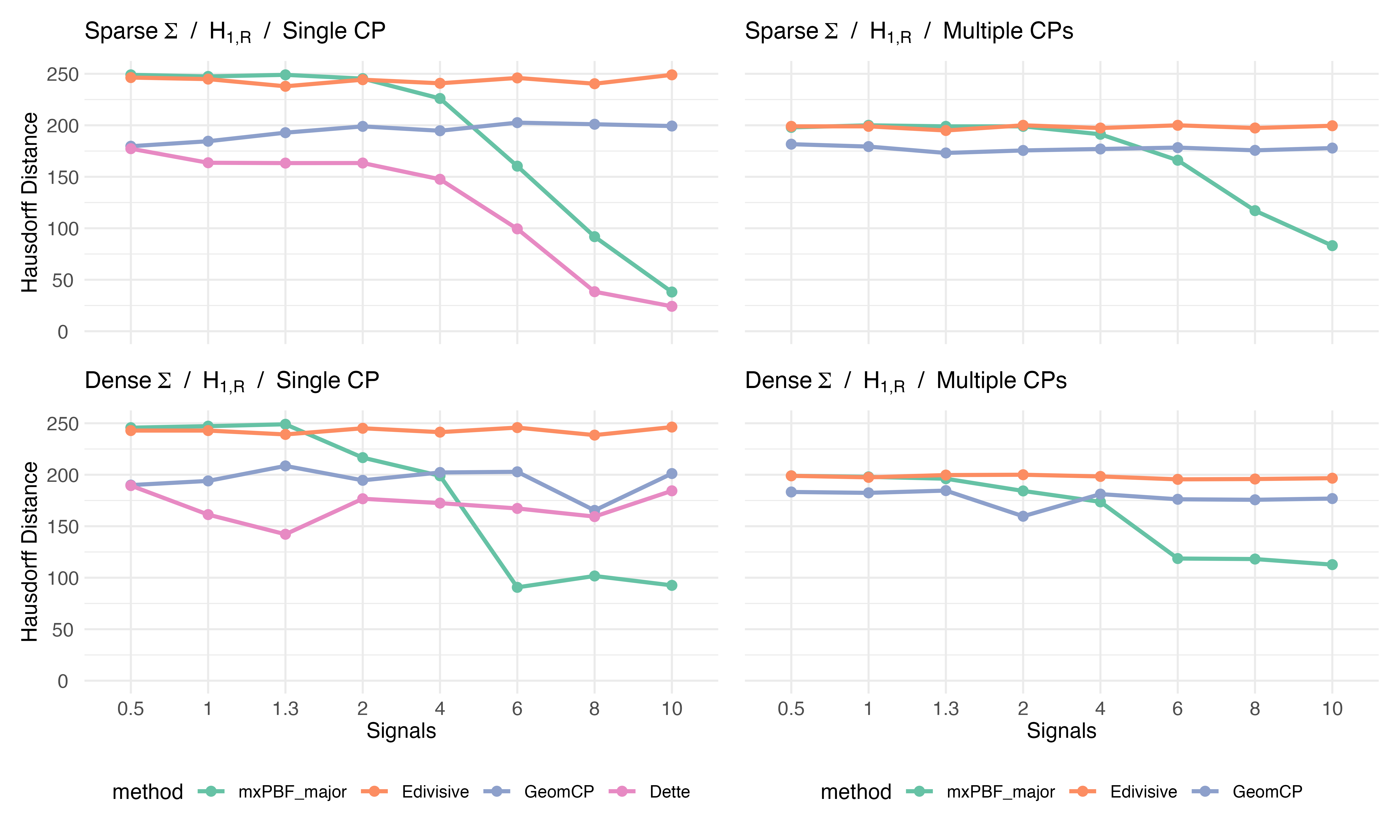}
				\caption{F1 scores and Hausdorff distances for change point detection in covariance structure based on 50 simulated datasets under \( H_{1,R} \) with \( p = 500 \).}
				\label{fig:FH_H1R_cov_500}
			\end{figure}
			\begin{figure}[!bt]
				\centering
				\includegraphics[width=0.97\textwidth]{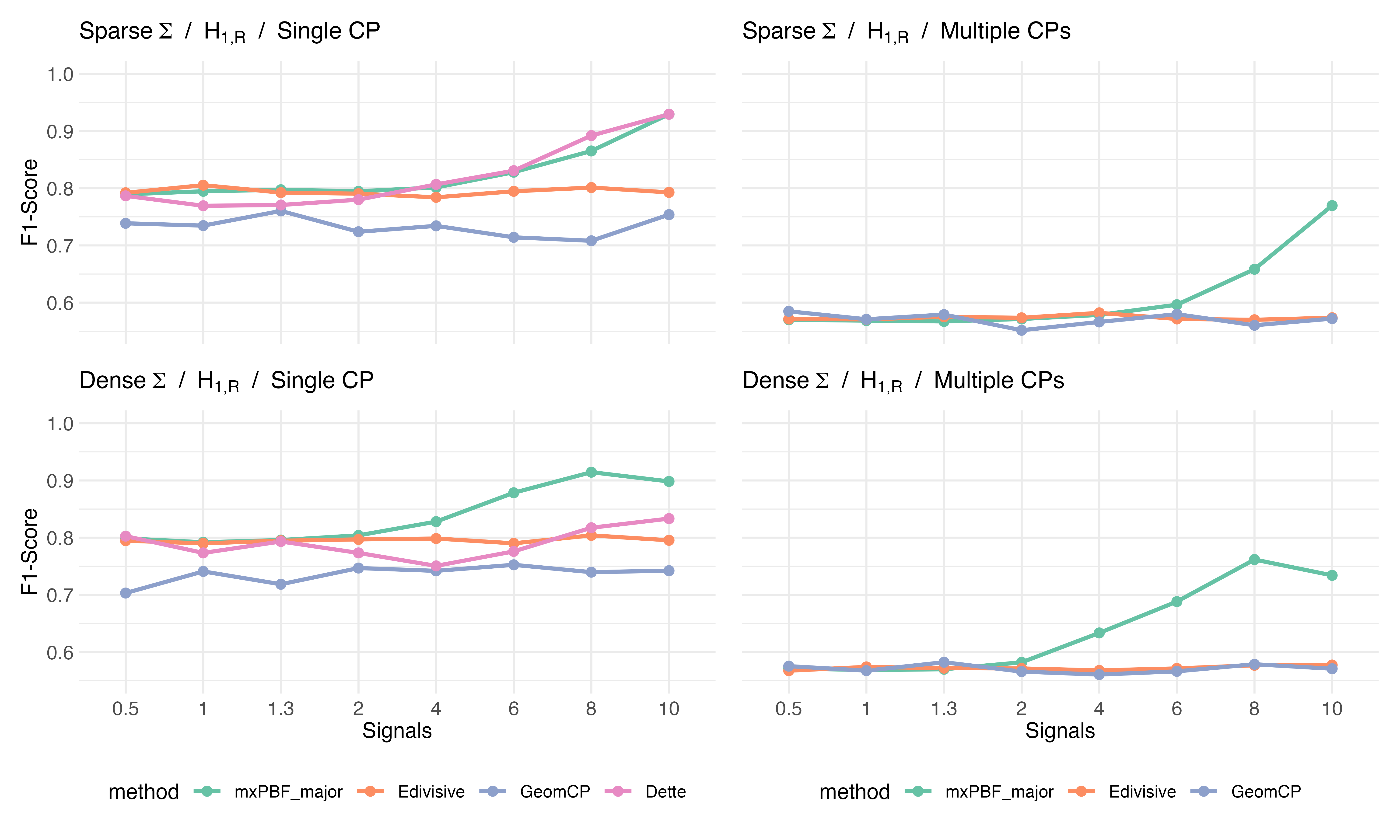}
				\includegraphics[width=0.97\textwidth]{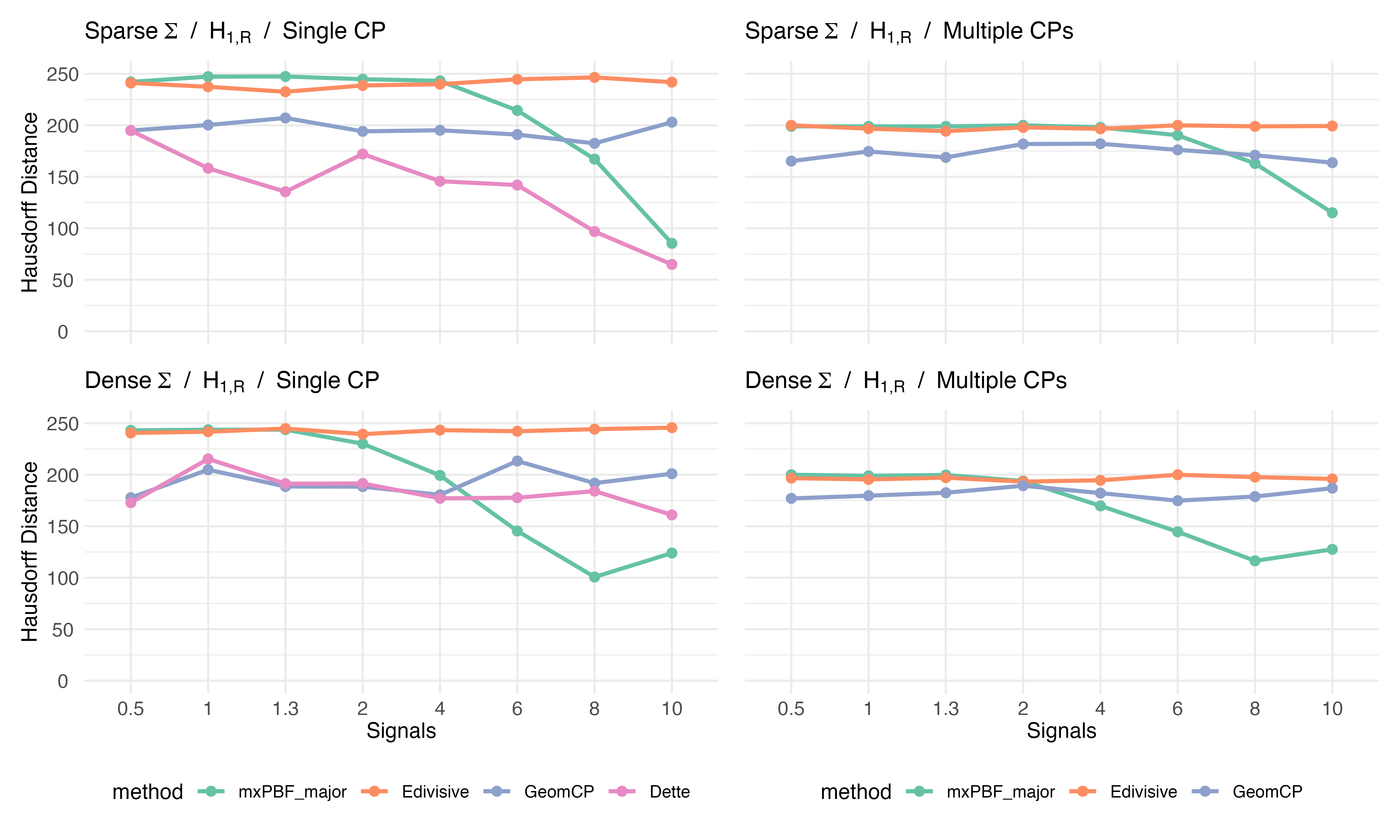}
				\caption{F1 scores and Hausdorff distances for change point detection in covariance structure based on 50 simulated datasets under \( H_{1,R} \) with \( p = 800 \).}
				\label{fig:FH_H1R_cov_800}
			\end{figure}
			Figures \ref{fig:FH_H1R_cov_500} and \ref{fig:FH_H1R_cov_800} display the suggested metrics based on 50 simulated datasets under ``rare signals'' scenario $H_{1,R}$, with $p = 500$ and $p = 800$, respectively.
			Similar to the case when $p = 200$, Edivisive and Geomcp struggle to detect change points, showing no improvement in performance as the signal size increases.
			Dette performs well in the sparse covariance scenario but struggles to detect changes when the covariance is dense.
			Conversely, mxPBF\_major outperforms the other methods in most cases, with its performance consistently improving as the signal size grows.

			\begin{figure}[!bt]
				\centering
				\includegraphics[width=0.97\textwidth]{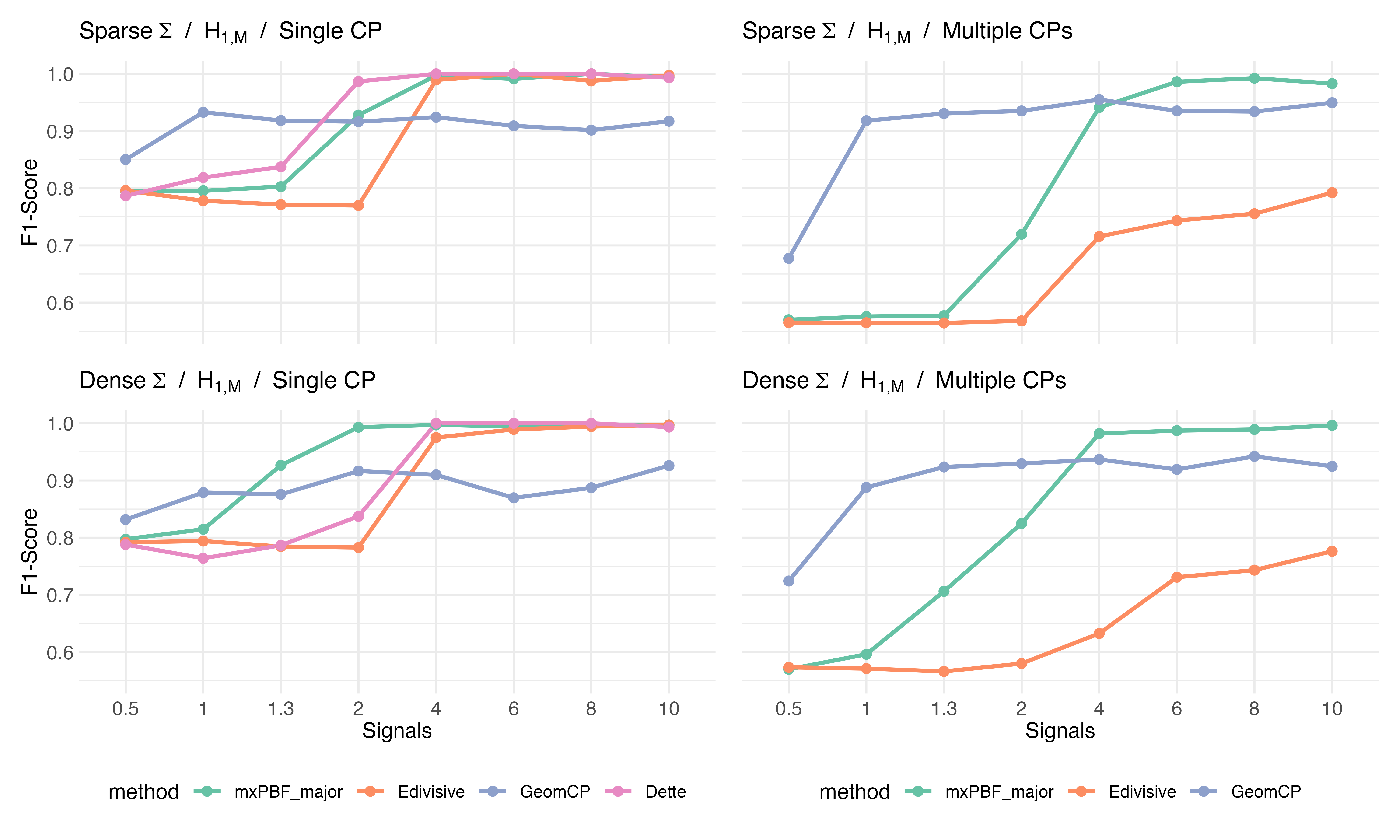}
				\includegraphics[width=0.97\textwidth]{Plots/haus_cov_500_h1r.png}
				\caption{F1 scores and Hausdorff distances for change point detection in covariance structure  based on 50 simulated datasets under \( H_{1,M} \) with \( p = 500 \).}
				\label{fig:FH_H1M_cov_500}
			\end{figure}
			\begin{figure}[!bt]
				\centering
				\includegraphics[width=0.97\textwidth]{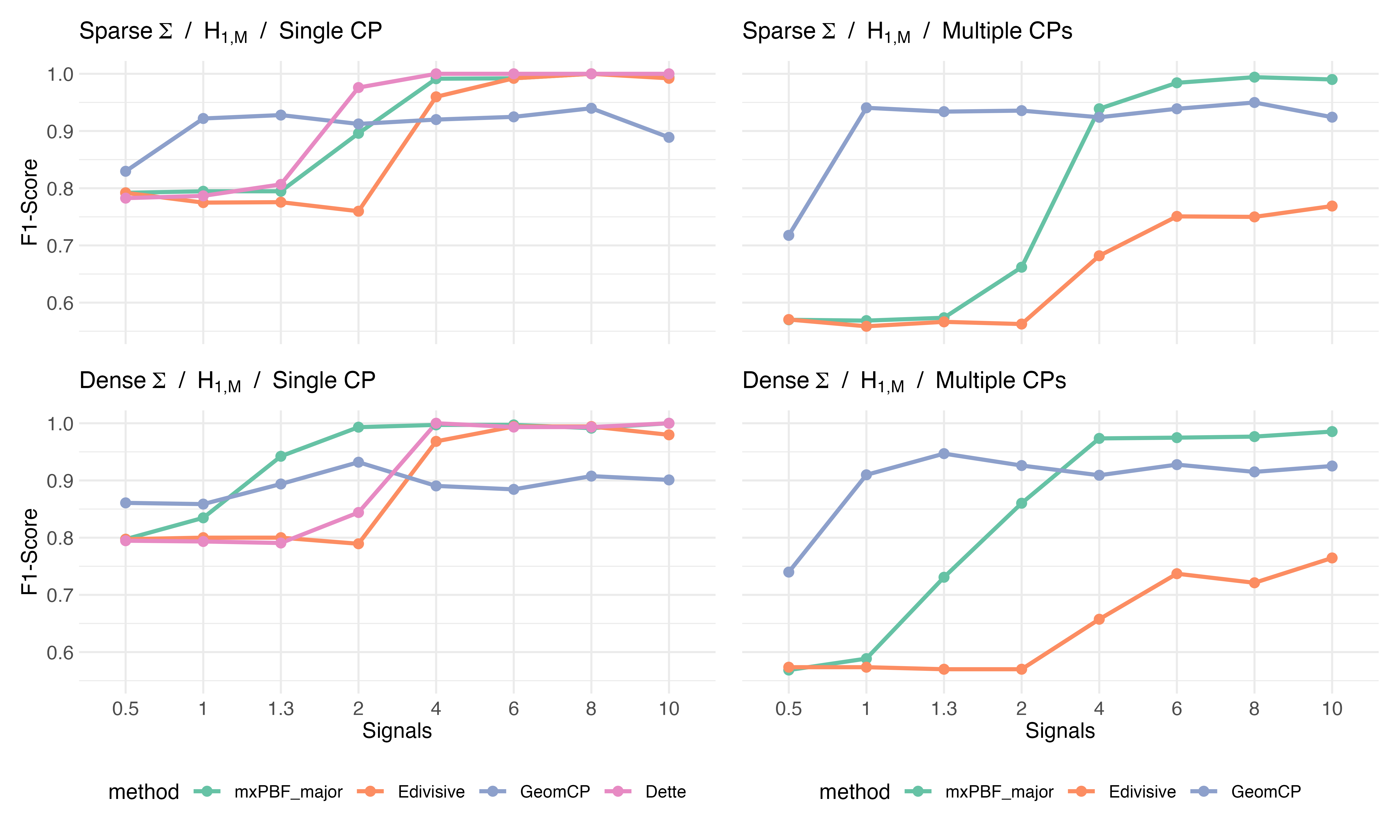}
				\includegraphics[width=0.97\textwidth]{Plots/haus_cov_800_h1r.png}
				\caption{F1 scores and Hausdorff distances for change point detection in covariance structure based on 50 simulated datasets under \( H_{1,M} \) with \( p = 800 \).}
				\label{fig:FH_H1M_cov_800}
			\end{figure}
			Figures \ref{fig:FH_H1M_cov_500} and \ref{fig:FH_H1M_cov_800} display the metrics derived from 50 simulated datasets under ``many signals'' scenario $H_{1,M}$, with $p = 500$ and $p = 800$, respectively. 
			The results show similar trends to those observed when $p = 200$.
			In scenarios with many signals, all methods show improved performance, likely due to the increasing number of signals with higher dimensions. 
			In scenarios with smaller signals, Geomcp outperforms the other methods. 
			However, its performance gains are limited by the detection of numerous false positives. 
			In the single change point scenario, mxPBF\_major, Dette and Edivisive effectively detect changes. 
			In the multiple change point scenario, Edivisive struggles to detect changes accurately, even with larger signals, while mxPBF\_major performs nearly perfectly when $\phi \geq 6$.

			\section{FPR-based method for selecting $\alpha$}\label{sec:alpha}
			
			In this section, we describe the false positive rate (FPR)-based method for selecting the hyperparameter $\alpha$ of mxPBF.
			Given an observed data matrix \( \bfX_n \), let \( \hat{\mu} \) and \( \hat{\Sigma} \) represent the sample mean and covariance matrix derived from \( \bfX_n \), respectively.
			If $\hat{\Sigma}$ is not positive definite, we make it positive definite by adding $\{-\lambda_{\min}(\hat{\Sigma}) + 0.1^3\}I_p$.
			We then generate a simulated dataset \( \bfX_{\text{sim}} = (X_{1,\text{sim}}, \ldots, X_{n,\text{sim}})^T \in \mathbb{R}^{n \times p} \), where each \( X_{i, \text{sim}} \) is a random sample from \( N_p(\hat{\mu}, \hat{\Sigma}) \). 
			Note that \( \bfX_{\text{sim}} \) can be considered a dataset simulated under the null hypothesis, with its mean vector and covariance matrix approximating those of the true data-generating distribution.
			For given a window size $n_w$ and a hyperparameter $\alpha$, we calculate the mxPBF based on $\bfX_{\text{sim}} $, denoted as \( B_{\max,10,\alpha}^{n_w}(\bfX_{\text{sim}}) \), 
			and reject the null if the mxPBF exceeds the threshold $C_{th}$.
			Note that $B_{\max,10,\alpha}^{n_w}(\bfX_{\text{sim}})$ can be either $B_{\max,10,\alpha}^{\mu, n_w}(\bfX_{\text{sim}})$ or $B_{\max,10,\alpha}^{\Sigma, n_w}(\bfX_{\text{sim}})$, and an mxPBF exceeding the threshold corresponds to a false positive.
			By generating $N$ simulated datasets $(\bfX_{\text{sim}}^{(s)})_{s=1}^N$ , we can calculate the following empirical FPR for each $\alpha$,
			\bea
			\widehat{\text{FPR}}_\alpha=N^{-1}\sum_{s=1}^{N}I(B^{n_w}_{\max,10,\alpha}(\bfX_{\text{sim}}^{(s)})>C_{th}).
			\eea
			We recommend setting the number of generations \(N\) between 300 and 500 in practice. 
			An experiment supporting this suggestion is described in Section \ref{subsec:Nsel}.
			
			Note that, after calculating the mxPBF for a given \( \alpha \), the mxPBF for a different \( \alpha^\star \neq \alpha \) can be easily computed as 
			\[
			B^{n_w}_{\max,10,\alpha^\star}(\bfX_{\text{sim}}) = B^{n_w}_{\max,10,\alpha}(\bfX_{\text{sim}}) - 0.5 \log\left(\frac{\gamma}{1+\gamma}\right) + 0.5 \log\left(\frac{\gamma^\star}{1+\gamma^\star}\right),
			\]
			where \( \gamma^\star = (n_w \vee p)^{-\alpha^\star} \) and \( \gamma = (n_w \vee p)^{-\alpha} \). 
			By forming a grid of values, such as \( \alpha \in \{0.01, 0.02, \dots, 15\} \), we can select the optimal value of \( \alpha \), denoted \( \widehat{\alpha} \), that achieves a prespecified FPR.
			
			\subsection{Sufficient number of simulated datasets}\label{subsec:Nsel}
			
			In this section, we provide empirical evidence to recommend a sufficient number of simulated datasets, $N$, for the stable selection of the hyperparameter $\hat{\alpha}$, which controls the empirical false positive rate.
			Since we generate simulated datasets using the sample mean and covariance, and select $\hat{\alpha}$ based on their quantiles, it is expected that generating more simulated datasets will lead to a more stable selection of $\hat{\alpha}$. 
			However, generating too many simulated datasets would require heavy computation.
			Therefore, we aim to find a proper range for generating samples that not only achieves stability in selecting \(\hat{\alpha}\), but is also computationally practical.
			Here, we design an experiment to empirically determine the sufficient range.  
			
			For simplicity, we fix \(n = 500\) and \(p = 100\), consider many signals and a dense precision or covariance matrix, and set \(\mu = 1\) and \(\psi = 8\) under the alternative hypotheses for the mean and covariance structure.  
			Additionally, we use a window size of \(n_w = 50\) and explore a wide range for the number of simulated datasets, \(N \in \{50, 100, 200, 300, 400, 500, 700, 1000\}\).
			For each dataset and number of simulated datasets $N$, we repeated the selection of $\hat{\alpha}$ 100 times.
			For other settings, we used those described in the paper.

			\begin{figure}[!bt]
				\centering
				\includegraphics[width=\textwidth]{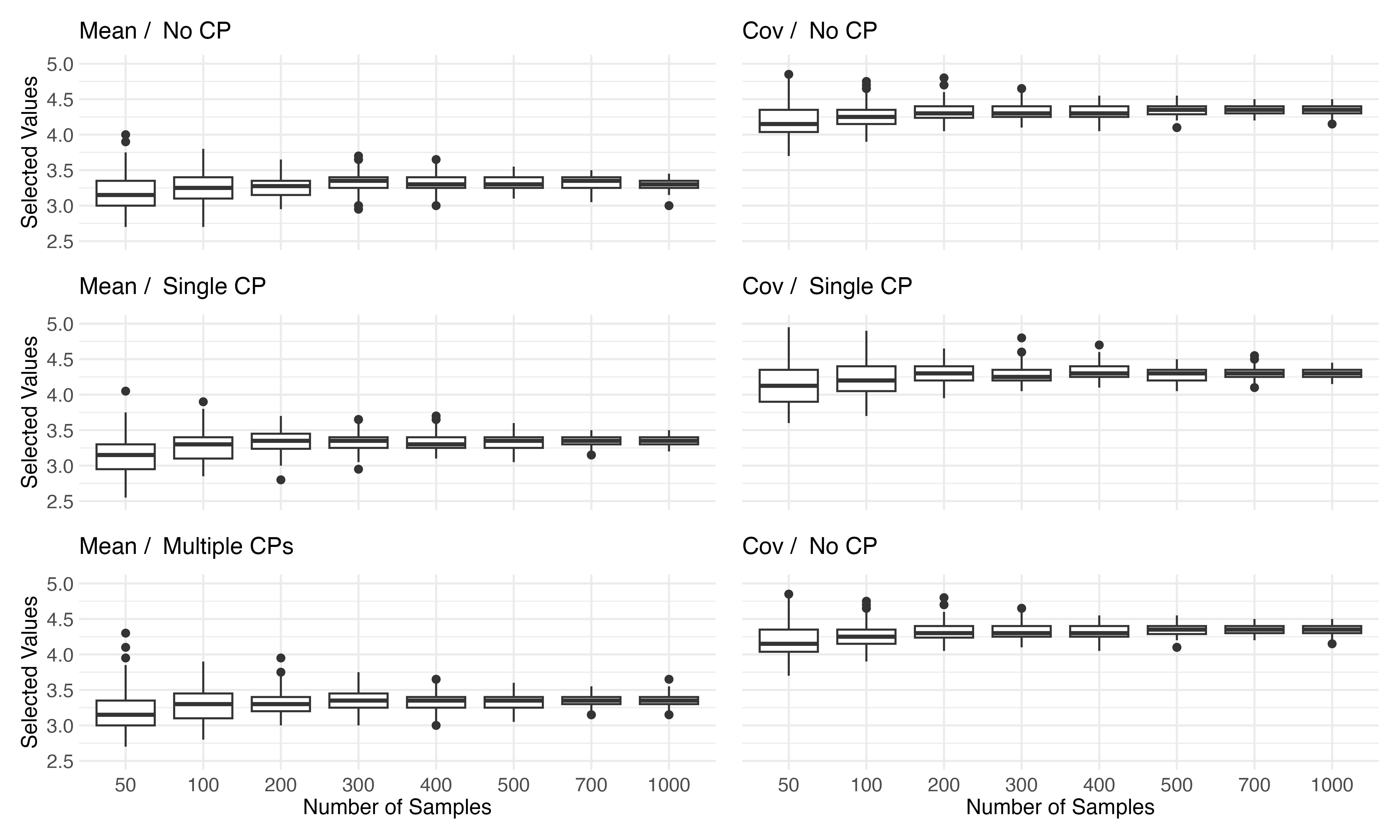}
				\caption{The selected $\hat{\alpha}$, which controlls the empirical false positive rate at 0.05, is based on 100 replications for both the mean and covariance methods, and for each of the following hypotheses: $H_{0}$ and $H_{1,M}$, with $p = 100$.}
				\label{fig:sufficientsample}
			\end{figure}
			Figure \ref{fig:sufficientsample} presents the distribution of the selected $\hat{\alpha}$, controlling the empirical false positive rate at 0.05 in each case. 
			With a smaller number of generations, the selected \(\hat{\alpha}\) appears unstable, exhibiting a wide range. 
			However, once the number of generations reaches 200 or 300, it appears to stabilize. 
			When $N\ge 300$, the range does not appear to become significantly narrower. 
			In every scenario, a similar distribution of the selected $\hat{\alpha}$ is observed, which suggests that more than $N=300$ generations are generally sufficient.

			\section{The multiscale method}
			In this section, we compare the performance of the proposed methods using various window settings, including both single and multiple windows, to highlight the advantages of the multiscale approach. 
			As explained in the paper, the proposed method requires careful selection of the window size, which can be challenging in many cases.
			We aim to demonstrate the impact of selecting an inappropriate window size and the effectiveness of the multiscale method.
			For simplicity, we fix $n=500$ and $p=200$, considering only F1 scores.
			We compared the performance of five proposed methods using different single windows against two versions of mxPBF\_major that utilize five and three windows, respectively.
			For mxPBF\_major with five windows, we use a set of window sizes \( \mathcal{N}_F = \{25, 40, 60, 80, 100\} \) and denote this method as Full.
			For mxPBF\_major with three windows, we use a set of window sizes \( \mathcal{N}_D = \{25, 60, 100\} \) and refer to this method as Default.
			For the methods using a single window, we use \( \mathcal{N}=25, \mathcal{N}=40, \mathcal{N}=60, \mathcal{N}=80, \mathcal{N}=100\), respectively, denoting these methods by their window sizes.
			Unless indicated otherwise, we use the same settings described in the paper.

			\subsection{Performance of single window methods in the mean structure}
			First, we compare the performance in detecting changes in the mean structure. 
			We consider signal sizes of \(\mu \in \{0.05, 0.1, 0.3, 0.5, 0.7, 1, 1.2\}\).
			\begin{figure}[!bt]
				\centering
				\includegraphics[width=0.95\textwidth]{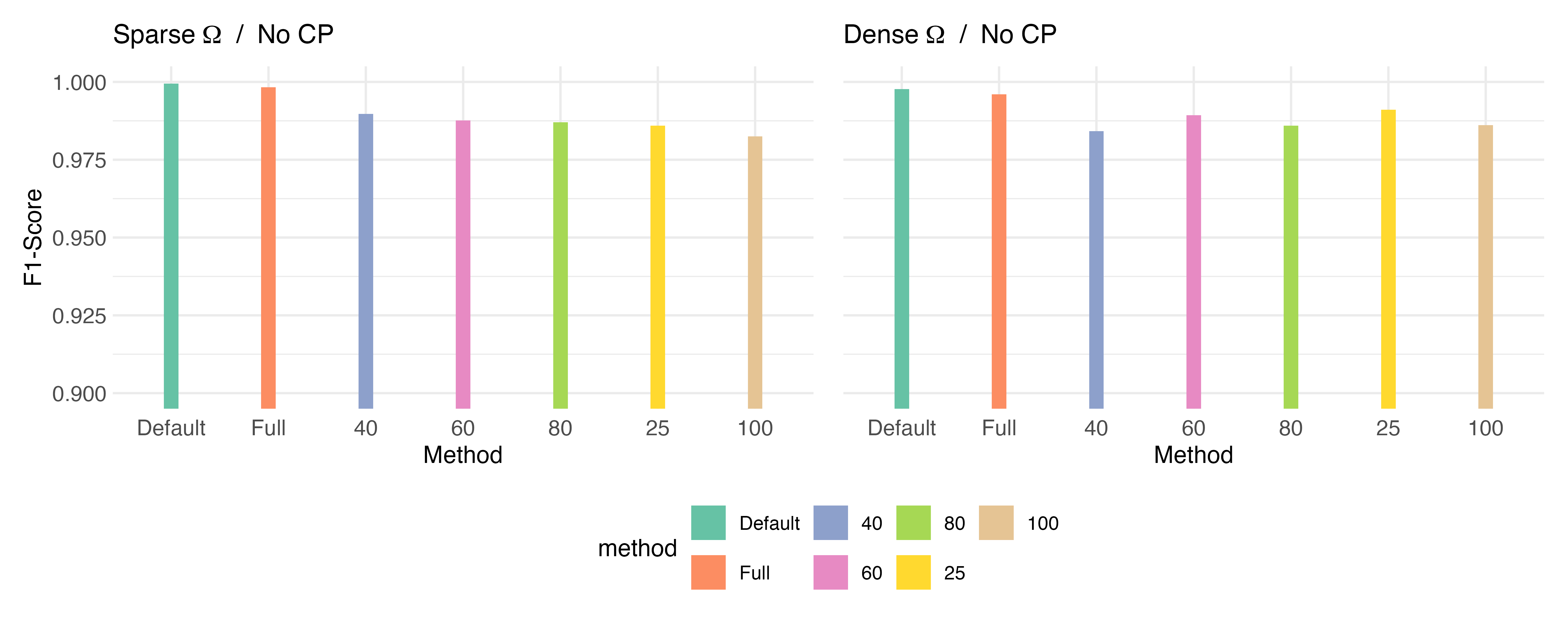}
				\caption{F1 scores for change point detection in mean structure are based on 50 simulated datasets for \( H_0 \) with \( p = 200 \). The methods \texttt{Full} and \texttt{Default} correspond to \texttt{mxPBF\_major} with sets of window sizes \(\mathcal{N}_F\) and \(\mathcal{N}_D\), respectively, while the methods denoted by numbers represent the proposed methods using a single window, with the numbers indicating the corresponding window sizes.}
				\label{fig:Fscore_single_null_mean}
			\end{figure}
			Figure \ref{fig:Fscore_single_null_mean} shows the F1 scores for each method, based on 50 simulated datasets under $H_0$ with $p = 200$. 
			Among the single-window methods, larger windows tend to detect more false positives.
			Both the Default and Full perform well in this regard, outperforming the single-window methods.

			\begin{figure}[!tb]
				\centering
				\includegraphics[width=0.95\textwidth]{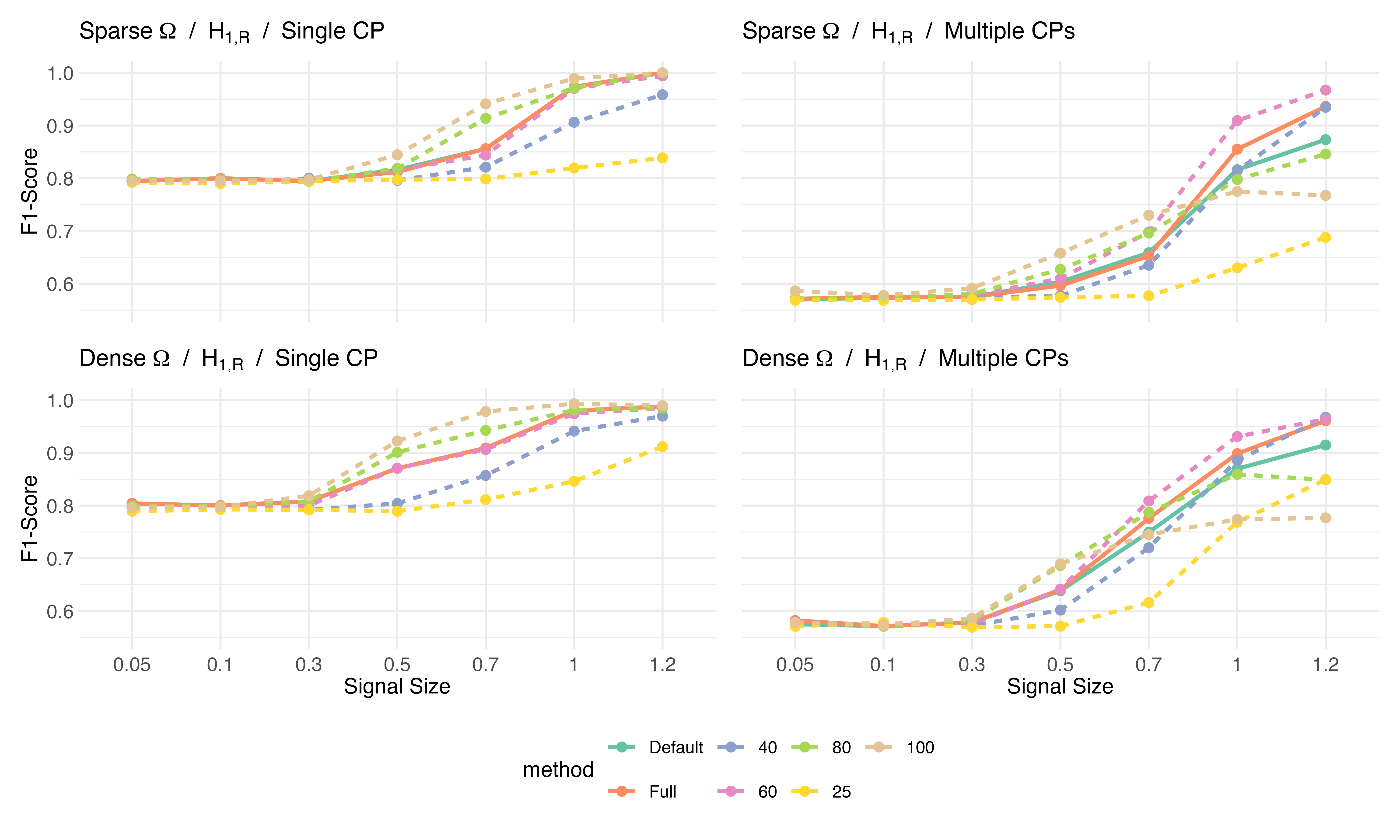}
				\caption{F1 scores for change point detection in mean structure  based on 50 simulated datasets under \( H_{1,R} \) with \( p = 200 \). Full and Default are prominently shown with solid lines, while single-window methods are depicted with dashed lines.}
				\label{fig:Fscore_single_H1R_mean}
			\end{figure}
			
			Figure \ref{fig:Fscore_single_H1R_mean} presents the F1 scores based on 50 simulated datasets under ``rare signals'' scenario $H_{1,R}$, with $p = 200$.
			Within the single-window methods, window size 25 demonstrates inferior performance in most scenarios, indicating that it is insufficient for capturing signals.  
			Additionally, window size 100 struggles to identify all changes in scenarios with multiple change points, as it fails to detect subsequent change points.
			In this regard, it can be said that window size 25 is suboptimal in most scenarios, particularly those with weak signals, while window size 100 is unsuitable for scenarios with multiple change points.
			Meanwhile, window sizes 40, 60, and 80 demonstrate decent overall performance, making them suitable options.
			Thus, Default and Full can be viewed as using one inappropriate window in single change scenarios and two in multiple change scenarios out of three and five windows, respectively.
			However, in most scenarios, both Full and Default demonstrate comparable performance to the appropriate window sizes.  
			Additionally, Full slightly outperforms Default in multiple change point scenarios.

			\begin{figure}[!tb]
				\centering
				\includegraphics[width=0.95\textwidth]{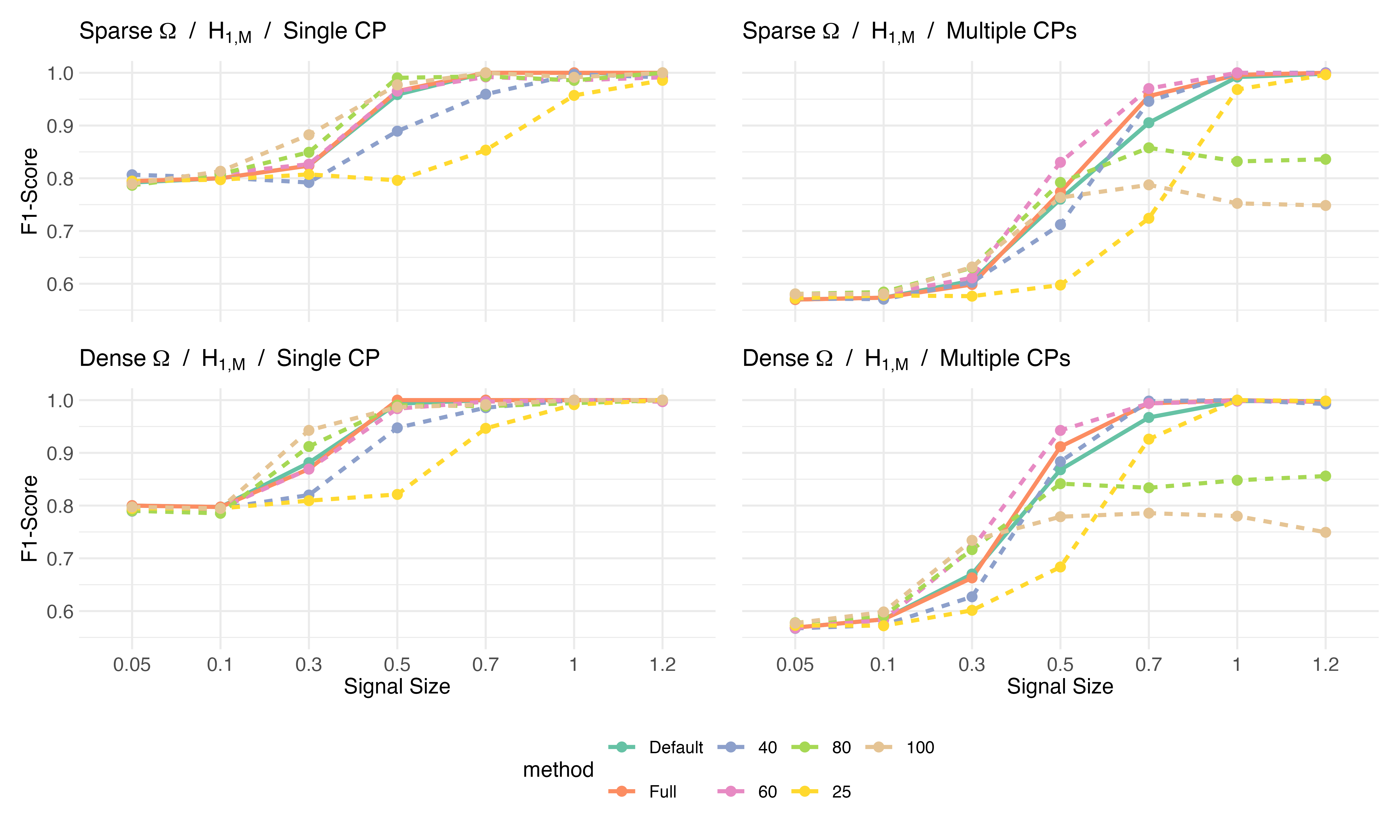}
				\caption{F1 scores for change point detection in mean structure  based on 50 simulated datasets under \( H_{1,M} \) with \( p = 200 \). Full and Default methods are illustrated using solid lines, while dashed lines represent single-window methods.}
				\label{fig:Fscore_single_H1M_mean}
			\end{figure}
			
			Figure \ref{fig:Fscore_single_H1M_mean} presents the F1 scores based on 50 simulated datasets ``many signals'' scenario $H_{1,M}$, with $p = 200$.
			In this setting, window size 25 does not appear to be suboptimal for \(\mu \geq 1\).  
			Additionally, in multiple change point scenarios, window sizes 80 and 100 struggle to identify all changes.  
			This suggests that the optimal window size depends on both the number and magnitude of the signals.
			In these scenarios, Full uses one inappropriate window in single-change scenarios where $\mu < 1$, two in multiple-change scenarios where $\mu \geq 1$, and three in multiple-change scenarios where $\mu < 1$, out of five windows.
			Similarly, Default uses one inappropriate window in both single-change scenarios where $\mu < 1$ and multiple-change scenarios where $\mu \geq 1$, and two in multiple-change scenarios where $\mu < 1$, out of three windows.
			However, similar to the ``rare signals" scenarios, both Full and Default show robust performance, with Full performing slightly better in multiple change point scenarios.

			\subsection{Performance of single window methods in the covariance structure}
			Next, we compare the performance in detecting changes in the covariance structure.
			We consider signal sizes of \(\psi \in \{0.5, 1, 3, 5, 8, 12, 15\}\).
			\begin{figure}[!bt]
				\centering
				\includegraphics[width=0.95\textwidth]{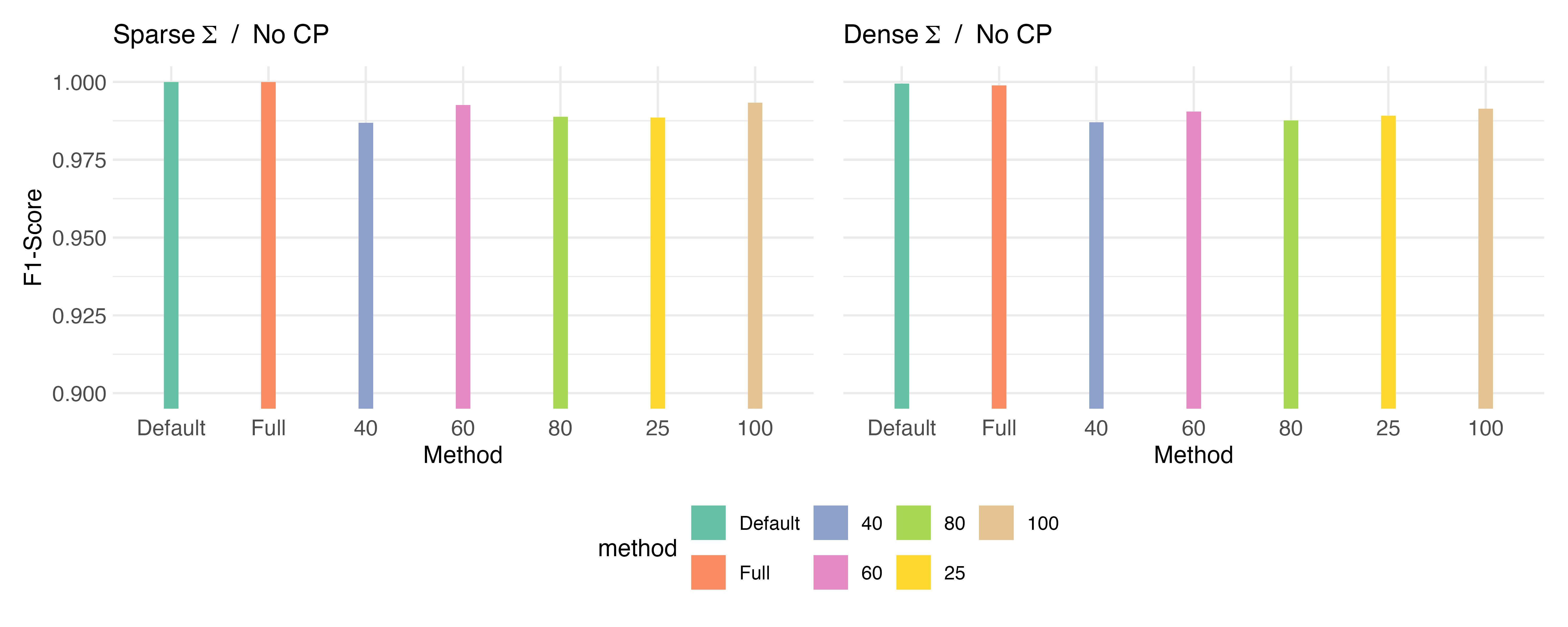}
				\caption{F1 scores for change point detection in covariance structure based on 50 simulated datasets for $H_0$ with $p = 200$.}
				\label{fig:Fscore_single_null_cov}
			\end{figure}
			Figure \ref{fig:Fscore_single_null_cov} shows the F1 scores for each method, based on 50 simulated datasets under $H_0$ with $p = 200$. 
			Similar to the results in mean change scenarios, both the Default and Full models outperform the single-window methods.

			\begin{figure}[!bt]
				\centering
				\includegraphics[width=0.95\textwidth]{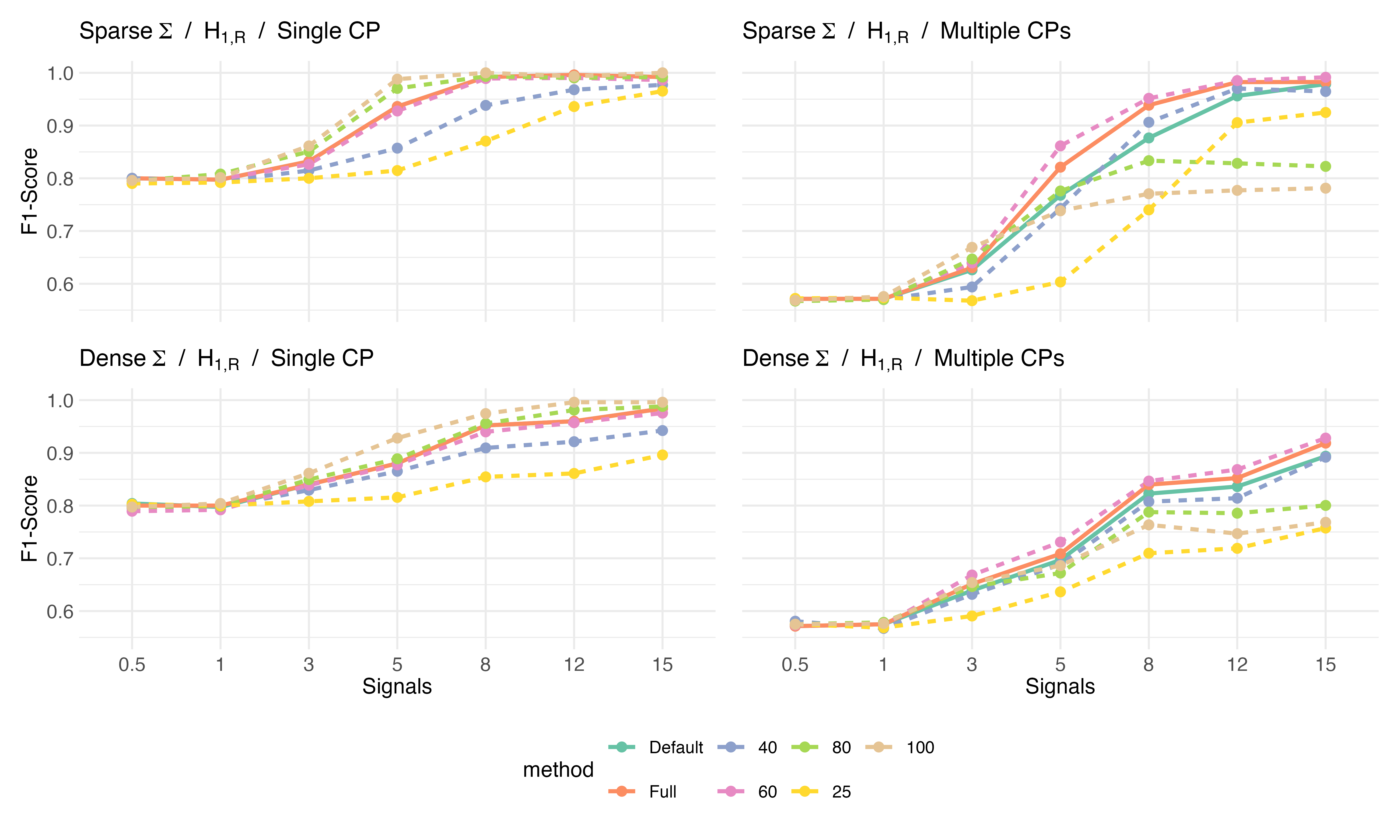}
				\caption{F1 scores for change point detection in covariance structure based on 50 simulated datasets under \( H_{1,R} \) with \( p = 200 \). Dashed lines depict single-window methods, while solid lines emphasize the Full and Default methods.}
				\label{fig:Fscore_single_H1R_cov}
			\end{figure}
			Figure \ref{fig:Fscore_single_H1R_cov} presents the F1 scores based on 50 simulated datasets under $H_{1,R}$, with $p = 200$.
			Consistent with the findings from previous results, Full and Default models demonstrate robust performance in most settings.
			Window size 25 appears suboptimal in most scenarios, particularly those with weak signals, while window sizes 80 and 100 are unsuitable for multiple change scenarios.

			\begin{figure}[!bt]
				\centering
				\includegraphics[width=0.95\textwidth]{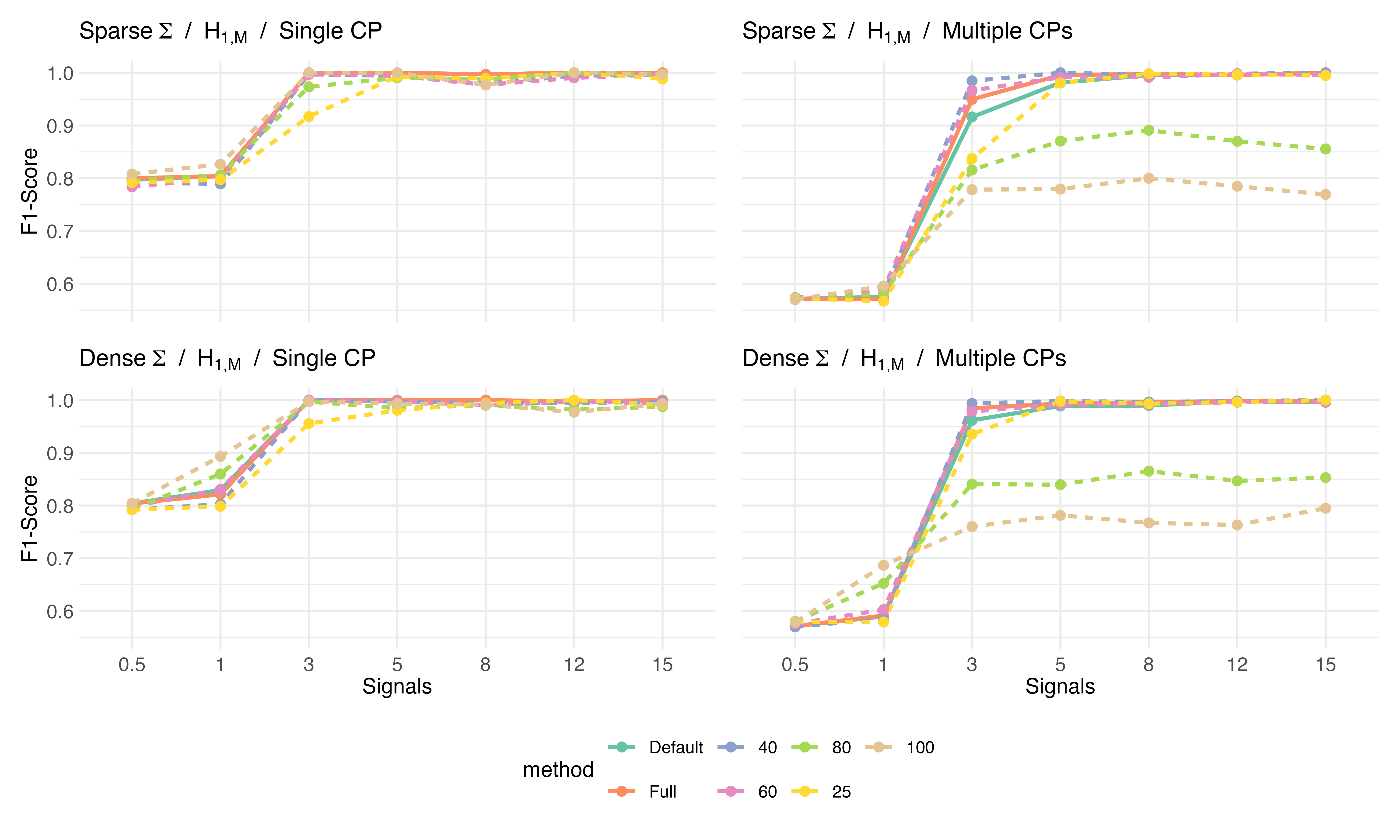}
				\caption{F1 scores for change point detection in covariance structure  based on 50 simulated datasets under \( H_{1,M} \) with \( p = 200 \). Solid lines highlight the Full and Default methods, whereas dashed lines denote the single-window approaches.}
				\label{fig:Fscore_single_H1M_cov}
			\end{figure}
			Figure \ref{fig:Fscore_single_H1M_cov} shows the F1 scores based on 50 simulated datasets under $H_{1,M}$, with $p = 200$.
			Full and Default show stable performance in most settings.
			In line with the previous results, window size 25 does not seem to be suboptimal for \(\psi \geq 5\), whereas window sizes 80 and 100 struggle to identify all changes in multiple change point scenarios.

			\section{Proofs of main results}\label{sec:proof}
			
			\subsection{Proofs of theorems in Section 2}
			
			\begin{proof}[Proof of Theorem 2.1]
				Suppose $H_{0}^{\mu}$ is true.
				Then,
				\bea
				&& \bbP_0 \left( \log B_{\max, 10}^{\mu, n_w} (\bfX_n)   \le - \big( \frac{\alpha}{2} - C_0' C \big) \log(n_w \vee p)  \right) \\
				&=& 1- \bbP_0 \left( \log B_{\max, 10}^{\mu, n_w} (\bfX_n)   > - \big( \frac{\alpha}{2} - C_0' C \big) \log(n_w \vee p)  \right) \\
				&\ge& 1- \sum_{l=n_w+1}^{n-n_w+1} \bbP_0 \left( \log B_{\max, 10}^{\mu, l, n_w} (\bfX_n)   > - \big( \frac{\alpha}{2} - C_0' C \big) \log(n_w \vee p)  \right) \\
				&\ge& 1- \sum_{l=n_w+1}^{n-n_w+1} \sum_{j=1}^p  \bbP_0 \left( \log  B_{10}^{\mu}( \bfX_{ (l-n_w):(l+n_w-1), j}  )     > - \big( \frac{\alpha}{2} - C_0' C \big) \log(n_w \vee p)  \right) \\
				&\ge& 1 - 2 (n-2n_w +1) p (n_w \vee p)^{-C}
				\eea
				for some constant $C_{2, \rm low}<C<C_2$ and $C_0' = [1 +2\{C_2 \log (n_w \vee p) \}^{-1/2}] / (1- 3\sqrt{C_2 \epsilon_{n_w}} )$, where the last inequality holds by the proof of Theorem 2.1 in \cite{lee2024bayesian} (Supplementary material, p.11).
				Since $(n-2n_w +1)p (n_w \vee p)^{-C_{2, \rm low}} = o(1)$, it completes the proof under the null $H_{0}^{\mu}$.

				Suppose $H_{1}^{\mu}$ is true, then there exist $K_0\ge 1$ change points $\{i_k\}_{k=1}^{K_0}$.
				Let the indices $k$ and  $j$ satisfy conditions (8) and $(i_{k}-i_{k-1})\wedge (i_{k+1}-i_{k}) \ge n_w$.
				Then, again by the proof of Theorem 2.1 in \cite{lee2024bayesian} (Supplementary material, p.12),
				\bea
				&& \bbP_0 \left( \log B_{\max, 10}^{\mu, n_w} (\bfX_n)   \ge  \frac{\alpha}{2}(C_1 -1) \log(n_w \vee p)  \right) \\
				&\ge& \bbP_0 \left( \log B_{\max, 10}^{\mu, i_k ,n_w} (\bfX_n)   \ge  \frac{\alpha}{2}(C_1 -1) \log(n_w \vee p)  \right) \\
				&\ge& 1 - 2 (n_w \vee p)^{-C}
				\eea
				for some constant $0<C<C_1$.
				It gives the desired result.	
				
			\end{proof}

			\begin{proof}[Proof of Theorem 2.2]
				Define 
				\bea
				\calS_0 &:=& \Big\{ n_w+1 \le l \le n-n_w +1 : |l-i_k|> n_w \text{ for all } 1\le k \le K_0   \Big\}, \\
				\calS_k &:=& \Big\{  i_k - n_w\le l \le i_k   \Big\}  \,\, \text{ for } k=1,\ldots,K_0,
				\eea
				where $\{i_k \}_{k=1}^{K_0}$ are the true change points.
				Then if 
				\bea
				\max_{l \in \calS_0}\log B_{\max,10}^{\mu, l,n_w}(\bfX_n) &\le& C_{\rm cp}
				\eea 
				and 
				\bea
				\max_{l\in \calS_k} \log B_{\max,10}^{\mu, l,n_w}(\bfX_n) &>& C_{\rm cp}
				\eea 
				for all $1\le k \le K_0$, it implies that $\what{K} = K_0$ and $\max_{1\le k \le K_0} |i_k - \what{i}_k| \le n_w$ by the definitions of $\what{i}_k$.
				Thus, the proof is completed if we show that 
				\bea
				\bbP_0  \left( \max_{l \in \calS_0}\log B_{\max,10}^{\mu, l,n_w}(\bfX_n) > C_{\rm cp} \,\text{ or }  \min_{1 \le k\le K_0} \max_{l\in \calS_k} \log B_{\max,10}^{\mu, l,n_w}(\bfX_n) < C_{\rm cp}  \right)  &\lra& 0
				\eea
				as $n\to\infty$.
				
				By the proof of Theorem 2.1,
				\bea
				\bbP_0 \Big( \max_{l \in \calS_0}\log B_{\max,10}^{\mu, l,n_w}(\bfX_n) > C_{\rm cp} \Big) 
				&\le& \sum_{l \in \calS_0} \bbP_0 \Big( \log B_{\max,10}^{\mu, l,n_w}(\bfX_n) > C_{\rm cp} \Big) \\
				&\le&  2 (n-2n_w+1) p (n_w \vee p)^{-C'}
				\eea
				for some constant $C_{2, \rm low}<C'<C_2$, and
				\bea
				&& \bbP_0 \Big( \min_{1 \le k\le K_0} \max_{l\in \calS_k} \log B_{\max,10}^{\mu, l,n_w}(\bfX_n) < C_{\rm cp} \Big) \\
				&=& \bbP_0 \Big( \max_{l\in \calS_k} \log B_{\max,10}^{\mu, l,n_w}(\bfX_n) < C_{\rm cp} \,\,\text{ for some } 1 \le k\le K_0 \Big) \\
				&\le& \sum_{1 \le k\le K_0}\bbP_0 \Big( \max_{l\in \calS_k} \log B_{\max,10}^{\mu, l,n_w}(\bfX_n) < C_{\rm cp}  \Big)  \\
				&\le& \sum_{1 \le k\le K_0}  \bbP_0 \Big( \log B_{\max,10}^{\mu, i_k,n_w}(\bfX_n) < C_{\rm cp}  \Big)  \\
				&\le& 2 K_0  (n_w \vee p)^{-C} 
				\eea
				for some constant $1<C< C_1$.
				Since we assume that $(n-2n_w+1) p (n_w \vee p)^{-C_{2, \rm low}} = o(1)$ and $K_0 (n_w \vee p)^{-C} = o(1)$ for some constant $1<C< C_1$, it completes the proof.	
			\end{proof}

			\begin{proof}[Proof of Theorem 2.3]
				We closely follow the proof of Proposition 3 in \cite{wang2018high}.
				Suppose $1/\delta_n \le \psi_{\min}^2 \le 1$. 
				Let  $P_1$ and $P_2$ correspond to $\{N_p(\mu_i^{(1)}, I_p)\}_{i=1}^n$ and $\{N_p(\mu_2^{(i)}, I_p)\}_{i=1}^n$, respectively,  $\mu_{i,j}^{(1)}  = \psi_{\min} I(j=1, i < i_1^{(1)})$ and $\mu_{i,j}^{(2)}  = \psi_{\min} I(j=1, i < i_1^{(2) } =i_1^{(1)}+\Delta)$ and $\Delta = \lfloor 1/\psi_{\min}^2 \rfloor$.
				Then, $P_1, P_2\in \calP_1^*$ and we have 
				\bea
				\inf_{\what{i}_1} \sup_{P \in \calP_1^*}  \frac{1}{n} \bbE_P |\what{i}_1- i_1| 
				&\ge& \inf_{\what{i}_1} \max_{P\in \{P_1, P_2\} }  \frac{1}{n} \bbE_P |\what{i}_1- i_1| .
				\eea
				Thus, if we show $ \inf_{\what{i}_1} \max_{P\in \{P_1, P_2\} }  \frac{1}{n} \bbE_P |\what{i}_1- i_1|  \ge 1/(16n \psi_{\min}^2)$, it will complete the proof.
				
				Note that 
				\bean
				\max_{P\in \{P_1, P_2\} }  \frac{1}{n} \bbE_P |\what{i}_1- i_1| 
				&=& \frac{\Delta}{n} \max_{P\in \{P_1, P_2\} }  \bbE_P \Big(\frac{|\what{i}_1- i_1| }{\Delta} \Big) \nonumber \\
				&=& \frac{1}{n} \max_{P\in \{P_1, P_2\} } \sum_{x \ge 1 }  x P \Big(  |\what{i}_1- i_1|    = x  \Big) \nonumber \\
				&\ge& \frac{1}{n} \max_{P\in \{P_1, P_2\} } \sum_{x \ge  \Delta /2 }  x P \Big(  |\what{i}_1- i_1|    = x  \Big) \nonumber \\
				&\ge&  \frac{\Delta}{2n} \max_{P\in \{P_1, P_2\} }  P \Big( |\what{i}_1- i_1| \ge \frac{\Delta}{2}  \Big)  \nonumber \\
				&=& \frac{\Delta}{n} \max \bigg\{  P_1  \Big( |\what{i}_1- i_1^{(1)}| \ge \frac{\Delta}{2}  \Big) , P_2  \Big( |\what{i}_1- i_1^{(2)}| \ge \frac{\Delta}{2}  \Big)    \bigg\} \nonumber  \\
				&\ge& \frac{\Delta}{2n} \max \bigg\{P_1  \Big( \what{i}_1 \ge i_1^{(1)} + \frac{\Delta}{2}  \Big)  ,  P_2  \Big( \what{i}_1 \le i_1^{(1)} + \frac{\Delta}{2}  \Big)  \bigg\}   \nonumber\\
				&\ge& \frac{\Delta}{2n} \frac{1 - d_{TV}(P_1 ,P_2) }{2}  , \label{lb_using_TV}
				\eean
				where $d_{TV}(P_1,P_2)$ is the total variation between $P_1$ and $P_2$.
				The last inequality holds because
				\bea
				&& 1 - 2 \max \bigg\{ P_1\bigg(\what{i}_1 \ge i_1^{(1)} + \frac{\Delta}{2}  \bigg) , P_2\bigg(\what{i}_1 \le i_1^{(1)} + \frac{\Delta}{2}  \bigg) \bigg\}  \\
				&\le& 1- P_1 \bigg(\what{i}_1 \ge i_1^{(1)} + \frac{\Delta}{2}  \bigg) - P_2\bigg(\what{i}_1 \le i_1^{(1)} + \frac{\Delta}{2}  \bigg) \\
				&\le& P_2\bigg(\what{i}_1 > i_1^{(1)} + \frac{\Delta}{2}  \bigg) - P_1 \bigg(\what{i}_1 > i_1^{(1)} + \frac{\Delta}{2}  \bigg)  \,\,\le\,\, d_{TV}(P_1, P_2) .
				\eea
				Furthermore, for the Kullback-Leibler divergence $KL(P_1\mid P_2) = \bbE_{P_1} \big(  \log \frac{dP_1}{dP_2} \big)$,
				\bean\label{TV_to_KL}
				d_{TV}^2(P_1,P_2) &\le& \frac{1}{2} KL(P_1 \mid P_2) \,\,=\,\, \frac{1}{4} \sum_{i=1}^n \|\mu_i^{(1)} - \mu_i^{(2)}\|_2^2  \,\,=\,\, \frac{\Delta}{4} \psi_{\min}^2 ,
				\eean
				where the inequality follows from \cite{pollard2002user} (p. 62) and the first equality holds due to  the Kullback-Leibler divergence  between two multivariate normal distributions.
				
				By applying (18) to (17), we have 
				\bea
				\inf_{\what{i}_1} \max_{P\in \{P_1, P_2\} }  \frac{1}{n} \bbE_P |\what{i}_1- i_1|  
				&\ge& \frac{\Delta}{4n} \Big( 1 - \frac{ \sqrt{\Delta} }{2} \psi_{\min} \Big) \\
				&\ge& \frac{1}{8n} \lfloor 1/\psi_{\min}^2 \rfloor 
				\,\,\ge\,\, \frac{1}{16n\psi_{\min}^2}    ,
				\eea
				which is the desired result.
			\end{proof}

			\subsection{Proofs of theorems in Section 3}
			\begin{proof}[Proof of Theorem 3.1]
				Suppose $H_{0}^{\sg}$ is true.
				Then, 
				\bea
				&&\bbP_0 \left( \log B_{\max,10}^{\sg, n_w}(\bfX_n)  < - \frac{1}{2}\big(\alpha - 6 C_3   + o(1)\big) \log(n_w\vee p)  \right)  \\
				&=& 1 -\bbP_0 \left( \log B_{\max,10}^{\sg, n_w}(\bfX_n)  > - \frac{1}{2}\big(\alpha - 6 C_3   + o(1)\big) \log(n_w\vee p)  \right)  \\
				&=& 1 -\bbP_0 \left( \max_{n_w+1 \le l \le n-n_w+1 }\log B_{\max,10}^{\sg, l,n_w}(\bfX_n)  > - \frac{1}{2}\big(\alpha - 6 C_3   + o(1)\big) \log(n_w\vee p)  \right)  \\
				&\ge& 1 - \sum_{l=n_w+1}^{n-n_w+1 }\bbP_0 \left( \log B_{\max,10}^{\sg, l,n_w}(\bfX_n)  > - \frac{1}{2}\big(\alpha - 6 C_3   + o(1)\big) \log(n_w\vee p)  \right)  \\
				&\ge& 1 - 6 (n- 2n_w + 1) p^2 (n_w \vee p)^{-C}  
				\eea
				for some constant $C$ such that $C_{3, \rm low} < C< C_3$, 
				where the last inequality follows from the proof of Theorem 3.1 in \cite{lee2024bayesian} (Supplementary material, p.19-20).
				The last term converges to 1 because the assumption $(n-2n_w +1) p^2 (n_w \vee p)^{-C_{3,\rm low}} = o(1)$.
				It completes the proof under the null $H_{0}^{\sg}$ because we assume $\alpha > 6C_3$.
				
				Suppose $H_{1}^{\sg}$ is true, then there exist $K_0\ge 1$ change points $\{i_k\}_{k=1}^{K_0}$.
				Let the index $k$ and pair $(i,j)$ satisfy $(i_{k}-i_{k-1})\vee (i_{k+1}-i_{k}) \ge n_w$ and condition (A3) or (A3$^\star$).
				Then, we have 
				\bea
				&& \bbP_0 \Big(  \log B_{\max,10}^{\sg, n_w}(\bfX_n) \ge   C_\alpha \log (n_w \vee p)     \Big) \\
				&=& \bbP_0 \Big(  \max_{n_w +1\le l \le n-n_w+1 }\log B_{\max,10}^{\sg, l,n_w}(\bfX_n) \ge  C_\alpha \log (n_w \vee p)     \Big)  \\
				&\ge& \bbP_0 \Big(  \log B_{\max,10}^{\sg, i_k ,n_w}(\bfX_n) \ge  C_\alpha \log (n_w \vee p)     \Big)  \\
				&\ge& 1 - 14(n_w\vee p)^{-C}
				\eea
				for some constant $0<C<C_1$, where $C_\alpha = C_{\rm bm}^2/8 - \alpha -1 + o(1)$ if condition (A3) is met, or $C_\alpha = C_{{\rm bm},a} - a_0- \alpha -1$ if condition (A3$^\star$) is met.
				Note that the last inequality follows from the proof of Theorem 3.1 in \cite{lee2024bayesian} (Supplementary material, p.21-22 and p.24).
				It gives the desired result.
			\end{proof}
			
			\begin{proof}[Proof of Theorem 3.2]
				Let $\{i_k\}_{k=1}^{K_0}$ be the true change points, $\calS_0 := \{ n_w+1\le l\le n-n_w+1: |l - i_k| > n_w \text{ for all } 1\le k \le K_0 \,\}$ and $\calS_k := \{  i_k -n_w \le l \le i_k \}$ for $k=1,\ldots,K_0$.
				Then if 
				\bea
				\max_{l \in \calS_0}\log B_{\max,10}^{\sg,l,n_w}(\bfX_n) &\le& C_{\rm cp}
				\eea 
				and 
				\bea
				\max_{l\in \calS_k} \log B_{\max,10}^{\sg, l,n_w}(\bfX_n) &>& C_{\rm cp}
				\eea 
				for all $1\le k \le K_0$, it implies that $\what{K} = K_0$ and $\max_{1\le k \le K_0} |i_k - \what{i}_k| \le n_w$.
				Thus, it suffices to show that
				\bea
				\bbP_0 \Big( \max_{l \in \calS_0}\log B_{\max,10}^{\sg, l,n_w}(\bfX_n) > C_{\rm cp}  \text{ or } \min_{1\le k \le K_0}\max_{l\in \calS_k} \log B_{\max,10}^{\sg, l,n_w}(\bfX_n) < C_{\rm cp} \Big) 
				&\lra& 0
				\eea
				as $n\to\infty$.
				
				Note that
				\bea
				\bbP_0 \Big( \max_{l \in \calS_0}\log B_{\max,10}^{\sg, l,n_w}(\bfX_n) > C_{\rm cp} \Big)
				&\le& \sum_{l \in \calS_0}  \bbP_0 \Big( \log B_{\max,10}^{\sg, l,n_w}(\bfX_n) > C_{\rm cp} \Big) \\
				&\le& 6 (n-2n_w+1) p^2 (n_w \vee p)^{-C'} \,\,=\,\, o(1)
				\eea
				for some constant $C'>0$ such that $C_{3,{\rm low}}<C'<C_3$, by the proof of Theorem 3.1
				Furthermore,
				\bea
				\bbP_0 \Big( \min_{1\le k \le K_0}\max_{l\in \calS_k} \log B_{\max,10}^{\sg, l,n_w}(\bfX_n) < C_{\rm cp} \Big) 
				&=&  \bbP_0 \Big( \max_{l\in \calS_k} \log B_{\max,10}^{\sg, l,n_w}(\bfX_n) < C_{\rm cp} \text{ for some } 1\le k \le K_0  \Big) \\
				&\le& \sum_{1\le k \le K_0}\bbP_0 \Big( \max_{l\in \calS_k} \log B_{\max,10}^{\sg, l,n_w}(\bfX_n) < C_{\rm cp}   \Big)  \\
				&\le& \sum_{1\le k \le K_0}\bbP_0 \Big(  \log B_{\max,10}^{\sg, i_k,n_w}(\bfX_n) < C_{\rm cp}   \Big)  \\
				&\le& 14 K_0  (n_w \vee p)^{-C} \,\,=\,\, o(1)
				\eea
				for some constant $2<C<C_1$, by the proof of Theorem 3.1.
				It gives the desired result.
			\end{proof}

			\begin{proof}[Proof of Theorem 3.3]
				We follow closely the line of the proof of Lemma 3 in \cite{wang2021optimal}.
				Suppose that $C^*\log p / (2\tilde{\sigma}_{\rm diff}^2) = \delta_n \le n/3$ and $0<\tilde{\sigma}_{{\rm diff}}< 3/4$, for some constant $0 < C^*<1$.
				For a given $\sigma>0$ and $1\le a \le p$, let $\sg_a = \sigma^2 \{I_p + 2\tilde{\sigma}_{\rm diff} u_a u_a^T\}$, where $u_a=(u_{a,1},\ldots, u_{a,p})$ is a unit vector with $u_{a,a} = 1$.
				Let $P_{0,a}^n$ be the joint distribution of random samples $(X_1,\ldots,X_n)$, where
				\bea
				X_1,\ldots,X_{\delta_n} \overset{i.i.d.}{\sim} N_p(0, \sg_a) \,\,\text{ and }\,\, X_{\delta_n+1},\ldots,X_n \overset{i.i.d.}{\sim} N_p(0, \sigma^2I_p),
				\eea
				and $P_{1,a}^n$ be the joint distribution of random samples $(X_1,\ldots,X_n)$, where
				\bea
				X_1,\ldots,X_{n-\delta_n} \overset{i.i.d.}{\sim} N_p(0, \sigma^2I_p) \,\,\text{ and }\,\, X_{n-\delta_n+1},\ldots,X_n \overset{i.i.d.}{\sim} N_p(0, \sg_a).
				\eea
				Define mixture distributions $P_i^n = p^{-1} \sum_{a=1}^p P_{i,a}^n$ for $i=1$ and $2$.
				Since $\sg_a - \sigma^2 I_p = 2 \sigma^2 \tilde{\sigma}_{\rm diff} u_a u_a^T$ and 
				\bea
				\frac{(2\sigma^2\tilde{\sigma}_{\rm diff} )^2}{ \sigma^4 + (2\sigma^2\tilde{\sigma}_{\rm diff})^2 } &\ge& \tilde{\sigma}_{\rm diff}^2
				\eea
				provided that $\tilde{\sigma}_{{\rm diff}}^2< 3/4$, it implies $\calP := \{ P_0^n , P_1^n \} \subset \calG(\tilde{\sigma}_{\rm diff} , C_d)$.	
				If we denote $ch(P_i^n)$ as the change point in the joint distribution $P_i^n$, we have $ch(P_0^n) = \delta_n+1$, $ch(P_1^n) = n-\delta_n+1$ and $ch(P_1^n) - ch(P_0^n) = n - 2\delta_n \ge n/3$.
				By Le Cam's Lemma (See Lemma 1 of \cite{yu1997assouad}),
				\bea
				\inf_{\what{i} } \sup_{P \in \calG(\tilde{\sigma}_{{\rm diff}},\, C_d)} \bbE_{P} |\what{i} - i_1|
				&\ge& \inf_{\what{i} } \sup_{P \in \calP} \bbE_{P} |\what{i} - i_1|  \,\,\ge\,\, \frac{1}{2}\cdot\frac{n}{3}\,\Big\{ 1 - \frac{1}{2}\|P_0^n - P_1^n\|_1 \Big\} ,
				\eea
				where $\|P_0^n - P_1^n\|_1 = \int | p_0^n(x) - p_1^n(x)| dx$ and $p_i^n$ is the joint density function corresponding to $P_i^n$.
				If we show that $\|P_0^n - P_1^n\|_1 \le 2/\sqrt{5}$, it completes the proof.
				
				Consider a rescaled covariance $\widetilde{\sg}_a := I_p + 2 \tilde{\sigma}_{\rm diff} u_a u_a^T = \sg_a /\sigma^2$.
				Let $\widetilde{P}_{0,a}^n$ be the joint distribution of random samples $(X_1,\ldots,X_n)$, where
				\bea
				X_1,\ldots,X_{\delta_n} \overset{i.i.d.}{\sim} N_p(0, \widetilde{\sg}_a) \,\,\text{ and }\,\, X_{\delta_n+1},\ldots,X_n \overset{i.i.d.}{\sim} N_p(0, I_p),
				\eea
				and $\widetilde{P}_{1,a}^n$ be the joint distribution of random samples $(X_1,\ldots,X_n)$, where
				\bea
				X_1,\ldots,X_{n-\delta_n} \overset{i.i.d.}{\sim} N_p(0, I_p) \,\,\text{ and }\,\, X_{n-\delta_n+1},\ldots,X_n \overset{i.i.d.}{\sim} N_p(0, \widetilde{\sg}_a).
				\eea
				Define mixture distributions $\widetilde{P}_i^n = p^{-1} \sum_{a=1}^p \widetilde{P}_{i,a}^n$ for $i=1$ and $2$.
				Since $\|\widetilde{P}_0^n - \widetilde{P}_1^n\|_1 = \|P_0^n - P_1^n\|_1$, it suffices to show that 	$\|\widetilde{P}_0^n - \widetilde{P}_1^n\|_1 \le 2/\sqrt{5}$.

				Let $\widetilde{P}_0^{\delta_n}$ and $\widetilde{P}_{1,a}^{\delta_n}$ be the joint distributions of $X_1,\ldots, X_{\delta_n} \overset{i.i.d.}{\sim} N_p(0, I_p)$ and $X_1,\ldots, X_{\delta_n} \overset{i.i.d.}{\sim} N_p(0, \widetilde{\sg}_a)$, respectively.
				Define the mixture distribution $\widetilde{P}_1^{\delta_n} := p^{-1}\sum_{a=1}^p \widetilde{P}_{1,a}^{\delta_n}$.	
				Note that $\|\widetilde{P}_0^n - \widetilde{P}_1^n\|_1 = 2 \|\widetilde{P}_0^{\delta_n} - \widetilde{P}_1^{\delta_n}\|_1 \le 2 \sqrt{\chi^2( \widetilde{P}_1^{\delta_n}, \widetilde{P}_0^{\delta_n}) }$, where
				\bea
				\chi^2(\widetilde{P}_1^{\delta_n}, \widetilde{P}_0^{\delta_n})
				&:=& \bbE_{\widetilde{P}_0^{\delta_n}}\Big( \frac{d \widetilde{P}_1^{\delta_n}}{d \widetilde{P}_0^{\delta_n}} - 1 \Big)^2
				\,\,=\,\, \frac{1}{p^2} \sum_{a=1}^p\sum_{b=1}^p \bbE_{\widetilde{P}_0^{\delta_n}}\Big( \frac{d\widetilde{P}_{1,a}^{\delta_n}}{d\widetilde{P}_0^{\delta_n}}\frac{d\widetilde{P}_{1,b}^{\delta_n}}{d\widetilde{P}_0^{\delta_n}} \Big) - 1
				\eea
				and the inequality follows from the Jensen's inequality.	
				By Lemma 5.1 in \cite{berthet2013optimal}, we have
				\bea
				\bbE_{\widetilde{P}_0^{\delta_n}}\Big( \frac{d\widetilde{P}_{1,a}^{\delta_n}}{d\widetilde{P}_0^{\delta_n}}\frac{d\widetilde{P}_{1,b}^{\delta_n}}{d\widetilde{P}_0^{\delta_n}} \Big)
				&=& \Big\{ 1 - (2 \tilde{\sigma}_{\rm diff} u_a^T u_b)^2 \Big\}^{-\frac{\delta_n}{2}}.
				\eea
				Thus,
				\bea
				\frac{1}{p^2}\sum_{a=1}^p\sum_{b=1}^p \bbE_{\widetilde{P}_0^{\delta_n}}\Big( \frac{d\widetilde{P}_{1,a}^{\delta_n}}{d\widetilde{P}_0^{\delta_n}}\frac{d\widetilde{P}_{1,b}^{\delta_n}}{d\widetilde{P}_0^{\delta_n}} \Big)  -1
				&=& \frac{1}{p^2}\Big\{ p\big(1 - 4\tilde{\sigma}_{\rm diff}^2 \big)^{-\frac{\delta_n}{2}} + p(p-1)  \Big\} - 1  \\
				&=& \frac{1}{p}\big(1 - 4\tilde{\sigma}_{\rm diff}^2 \big)^{-\frac{\delta_n}{2}} - \frac{1}{p} \\
				&=& \frac{1}{p}\big(1 - 4\tilde{\sigma}_{\rm diff}^2 \big)^{-\frac{1}{4\tilde{\sigma}_{\rm diff}^2} \,2\delta_n \tilde{\sigma}_{\rm diff}^2 } - \frac{1}{p} \\
				&\le& \frac{6}{5p} e^{ 2\delta_n \tilde{\sigma}_{\rm diff}^2 } - \frac{1}{p} \\
				&=& \frac{6}{5}p^{C_d^*-1} - \frac{1}{p} \,\,\le\,\, \frac{1}{5}  ,
				\eea
				because we choose $2\delta_n \tilde{\sigma}_{\rm diff}^2 = C^*\log p$ for some constant $0<C_d^*<1$, where the first inequality holds for all large $p$.
				It implies that $\|\widetilde{P}_0^n - \widetilde{P}_1^n\|_1 \le 2/\sqrt{5}$, which gives the desired result.		
			\end{proof}

			\begin{proof}[Proof of Theorem 3.4]
				We follow closely the line of the proof of Lemma 4 in \cite{wang2021optimal}.
				Suppose that $C_d\log p /(2\tilde{\sigma}_{\rm diff}^2) \le \delta_n \le n/3$ and $C_d >2$.
				Similar to the proof of Theorem 3.3, let $P_0^n$ be the joint distribution of 
				\bea
				X_1,\ldots, X_{\delta_n} \overset{i.i.d.}{\sim} N_p(0, \sigma^2 I_p) \text{ and } X_{\delta_n+1},\ldots, X_n \overset{i.i.d}{\sim} N_p(0, \sg_a),
				\eea 
				and $P_1^n$ be the joint distribution of
				\bea
				X_1,\ldots, X_{\delta_n+ t} \overset{i.i.d.}{\sim} N_p(0, \sigma^2 I_p) \text{ and } X_{\delta_n+t+1},\ldots, X_n \overset{i.i.d}{\sim} N_p(0, \sg_a) 
				\eea 
				for some $0<t<n/3$.
				Then $\calQ := \{ P_0^n, P_1^n \} \subset \calG(\tilde{\sigma}_{\rm diff} , 3)$ and $ch(P_1^n) - ch(P_0^n) \ge t$.
				By Le Cam's Lemma, 
				\bea
				\inf_{\what{i} } \sup_{P \in \calG(\tilde{\sigma}_{{\rm diff}},\, C_d)} \bbE_{P}|\what{i} - i_1| 
				&\ge& \inf_{\what{i} } \sup_{P \in \calQ} \bbE_{P}|\what{i} - i_1| \,\,\ge\,\, \frac{1}{2}\cdot t \,\Big\{ 1 - \frac{1}{2}\|P_0^n - P_1^n\|_1 \Big\} .
				\eea
				By the same arguments used in the proof of Theorem 3.3, we have
				\bea
				\frac{1}{2}\|P_0^n - P_1^n\|_1 
				&=&  \| \widetilde{P}_0^{t} - \widetilde{P}_1^{t}\|_1,
				\eea
				where $\widetilde{P}_0^{t}$ and $\widetilde{P}_1^{t}$ are  joint distributions of $X_1,\ldots ,X_t \overset{i.i.d.}{\sim} N_p(0, I_p)$ and $X_1,\ldots ,X_t \overset{i.i.d.}{\sim} N_p(0, \widetilde{\sg}_a)$, respectively.
				Note that
				\bea
				\| \widetilde{P}_0^{t} - \widetilde{P}_1^{t}\|_1
				&\le& 2 \sqrt{ \chi^2 (\widetilde{P}_1^{t}, \widetilde{P}_0^{t}) } \\
				&\le& 2 \sqrt{(1- 4\tilde{\sigma}_{\rm diff}^2)^{-\frac{t}{2} } - 1} \\
				&\le& 8 \sqrt{ \tilde{\sigma}_{\rm diff}^2 t },
				\eea
				where the last inequality follows from Lemma 15 in \cite{wang2021optimal}.
				If we take $t = \tilde{\sigma}_{\rm diff}^{-2}/2^8$, it gives
				\bea
				\inf_{\what{i} } \sup_{P \in \calQ} \bbE_{P}|\what{i} - i_1| 
				&\ge& \frac{1}{4} \cdot \frac{\tilde{\sigma}_{\rm diff}^{-2}}{2^8} .
				\eea

			\end{proof}

			\bibliographystyle{dcu}
			\bibliography{references}
			
		\end{document}